# Nanomaterials for Quantum Information Science and Engineering

Adam Alfieri, Surendra B. Anantharaman, Huiqin Zhang, Deep Jariwala*

Electrical and Systems Engineering, University of Pennsylvania, Philadelphia, PA 19104, USA

*Corresponding Author: dmj@seas.upenn.edu

## Abstract

Quantum information science and engineering (QISE) which entails generation, control and manipulation of individual quantum mechanical states together with nanotechnology have dominated condensed matter physics and materials science research in the 21st century. Solid state devices for QISE have, to this point, predominantly been designed with bulk material as their constituents. In this review, we consider how nanomaterials or low-dimensional materials i.e. materials with intrinsic quantum confinement - may offer inherent advantages over conventional materials for QISE. We identify the materials challenges for specific types of qubits, and we identify how emerging nanomaterials may overcome these challenges. Challenges for and progress towards nanomaterials-based quantum devices are identified. We aim to help close the gap between the nanotechnology and quantum information communities and inspire research that will lead to next-generation quantum devices for scalable and practical quantum applications.

## Table of Contents

**1.0.0 Introduction**

### 1.1.0 *Quantum Information Science and Engineering*

Quantum information science and engineering (QISE) encompasses the use and dissemination of qubits – bits of information encoded in quantum, two-level physical systems with a finite energy splitting- for computation and measurement. For this review, we divide the broad field of QISE into three branches: (i) quantum computing, (ii) quantum communications, and (iii) quantum sensing. We start by introducing two concepts that are universal in QISE: qubits and entanglement.

While classical bits of information correspond to continuous variables in macroscopic physical systems, qubits are stored in the quantum state of a two-level physical system with a finite energy difference or any linear superposition of these states[1]. These quantum states can be labeled $|0\rangle$ and $|1\rangle$ (Fig. 1a). The Bloch sphere (Fig. 1b) is a useful representation of a single qubit. Any arbitrary quantum state in a two-dimensional Hilbert space, $|\psi\rangle = \alpha|0\rangle + \beta|1\rangle$ where $|\alpha|^2 + |\beta|^2 = 1$, can be represented as a unit vector on the Bloch sphere. In a general N-qubit quantum system, the state exists in a superposition of all possible permutations of the system: $|\psi\rangle = \sum_{n=|00...0\rangle}^{|11...1\rangle} c_n |n\rangle$ where $\sum_n |c_n|^2 = 1$. While qubit is a two-level quantum system, in principle it can be extended to qutrits (three-level system) and further to qudits (any d > 2) which are many level systems[2]. Due to its multilevel nature, a qudit provides a larger state space to store and process information. There have been limited physical demonstrations of qudits, so we focus on qubits in this review.

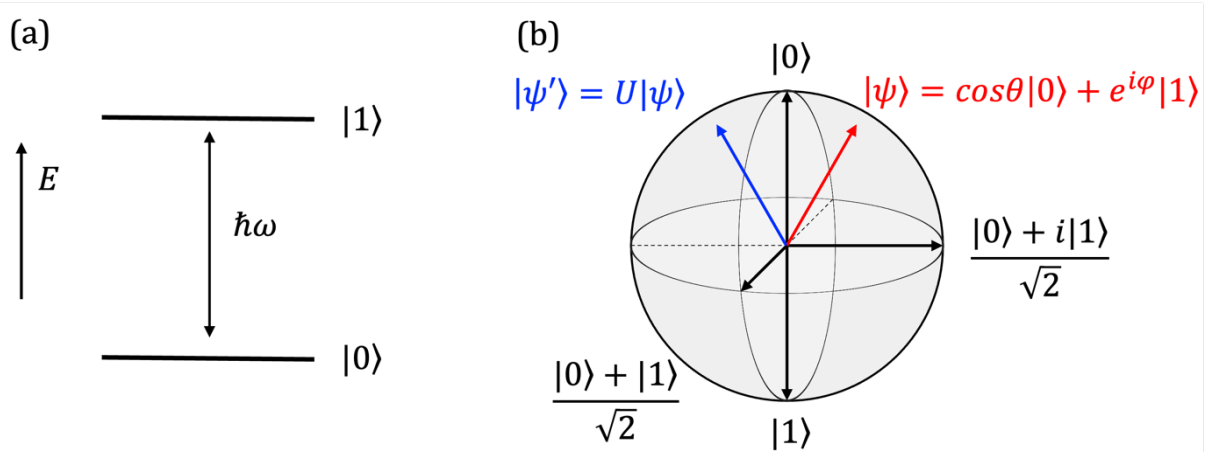

*Figure 1. An introduction to quantum information processing and comparison with classical computation.*
(a.) A qubit with an energy splitting $\hbar\omega$. (b.) The Bloch sphere representation of a single qubit. A single-qubit gate, $U$, rotates the qubit from state $|\psi\rangle$ (red) to $|\psi'\rangle$ (blue).

Quantum entanglement - a phenomenon that arises between states in an inseparable composite quantum system - is one of the core elements of quantum information processing[3]. Entanglement is possibly the most 'non-classical' tenet of quantum physics and can connect two or more states anywhere in space-time; it is impossible to act on one of these states without affecting the other(s). Entanglement can be used to accelerate quantum computing[4]; assist and secure the transmission of qubits[5]; and improve quantum sensing resolution[6,7].

*1.1.1 Quantum computing*

Quantum computing offers several potential advantages over classical computing:

- Classical algorithms cannot adequately model the dynamics of physical quantum systems[8], particularly systems with multi-partite entanglement and other highly "non-classical" behavior. Quantum computers are therefore an excellent platform to simulate physical systems that behave non-classically.
- Quantum computers enable factorization algorithms[9,10] and eigensolvers[11] for problems that would be too slow or otherwise intractable on a classical computer.
- Quantum computers have a larger computational space per bit: an N-bit classical system has a computational space of size N, while an N-qubit quantum system has a computational space of size $2^N$.

We have reached the beginning of the Noisy Intermediate-Scale Quantum (NISQ) technology era[12], with 50-100 qubit processors that are not fault tolerant. In 2019, "quantum supremacy," was first demonstrated when a 53 qubit computer based on superconducting qubits was used to sample the output of a pseudo-random quantum circuit: the quantum processor was shown to take 200 seconds to sample the output of the superconducting qubit circuits one million times to produce a set of random bit-strings with high fidelity (i.e. a high degree of randomness) - a task that would have taken a classical supercomputer 10,000 years with a similar degree of randomness[13]. This was a very specific task tailor-made for a quantum computer, but it still highlights both the progress made on quantum information processing and its future potential. While it is unlikely that quantum computation will replace conventional computation, it is indisputable that it will eventually be extremely useful as a complementary technology to solve specific problems more efficiently.

Quantum computation requires the use of quantum logic gates that act on qubits, analogous to classical logic gates. Unlike classical logic gates, which are Boolean operators, quantum logic gates are unitary operators that act on single or multiple qubits. Operations on single qubits can be represented as rotations on the Bloch sphere. When the system is driven at the Larmor frequency by applying an external field, the qubit oscillates between the two possible states at the Rabi frequency, $f_{Rabi}$. A state can therefore be rotated a controlled amount by applying a pulsed field at the Larmor frequency. The larger $f_{Rabi}$, the faster gate operations can be performed. A logic gate that acts on a $2^N$ Hilbert space is a unitary transformation that can be represented as a $2^N$ x $2^N$ matrix. A set of quantum gates is considered universal if any possible operation can be performed using some combination of the gates. For an arbitrarily large number of qubits, any operation can be performed using a combination of the 2-qubit controlled-NOT (CNOT) gate and single-qubit unitary operators.

The physical implementation of a quantum computer requires a scalable system of qubits that meet DiVincenzo's criteria[14]: (i) the state has discrete energy levels with some energy splitting (ii) the qubit can be initialized to a known state, (iii) the decoherence time of the qubit is longer than the time it takes to perform a quantum gate operation on it, (iv) a set of universal quantum gates exist for the qubit(s), (v) error correction is possible, and (vi) the state can be read-out.

Several metrics are used to evaluate the performance of qubits and the gates that operate on them. The relaxation time, $T_1$, is a measure of the qubit lifetime. For example, it is the time it takes for an excited state to relax to the ground state. The coherence time of a qubit, $T_2$, is the most important figure of merit for individual qubits. The coherence time is the duration for which a qubit "survives" in a pure quantum state. There are various kinds of $T_2$ measurements. The time it takes for a qubit to dephase is denoted $T_2^*$. This can be extended by the Hahn echo sequence and the Carr-Purcell-Meiboom-Gill (CPMG) sequence. We list the coherence time as the maximum coherence time unless otherwise specified. Quantum gates are characterized by their speed and fidelity. Gate speed is characterized by the time it takes to perform a certain gate operation and is limited by the Rabi frequency, $f_{Rabi}$, as mentioned above. It is desirable to maximize the number of gate operations that can be performed while a qubit is coherent (i.e. within $T_2$). Therefore, when considering different qubit types and platforms, it is the ratio of $T_2/T_{Gate}$ that should be focused on instead of the individual values. The fidelity of a quantum gate, $F$, is how accurately the gate projects the actual state $|\psi\rangle$ onto the target state $|\psi_0\rangle$: $F = |\langle\psi|\psi_0\rangle|^2$. In practice, quantum tomography is used to reconstruct the state using a series of measurements on an ensemble of qubits, allowing a density matrix $\rho = |\psi\rangle\langle\psi|$ to be measured. $F$ is extracted from comparing the measured density matrix with the density matrix of the target state $\rho_0 = |\psi_0\rangle\langle\psi_0|$:

$$F = (\text{tr}(\sqrt{\sqrt{\rho_0}\rho\sqrt{\rho_0}}))^2$$

Where tr is the trace of the matrix. Because $F$ is a measure of probability, its ideal value is 1.

The ultimate goal of quantum computing research is to develop a fault-tolerant system. Various sources of errors exist for quantum computers (and classical for that matter), notably coherent gate inaccuracies and decoherence due to the environment. While errors should be minimized, a non-zero error rate is inevitable. A fault-tolerant computer can withstand these mistakes if the error rate is below some threshold[15]. Analogous to the redundant coding concept in classical error correction, quantum error correction (QEC) codes have been developed that use a string of entangled physical qubits to encode one logic qubit[16–18]. Encoding of logical qubits has been experimentally demonstrated in superconducting circuits[19] and ion trap processors[20]. These logical qubits, of course, require gates that act on them in addition to the physical qubits that they are composed of. Circuits for error correction and gates acting on logical qubits become increasingly complex and are beyond the scope of this review. The interested reader is encouraged to read ref[21]. The complexity of these circuits and the importance of error correction should underscore the importance of fast, high fidelity quantum gate operations and long coherence times.

Quantum memories are a crucial component in QIP. In quantum computing, like classical computing, it is necessary to store data long enough for various operations and QEC[22]. Quantum memories must have long coherence times, and states must be able to be swapped into and read out of the memory with high fidelity. It is also ideal to use quantum memories that can be optically addressed so they can also serve as a spin-photon interface[23]. Error correction codes must also be implemented for the memory to prevent against decoherence and other errors[16].

The major platforms currently being explored for quantum computing are semiconductor quantum dots (QDs)[24], spin defects in wide-bandgap materials[25], superconducting qubits[26], topologically protected solid-state systems[27], Rydberg atoms[28], ion trap systems[29], and photonic quantum computing[30]. In this review, we introduce these qubit technologies (except for ion trap qubits and Rydberg atoms as these are not solid state and hence emerging low-dimensional materials do not apply) and discuss the current materials systems used to implement these qubits before discussing how low-dimensional materials may be used in them and what advantages or benefits do they offer.

*1.1.2 Quantum communications*

Quantum communications aims to transfer a quantum state between two spatially separated nodes with high fidelity and security. The quality of a quantum communications system is characterized by fidelity of the transferred state as compared to the initial state. Photons are widely accepted as the best platform for travelling qubits[31,32] as they interact minimally with their environment and can be transmitted through free space or existing fiber-optic infrastructure, covering long distances within their coherence times. Quantum states are commonly encoded in the photon polarization[33,34], but states can also be encoded in the orbital angular momentum[35,36] or – for energy-time entangled photons[37] – the time-bin or phase of the photon[38]. The ultimate goal of quantum communications is to realize a distributed quantum internet of spatially separated nodes, analogous to our existing internet and - potentially - utilizing its existing infrastructure. A quantum internet has two main advantages over the classical counterpart. First, the size of the state space for a network of $k$ quantum nodes, each with $n$ qubits, is $2^{nk}$ when connected by quantum channels, compared to $k2^n$ when connected by classical channels[31]. This enables greater computational power in the form of distributed quantum computing. The second advantage is security. One of the most widely used classical security protocols relies on the difficulty of factoring a large number into two prime numbers[39]. However, Shor demonstrated that this kind of factorization is tractable on a quantum computer[9], rendering this form of public key encryption defenseless against an attack by a quantum computer. In fact, no known classical security protocol is unconditionally secure against an attack by a quantum computer[1]. Thankfully, the field of quantum cryptography has emerged and grown alongside quantum computation, and protocols exist that are secure against attacks by quantum computers.

Quantum cryptography is a major focus of quantum communications and leverages the unique laws of quantum physics to secure information against eavesdroppers bound by those same laws. Quantum key distribution (QKD) is the most well-known and widely used quantum cryptographic scheme. The initial QKD protocol was first developed by Bennett and Brassard in 1984 (BB84)[40]. Modern implementations of QKD[41–43] commonly use a decoy-state protocol[44], which is a modified version of BB84 that overcomes a photon-splitting attack[45]. QKD is analogous to public key distribution, but information is encrypted and secured by basic principles of quantum mechanics: (i.) any measurement of a quantum system affects the system, (ii.) the no-cloning theorem[46] – a quantum state cannot be copied – prevents an eavesdropper from obtaining a perfect copy of a qubit, and (iii.) the uncertainty principle limits the accuracy at which conjugate continuous variables (e.g., position/momentum, energy/time, etc.) can be determined simultaneously. Consequently, an adversary cannot eavesdrop on a signal without being detected.

Entangled photon pairs (also called EPR pairs) created in one of four maximally entangled Bell states:

$$|\Psi^\pm\rangle = (|01\rangle \pm |10\rangle)/\sqrt{2} \text{ or } |\Phi^\pm\rangle = (|00\rangle \pm |11\rangle)/\sqrt{2}$$

are a resource for quantum communications and enable protocols that can aid in the transmission of a state or improve information density[1,3]. The quality of entanglement, or entanglement fidelity, is given by the overlap between the density matrix of the state, $\rho = |\psi\rangle\langle\psi|$, with the density matrix of some Bell state, $\rho_B = |B\rangle\langle B|$ where $|B\rangle = |\Psi^\pm\rangle$ or $|\Phi^\pm\rangle$:

$$F = (\text{tr}(\sqrt{\sqrt{\rho_B}\rho\sqrt{\rho_B}}))^2$$

A Bell state measurement (BSM) is a measurement performed on two entangled particles to determine which of the four Bell states the particles are in. The key hardware component in a BSM is a photodetector sensitive enough to detect a single photon. BSMs are a crucial part of any quantum communications protocol that uses entangled photon pairs. A reliable single photon detector is therefore a necessity.

Quantum teleportation, introduced by Bennett *et al.* in 1993[5], is a cornerstone of quantum communications and is particularly useful when direct transmission cannot preserve the coherence of a state. In a traditional teleportation scheme, Alice begins with particle 1 in some state, $|\psi\rangle = \alpha|0\rangle + \beta|1\rangle$, that she wishes to send to Bob. A pair of entangled states (particles 2 and 3) is created at a third location in the Bell state $|\Psi^-\rangle$, and particle 2 is sent to Alice while 3 is sent to Bob. Alice performs a BSM on particles 1 and 2. At this point, particle 3 can be in one of 3 states. Alice must now communicate the results of her BSM via a classical channel. Bob then performs a unitary operation on particle 3, bringing it into the initial state $|\psi\rangle$. Alice no longer maintains the initial state because particle 1 is in an entangled state. $|\psi\rangle$

has therefore disappeared from Alice and appeared at Bob, hence the name teleportation. In this process, no information is revealed about the initial state $|\psi\rangle$, and Alice does not maintain a copy of $|\psi\rangle$; the no-cloning theorem is therefore not violated. Because the process requires a classical channel for verification, information is not communicated faster than the speed of light allows. Teleportation was first experimentally demonstrated in 1997 by Bouwmeester *et al.* using polarization-entangled photons[47]. Since, teleportation has been demonstrated via optic fibers and free space connections at distances >100 km[48,49], and quantum teleportation by satellite has also been shown[50].

Photon loss and other sources of decoherence in optical fibers limit the scalability of quantum networks. Photon losses, and therefore error probabilities, increase exponentially with channel length[51]. Channel lengths significantly longer than the coherence length of a travelling qubit reduce the fidelity of state transfer to levels below those that can be purified. Due to the no-cloning theorem, a weak signal cannot be amplified. Instead, a device called a quantum repeater (QR) must be used. A QR is an intermediate node, B, inserted into a quantum channel (between A and C) that, in addition to effectively relaying information from node A to C, actively corrects for photon losses and other errors that deteriorate the state[52]. QRs use protocols based on either entanglement generation/purification[51,53,54] or QEC[55,56] to correct for photon losses and operational errors. The former requires a quantum memory while the latter does not. Both approaches require a few-qubit processor and a single photon source.

The practical realization of quantum communications clearly relies on the development of single photon sources, EPR pair sources, reliable quantum repeaters (with or without an optically addressable quantum memory), and single photon detectors.

### 1.1.3 Quantum sensing

Quantum states are often highly sensitive to their environment. For quantum computing and communications, this is a negative because it leads to decoherence, but this attribute can be exploited for sensors with better resolution than classical devices. Quantum sensors measure some physical quantity $V(t)$ that makes some contribution, $H_V(t)$, to the total Hamiltonian:

$$H_{tot}(t) = H_0 + H_{ctrl}(t) + H_V(t)$$

Where $H_0$ is the intrinsic Hamiltonian of the qubit and $H_{ctrl}(t)$ is the control Hamiltonian related to qubit manipulation. A simplified quantum sensing procedure initializes a qubit or ensemble of qubits in a known state, $|\psi_i\rangle = c_0|0\rangle + c_1|1\rangle$. $|\psi_i\rangle$ is then transformed via a unitary operation to a state convenient for a certain measurement: $|\psi'\rangle = U_1|\psi_i\rangle$. The state then evolves under the influence of $V(t)$ for some time $t$:

$$|\psi'(t)\rangle = U(V(t), t)|\psi'(t = 0)\rangle$$

$|\psi'(t)\rangle$ is then transformed into some superposition of observable states:

$$|\psi_f\rangle = U_2|\psi'\rangle = c'_0|0\rangle + c'_1|1\rangle$$

The state is then projectively read out as either $|0\rangle$ with probability $|\langle 0|\psi_f\rangle|^2$ or as $|1\rangle$ with probability $|\langle 1|\psi_f\rangle|^2$. Repeating this process, the state $|\psi_f\rangle$ is reconstructed. The probability of a transition, corresponding to the measurement of some physical parameter, is extracted from the overlap of the final state with the initial state[7].

The sensitivity is inversely proportional to the transduction parameter and the square root of the coherence time: $s \propto (\gamma\sqrt{T_2})^{-1}$. Therefore, like the aforementioned branches of QIP, minimizing decoherence is critical. Additionally, the transduction parameter must be maximized. Qubits generally best transduce the same kinds of fields that can be used to manipulate them: spin qubits are most sensitive to magnetic fields, charge qubits are most sensitive to electric fields, etc. For physical implementation, DiVincenzo's criteria related to discrete energy levels, state initialization, and state readout apply to quantum sensors.

Common quantum sensing platforms include atomic vapors[57], trapped ions[58], superconducting qubits[59], superconducting quantum interference devices (SQUIDs)[60], spin defects in solids[61], among others. Some of these technologies - such as SQUIDs and nuclear spin ensembles[62] - date back over 50 years, while others - namely spin defects in solids - are emerging. Photonic quantum sensing is another related, broad field that largely uses interference measurements between photons. The only components for photonic quantum sensing that are relevant to this review are quantum light sources and single photon detectors, which we discuss primarily in the context of quantum communications. Therefore, our discussion of quantum sensing in this paper focuses on the use of stationary qubits.

Entanglement can be utilized to achieve improved resolution in quantum metrology. An ensemble of $N$ unentangled states has a sensitivity that improves with $\sqrt{N}$, but ensembles can also introduce decoherence due to state-state interactions; an ensemble of $N$ entangled states improves the sensitivity by a factor of $N$, improving the resolution to below the shot noise limit and approaching the Heisenberg limit[7]. Some photonic quantum sensing techniques make use of photonic entanglement to enable resolution below the Rayleigh limit[6].

For quantum sensing, qubits that have both high transduction and long coherence times are needed. Additionally, the ability to operate entangled ensembles of states provides enhanced sensitivity. Finally, single-shot readout by an application-suitable means is crucial.

### *1.2.0 Nanomaterials*

Sources of decoherence are clearly undesirable when it comes to QIP. It is therefore imperative to have high purity materials with interfaces free of defects and trap states. In solid state quantum systems, it is somewhat intuitive to consider new materials with intrinsic quantum properties themselves.

We define nanomaterials as materials that have intrinsic quantum confinement in at least one dimension which is also reflected in its electronic structure and other physical properties as well. Materials are classified according to the number of dimensions along which they do not experience quantum confinement (i.e., a 1D material is quantum confined in two dimensions). In addition to inherent quantum confinement, one immediate benefit of nanomaterials is that the surfaces are intrinsically or extrinsically (via capping ligands) passivated, allowing for high purity interfaces. Below we consider emerging nanomaterials that are potentially useful for QIP.

### *1.2.1 0D materials*

Zero-dimensional semiconducting quantum dots (QDs) are confined in all dimensions and exhibit atomic-like optical transitions due to the reduced density of states, which discretizes the energy levels. The discrete energy levels are dependent on the size of the QD due to "particle in a box" effects, where the level spacing is inversely proportional to the dimensions of the system. The ultra-sharp and size-tunable optical properties make QDs particularly popular for photonic applications. For this section, we only consider intrinsically 0D materials, so we ignore lithographically defined dots and electrostatically defined QDs.

We divide QDs into two classes based on the synthetic method: epitaxially grown (also called self-assembled) quantum dots (eQDs) and colloidal quantum dots (cQDs). eQDs are most commonly II-VI and III-V compound semiconductor materials grown as epitaxial islands on a substrate by molecular beam epitaxy (MBE) or metal organic vapor phase epitaxy (MOVPE)[63]. The most common eQDs are based on III-V heterostructures such as GaAs/AlGaAs and InAs/GaAs[64]. cQDs, conversely, are solution-processed and capped with ligands. The most common cQDs are II-VI semiconductors and lead-halide perovskites.

III-V eQDs are grown to have band alignments that lead to the formation of quantum wells. The emission from eQDs is tunable by stoichiometry of alloys within the III-V system and quantum well dimensions. III-V eQDs are commonly used for diode lasers[65] and other optoelectronic devices[66,67].

cQDs have high quantum efficiencies, strong and tunable absorption, inexpensive and highly scalable processing, and emission properties that can be tuned by size[68,69]. II-VI semiconductors are often grown as core-shell nanoparticles, in which a II-VI semiconductor shell is grown around a quantum dot core composed of a different II-VI material. Tuning the band structure of these heterostructures allows control of the optoelectronic properties. These heterostructures can also be grown as nanoplatelets[70], which have a layered 2D structure but still have nanoscale lateral dimensions. Lead halide perovskites are another class of cQDs. This includes inorganic $CsPbX_3$ (X= Cl, Br, I) perovskites[69] and hybrid organic-inorganic perovskites (HOIP) such as formadinium lead halide perovskites ($FAPbX_3$) and methylammonium lead halide perovskites ($MAPbX_3$)[71]. Both II-VI cQDs and perovskites are exciting for light emission[72] – both lasing[73] and light emitting diodes (LEDs)[74]. They have also been explored for photovoltaics[75], photodetection, and photocatalysis[76].

*1.2.2 1D materials*

Semiconductor nanowires (NWs) and nanotubes are 1D structures. The 1D structure leads to highly anisotropic electronic, optical, and mechanical properties. Carbon nanotubes (CNTs) are one of the most well-known and most-studied nanomaterials. Semiconducting CNTs have found a wide range of applications including nanoelectronics[77], photonics[78], and sensors[79,80]. Semiconducting III-V nanowires grown by MBE and MOVPE exhibit high carrier mobilities and have been extensively studied for nanoelectronics[81], optoelectronics[82,83], and – as we will discuss - quantum applications. Group IV nanowires have similarly been heavily studied for electronics[84,85].

*1.2.3 2D materials*

Since the discovery of graphene[86], an atomically thin sheet of $sp^2$-hybridized carbon with outstanding electrical[87] and optical properties[88–90], the prediction, synthesis, and use of novel 2D materials has quickly become one of the most active fields in condensed matter and materials science research. In addition to graphene, other elemental 2D materials, often referred to as x-enes, have been studied, notably phosphorene (also known as black phosphorus, or BP), a direct bandgap semiconductor with a high carrier mobility and layer-dependent bandgap[91,92].

In 2010, monolayer MoS$_2$, a direct bandgap semiconductor in the monolayer limit, was isolated[93]. Transition metal dichalcogenides (TMDs) have been widely explored in the past decade due to their promise for electrical[94–97], optical[98–105], and electrochemical[106,107] applications. While 2H-phase, group VI TMDs (MX$_2$: M = Mo, W; X = S, Se, Te) are the prototypical TMDs and the most studied, the chalcogenides of transition metals from other groups and other classes of 2D material have also recently garnered interest for plasmonics[108], charge density waves[109,110], and superconductivity[111], among other properties/applications[105–107,112]. 2D (anti)ferromagnets[112–117], ferroelectrics[118–120], multiferroics[121], topological insulators[122], and quantum Hall materials[123] have been discovered.

Small flakes of 2D materials are commonly prepared via mechanical exfoliation from bulk crystals, but techniques for synthesizing monolayer/few-layer films via metal-organic chemical vapor deposition (MOCVD)[124,125], CVD[126,127], molecular beam epitaxy (MBE)[128], etc. have been developed, and synthesis of large area 2D materials and their heterostructures has been an active field of research.

The properties of the materials used in quantum devices - or any device, for that matter - create inherent limitations on device performance. Materials synthesis poses additional limitations on the types of devices that can be realized and the degree to which they can be scaled. In this review, we consider materials-related challenges to the realization of devices for quantum information science, and we identify how emerging nanomaterials may be able to address these challenges. Further, we identify recent advances in nanomaterial-based quantum devices and nanomaterials synthesis that may enable future devices, and we consider nanomaterials-specific challenges for future quantum devices.

## 2.0.0 Quantum Dot Qubits

Carriers confined in semiconductor quantum dots (QDs), typically 10-100 nm in diameter, are a potential platform for quantum computing. QD-based quantum computers are a strong platform for quantum simulation of many-body systems[129], and QD qubit arrays have recently been used to simulate Fermi-Hubbard models[130] and Nagaoka ferromagnetism[131]-- problems that are intractable on modern classical computers. Single electrons (or holes) can be confined geometrically or electrostatically; the latter is the more common approach to form planar QDs. Electrostatic confinement requires a high mobility semiconductor with a high mobility and clean interfaces[132]. Various qubit types can be realized in QD platforms.

Charge qubits – encoded in the presence of an electron in the left, $|L\rangle$, or right, $|R\rangle$, dot in a double quantum dot (DQD) system, then forming bonding and antibonding states when there is an interdot tunnel coupling – are the simplest type of semiconductor QD qubit[133]. Charge qubits can be rapidly controlled by voltage pulses, but charge coherence is highly sensitive to charge noise, limiting coherence times to several ns to hundreds of ns. The conversion of other types of qubits to charge qubits is useful for readout.

Electron spins in QDs, proposed by Loss and DiVincenzo in the late 1990s[134], are the most studied qubit variety. In this scheme, qubits are encoded in the spin states of the single electron QDs: spin-up, $|\uparrow\rangle$, or spin down, $|\downarrow\rangle$. In the presence of an applied magnetic field, the two states are split by an energy, $\Delta$, due to the Zeeman effect. The spins in neighboring single-electron QDs are separated by a voltage-dependent tunnel barrier, leading to a transient

Heisenberg exchange coupling between the two spins, contributing the term $J(V(t))\boldsymbol{S_1} \cdot \boldsymbol{S_2}$ to the overall Hamiltonian. Two electron spins in a DQD can also be represented as spin singlet-triplet qubits:

$$|S\rangle = \frac{1}{\sqrt{2}}(|\uparrow\downarrow\rangle - |\downarrow\uparrow\rangle) \text{ and } |T_0\rangle = \frac{1}{\sqrt{2}}(|\uparrow\downarrow\rangle + |\downarrow\uparrow\rangle)$$

Gate operations use a combination of exchange coupling and electron spin resonance (ESR), a process in which in-plane, AC electromagnetic pulses with frequency $\omega$ such that $\hbar\omega = \Delta$ are applied to drive a spin precession. In systems with strong spin-orbit coupling (SOC), electric dipole spin resonance (EDSR) can instead be employed for fast, all-electrical control[135]. However, SOC can also be a source of decoherence for spin qubits[136].

In semiconductors that lack inversion symmetry, the valley pseudospin degree of freedom can be exploited as an alternative to the spin degree of freedom. Broken inversion symmetry leads to degenerate valley states in the conduction band. Lifting the valley degeneracy enables the use of the valley index as a qubit[137]. Spin-valley hybrid qubits (also called Kramers qubits) can also be encoded in the states resulting from spin-valley coupling in systems with sufficiently strong coupling.

Three-spin qubits are another class of QD qubits that offer all-electrical control and protection from sources of decoherence. Three-spin qubits can be implemented as exchange only (EO), spin-charge hybrid, and resonant exchange (RX) qubits[138]. EO qubits are implemented in three-electron, triple quantum dot (TQD) systems. Rotations of EO qubits can be performed about two axes and are driven by the voltage-dependent exchange interaction. While EO qubits allow all-electrical control, complex gate protocols and charge noise sensitivity remain issues[133]. Charge noise can be mitigated by operating EO qubits at "sweet spots" and by using dynamical decoupling protocols. RX qubits are a type of EO qubit in which the exchange interaction is always turned on and the qubits are manipulated on both axes of the Bloch sphere by resonantly modulating the exchange interaction[139,140]. RX qubits can be coupled to microwave photons in superconducting microcavities[141], which is useful for qubit manipulation, readout, and long-distance coupling[142]. Operating RX qubits in the symmetric regime can reduce sensitivity to noise[143]. Spin-charge hybrid qubits - implemented in three-electron, DQD systems - are similar to single-triplet qubits but combine the speed of charge qubits with the coherence times of spin qubits[144]. Spin-charge hybrid qubits offer fast electrical control with more efficient operation than EO qubits. Moreover, the two-dot device structure reduces the optimization of the charge noise "sweet spot" to one parameter. Russ and Burkhard cover triple-electron-spin qubits in greater detail[138].

Early implementations of QD electron spin qubits were demonstrated in gate-defined, lateral quantum dots in a two-dimensional electron gas (2DEG) in III-V heterostructures[145,146]. 2DEGs in III-V heterostructures have high mobilities and high quality III-V heterostructures can be grown with relative ease, but hyperfine interaction between electron spins and the non-zero nuclear spins of group III and V elements leads to decoherence in III-V quantum dots[147]. Nevertheless, coherent control of spin qubits in multiple[148] III-V QDs has been demonstrated, albeit with limited $T_2$.

Si/SiGe QD devices have largely replaced III-V heterostructures as the leading technology over the past several years. The low abundance (~4.7%) of spin-½ $^{29}$Si isotopes in natural silicon and the ability to synthesize isotopically pure layers of spin-0 $^{28}$Si nuclei[149] help Si-based devices overcome the main decoherence issue of III-V QDs and achieve coherence times of $T_2 = 28$ ms[150]. The bonus of CMOS-compatible processing enables greater scalability of Si/SiGe devices. Single qubit gate fidelities >99.9%[151] and two-qubit gate fidelities >99%[152] have been demonstrated in Si/SiGe QDs. While Si/SiGe spin qubit gates have been demonstrated with high fidelity, required micromagnets or integrated striplines for ESR are obstacles for scaling. Importantly, ESR-driven systems can be difficult to scale because of the electromagnetic fields cannot be localized to single qubits[153]. Additionally, the process of ESR is slow compared to electrical manipulation. All-electrical control of spin qubits under a static magnetic field is thus significantly more practical for scalable quantum computation.

Hole spins in Ge have emerged as a promising alternative to III-V- and Si- based platforms. Semiconductor holes have high SOC, enabling fast, all-electrical manipulation of spins via EDSR. Moreover, the use of hole states eliminates complications resulting from valley degeneracies in the conduction band. Silicon's band alignment is not conducive to confining holes, but strained Ge/SiGe quantum wells can be formed with relative ease, and confined holes exhibit carrier mobilities of ~$10^6$ cm$^2$/Vs[154]. Germanium maintains the benefit of possible isotopic purification for zero nuclear spins. All-electrical, universal control of qubits with Rabi frequencies exceeding 100 MHz[155] and coherence times up to 150 μs[156] have been demonstrated. Recently, a four-qubit Ge-based quantum processor was demonstrated with all-electrical logic gates and single qubit gate fidelities up to 99.9%[157], comparable to Si-based technologies. Ge qubit technologies are less mature than Si/SiGe but show promise. Ge quantum technologies are reviewed in more detail in ref[158].

### 2.1.0 *2D materials for QD Qubits*

#### 2.1.1 *Graphene quantum dots (GQDs)*

While electron and hole based 2D charge carrier gases in confined III-V and group V quantum wells are attractive, electrostatic confinement of the wavefunction in thickness dimension is still limited by the physical layer thickness. In contrast graphene is an atomically thin sheet of sp$^2$-hybridized carbon and is a type-I Dirac semimetal with highly tunable electrical properties and a high mobility thereby making it the thinnest 2DEG system[87,159]. Graphene's low intrinsic spin-orbit coupling and negligible hyperfine interaction make it a promising material for quantum dot spin qubits with potentially long coherence times[160]. Like silicon, graphene can be isotopically purified to further reduce hyperfine interactions. For graphene to be used as a host for spin qubits, two conditions must be satisfied: (i.) a bandgap must be opened, and (ii.) the valley degeneracy must be lifted. Graphene nanoribbons (GNRs) with armchair boundaries are semiconducting and lack the valley degeneracy of bulk graphene[160]. Alternatively, a gate-tunable bandgap can be created in bilayer graphene (BLG) with the application of an out-of-plane electric field[161] while an out-of-plane magnetic field must be applied to lift the valley degeneracy[160].

In addition to the spin degree of freedom, graphene offers a binary valley degree of freedom that emerges due to broken inversion symmetry. Spin-valley coupling results in two Kramer pairs: $|\uparrow, K\rangle, |\downarrow, K'\rangle$ and $|\uparrow, K'\rangle, |\downarrow, K\rangle$. A large, tunable valley splitting is needed for practical realization of valley qubits. The valley splitting under an applied magnetic field, $B_\perp$, is given by $\Delta E_{K,K'} = g_v \mu_B B_\perp$, where $g_v$ is the valley analogue of the spin Lande g-factor, $g_s$. The $g_v$ factor in GQDs has been demonstrated to be electrostatically tunable from 20 to 90[162], resulting in a valley splitting an order of magnitude greater than Si/SiGe systems[163]. One challenge for spin-valley qubits in graphene is the low intrinsic SOC. The spin-valley coupling in single-electron GQDs was recently determined to be dominated by Kane-Mele SOC[164], with a coupling of approximately 60 $\mu$eV. This suggests that the spin-valley coupling can be enhanced by boosting the weak intrinsic SOC of graphene by the proximity effect[165].

Schematics of a typical device structure for a graphene double quantum dot (DQD) device and the corresponding band structure is shown in Fig. 2a. Side split-gate electrodes laterally confine carriers to a narrow channel region in the host material while top finger gate electrodes control the formation of QDs. In single QDs, three finger electrodes are needed: two side barrier electrodes for confinement and a center "plunger" electrode that controls the accumulation or depletion of electrons. In a double quantum dot (DQD) device, there is one plunger for each dot, two side barrier electrodes, and a middle barrier electrode that can be used to tune the tunnel coupling, $J(V_c(t))$, between dots. In Fig. 2a, the red dotted line shows the barrier lowered, increasing the coupling, while the blue line shows the coupling decreased. While the schematic shows the DQD device for electrons in the conduction band, hole QDs in the valence band can also be generated.

Recent results have fulfilled important prerequisites for the realization of GQD spin and valley qubits. Banszerus et al.[166] demonstrated single electron DQDs, while Tong et al.[162] have since realized ambipolar single-carrier confinement in DQDs, seen in the charge stability diagram in Fig. 2b. Recently, gate-tunable tunnel coupling in a BLG DQD device was also reported for the (2,0)-(1,1) transition and the (4,2)-(3,3) transition, where (x,y) indicates the number of electrons in the left and right dots in a DQD device[167]. This control over tunnel coupling is a crucial step towards various qubit types in GQDs. Understanding the complex interplay between spin and valley states, particularly in coupled dots, is another important task. It's been determined that in two-carrier GQDs, the ground state is best expressed as a spin-triplet valley-singlet state at low magnetic fields[168]. In the high field regime, however, it was recently reported that the ground state switches to the spin-singlet valley-triplet state[169]. Finally, a crucial prerequisite is the ability to read-out and manipulate spins. One common method of readout for spin states is spin-to-charge conversion via Pauli blockade. Spin and valley blockade was recently shown in a graphene DQD system, as seen in the charge stability diagram in Fig. 2c[169]. Dispersive charge readout was also recently shown for graphene charge states[170]. The combination of these two advances could allow readout of spin or valley states via spin-to-charge conversion.

While most studies have focused on gate-defined GQDs, the bottom-up synthesis of graphene nanostructures is another approach to generate quantum confinement in graphene. Recent theoretical work has suggested the formation of Moiré potential-induced quantum dots in twisted bilayer graphene (tBLG) at low twist angles (Fig. 2d)[171]. The CVD synthesis of tBLG with controlled twist angle has been demonstrated[172], and could allow the experimental realization of intrinsic GQDs in tBLG. Another approach utilizing geometric confinement is the formation of in-plane hBN-G heterostructures with clean edges synthesized by catalytic conversion of hBN to graphene with patterned Pt layers[173]. Kim et al.[174] used this approach to synthesize lateral heterostructures of sub-15 nm GQDs embedded in an hBN host layer using dense arrays of self-assembled Pt nanoparticles as the catalyst (Fig 2g). This enabled the fabrication of vertical single-electron tunneling transistors of the structure 30 nm hBN/G/2L-hBN/GQD-hBN/2L-hBN/G/20 nm hBN on $SiO_2$, the band diagram of which is shown in Fig. 2h. The conductance plot in Fig. 2i verifies the single electron transport for a low-density array of GQDs. The coulomb diamonds overlap more as the GQD array density increases. Demonstrating and controlling tunnel coupling between neighboring dots in a similar matrix structure could eventually allow the synthesis of a scalable many-qubit system useful for futuristic quantum computation, simulation, or basic scientific research on many-body systems in coupled GQDs.

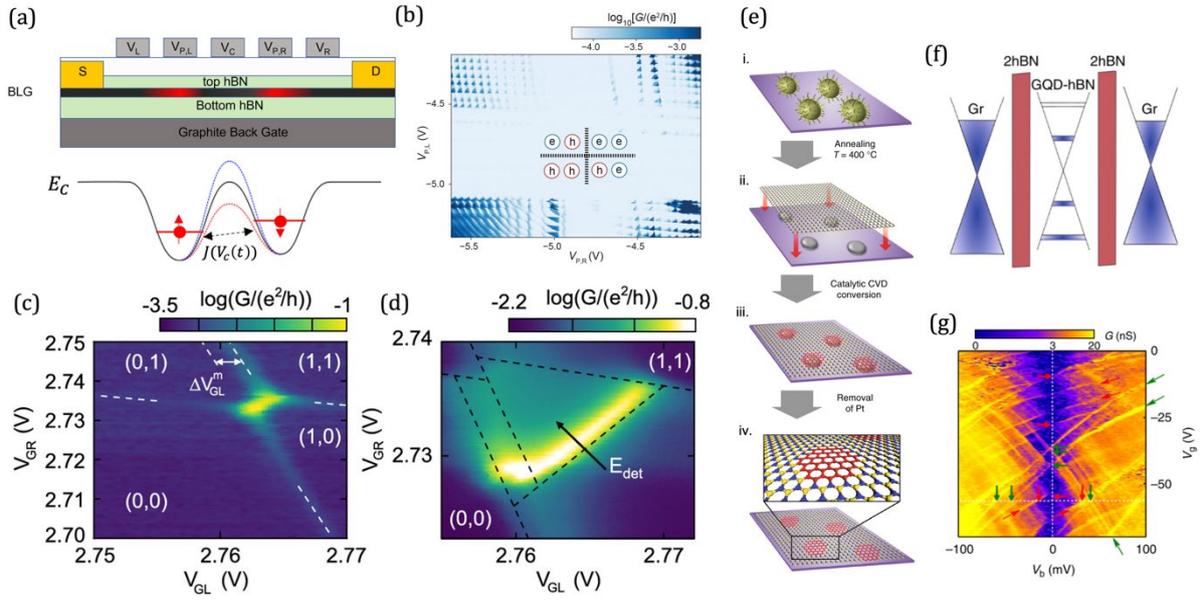

*Figure 2. Graphene quantum dot devices for quantum information processing.* (a) Schematic structure of a double graphene quantum dot (GQD) device and the corresponding band structure. (b) Charge stability diagram of double quantum dot (DQD) device displaying ambipolar carrier confinement with single level transport. Adapted with permission from ref[162]. Copyright 2021 American Chemical Society. (c.) Charge stability diagram at a bias of 0.1 mV demonstrating single carrier confinement in a graphene DQD device and (d.) a further zoomed in view of the charge stability diagram at a bias of 1 mV. The arrow shows the energy detuning along the axis of transition from then (0,1) to (1,0) state. Panels c, d adapted with permission from ref.[166] Copyright 2020 American Chemical Society. (e) Synthesis of in-plane GQD-hBN heterostructures: i. Pt nanoparticles (NPs) are assembled using block copolymers, and the polymer is then removed by annealing; ii. Monolayer hBN is transferred onto the NPs; iii. Catalytic conversion of hBN to graphene occurs only in areas where there are NPs; iv. NPs are removed. (f) Vertical single-electron tunneling transistor energy band diagram. (g) Conductance plot of low-density array of GQDs demonstrating single level transport. Panels e, f, g adapted from ref[174].

Semiconducting graphene nanoribbons (GNRs) are another potential system for GQD qubits. Several early studies used lithographically defined GNRs to form GQDs. However, charge localization at impurities along etched GNR edges causes decoherence[175]. GNRs, however, offer their own advantages: the intrinsic lifting of the valley degeneracy allows the operation of qubits without the need for an applied magnetic field to lift the degeneracy, offering greater potential scalability; widths as small as 10 nm[176] also offer greater lateral confinement. Recent advances in the bottom-up (i.e., lithography free) growth of GNRs with pristine armchair edges and controllable placement/orientation make the development of GQD devices in GNRs more feasible[177–180].

The next milestones for GQD-based qubits are the measurement of coherence times for spin/valley states and the coherent control of qubits. The lower bound of the relaxation time for charge states[181] and spin states[182] have been estimated using pulsed-gate spectroscopy to be on the order of hundreds of ns, likely limited by the measurement technique. Single-shot readout is needed to more accurately measure lifetimes and coherence times of states in GQDs[182].

*2.1.2 Transition metal dichalcogenide (TMD) QDs*

Monolayer and bilayer TMDs and their heterostructures have been identified as other promising 2D material hosts for gate-defined QD spin-valley qubits[183]. Recent theoretical studies have further described gate operations on single spin qubits[184] and both one- and two-spin-valley[185] and valley[186] qubit systems in TMDs. Fig. 4a shows the atomic structure of a prototypical TMD, 2H-$MoS_2$. TMDs exhibit a strong intrinsic SOC due to the metal atoms' $d$ orbital electrons; strong SOC offers the potential for all-electrical control of qubits via EDSR. TMDs also demonstrate Rashba SOC that can be tuned by an electric field[183]. The strong SOC also causes strong spin-valley coupling. Monolayer group-IV TMDs ($MX_2$ where M = Mo, W and X = S, Se) have a direct bandgap at the $K$ and $-K$ valleys and have an intrinsic broken inversion symmetry which results in valley-dependent optical selection rules for circularly polarized light[187,188]. Fig. 4b shows this phenomenon schematically. $\sigma^+(\sigma^-)$ polarized light has allowed transitions at the $K(-K)$ points. The valence and conduction bands have spin-dependent splitting at the two $K$ points under an applied B-field due to the Zeeman effect. The SOC and valley-dependent optical selection rules importantly combine to allow the selective optical initialization and read-out of spins[189]. This intrinsic spin-photon interface is one of the primary benefits of TMD spin-valley qubits.

Gate-defined QDs have been demonstrated in $WSe_2$[190,191], $WS_2$[192], and $MoS_2$[193–195] monolayers and $MoS_2$ nanotubes[196]. Fig 4c depicts an example of a gate-defined QD in $WSe_2$ sandwiched between layers of hBN. The coulomb blockade in the conductance plot for the device (Fig. 4d) confirms single-level transport. Despite the confirmation of single-level transport, confinement down to the single electron/hole limit in TMD QDs has not yet been achieved.

There are potential approaches beyond electrostatic gating to confining carriers in TMDs. Quantum dots of TMDs with radii small enough to only allow single carrier confinement and coupled via tunnel barriers, optical channels, or other means could be realized with sufficient advances in materials processing. Recently, sub-20 nm $MoS_2$ quantum dots were created by rapid thermal annealing[197], and heterostructures of $ReS_2$ quantum dots in a $MoS_2$ matrix were realized[198]. We imagine that confinement of single spins in bottom-up synthesized TMD quantum dots could eventually be useful for QIP. Like in graphene, moiré potentials formed in twisted bilayers of TMDs or TMD heterostructures offer another route to charge confinement in TMDs.

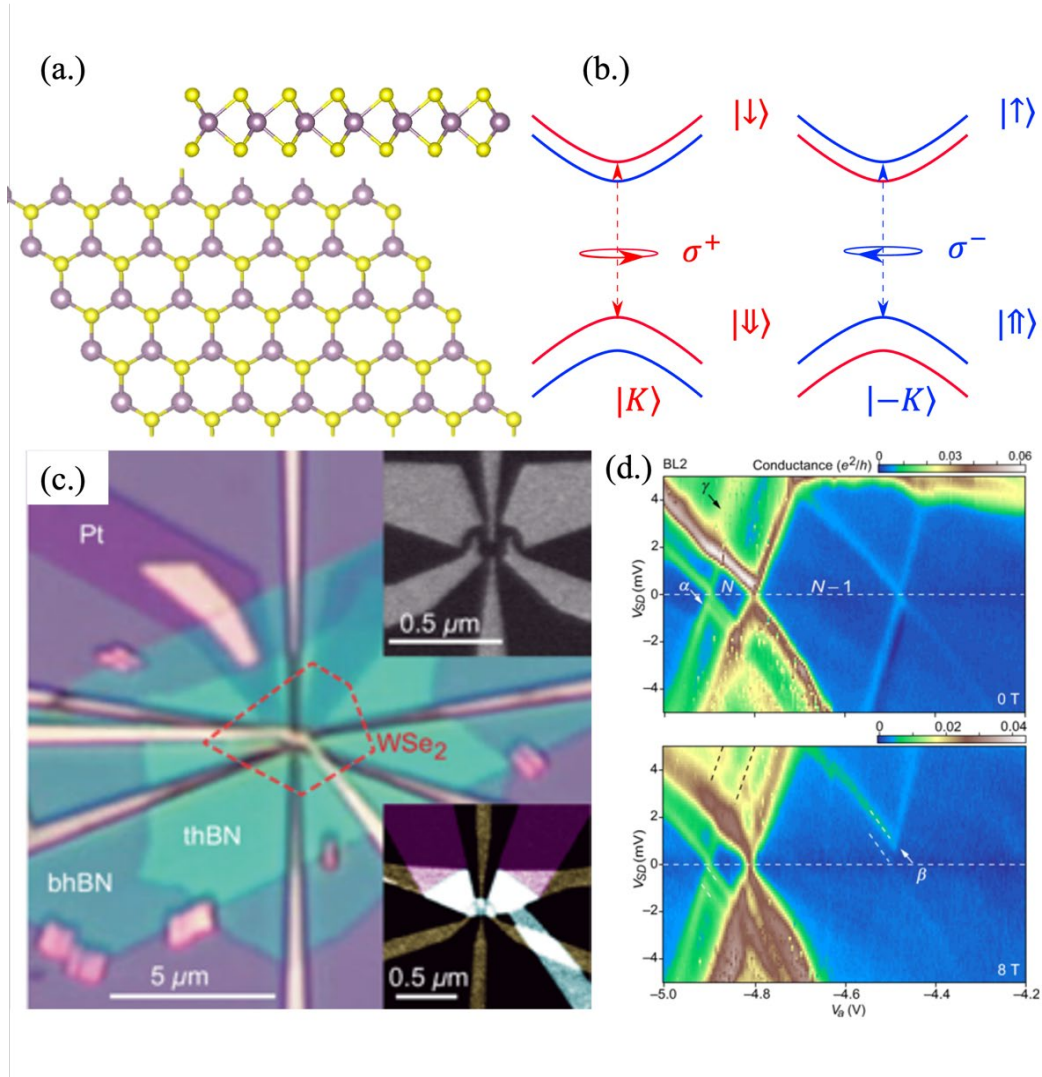

*Figure 3. Towards spin-valley qubits in transition metal dichalcogenide quantum dots.* (a.) Structure of 2H-$MoS_2$, a prototypical TMD. Yellow atoms are sulfur, light purple atoms are molybdenum. Model created with ref.[199] (b.) Valley dependent optical selection rules in TMDs. (c.) Gate-defined quantum dot device structure in a hBN-encapsulated $WSe_2$ flake and (d.) the corresponding single-level charge transport, evidenced by the Coulomb diamonds in the conductance plot. Fig. c, d reprinted from ref.[191] with permission from American Physical Society.

An active field of research within TMD spin-valley devices is using ferromagnetic and/or ferroelectric heterostructures to lift the valley degeneracy of the TMD layer, inducing a spontaneous valley polarization. In Table 2, we compare theoretical and experimental 2D materials and heterostructures that demonstrate an intrinsic valley splitting. Promising heterostructures have been identified using *ab initio* methods, including $WSe_2/CrI_3$[200] and $WSe_2/CrSnSe_3$[201]. "Ferrovalley" 2D materials - direct bandgap semiconductors that are also ferromagnetic or ferroelectric, leading to an intrinsic valley polarization - have also been predicted theoretically[202–204], notably 2H-$VSe_2$, a ferromagnetic semiconductor that can be controlled electrically by heterostructuring it with $Sc_2CO_2$[204]. While *ab initio* predictions look promising for 2H-$VSe_2$, the 1T phase is the most thermodynamically stable. Synthesis of 2H phase $VSe_2$ monolayers must be accomplished before any experimentation on it for valleytronics or valley qubits can be done.

Research towards spin-valley qubits in TMD-based gate-defined QDs is still in its infancy. Confinement of single free electrons (holes) has yet to be demonstrated in TMD QDs, and spin and valley state coherence times have not been determined. Blockade measurements on TMD QDs are made challenging by the difficulty in making ohmic contacts to TMDs at low temperatures[205]. Finding suitable 2D/3D ohmic contacts is therefore a crucial materials challenge for TMD-based QD devices. Optical readout may be an alternative. An additional challenge in TMDs is the synthesis of high quality large-area TMD monolayers. This is not currently a limiting factor as exfoliated flakes have been used instead in single devices, but future scalability of TMD quantum devices will require large-area, single-crystalline monolayers with minimal defects.

### *2.1.3 Other 2D Materials*

Silicene is the 2D allotrope of silicon, and it therefore has minimal nuclear spins and can be isotopically purified. Silicene and other non-graphene 2D materials, germanene and stanene, have a slightly buckled, nonplanar structure, leading to strong intrinsic SOC[206]. The attraction of long coherence times due to minimal nuclear spins and SOC-enabled electrical control have inspired theoretical research on electrostatic QD qubits in silicene[207,208]. Despite the experimental synthesis of silicene[209,210], silicene QD devices have- to our knowledge - not yet been experimentally realized.

### *2.2.0 1D Materials for QD Qubits*

Semiconducting nanowires (NWs) and nanotubes (NTs) have highly anisotropic electronic structures and strong spin-orbit interaction. The intrinsic lateral confinement of carriers simplifies device design, and the strong SOC enables fast, all-electrical control. NWs and NTs can also be (relatively) easily integrated with superconducting microcavities. Circuit cavity quantum electrodynamics (cQED) uses superconducting microwave resonators to entangle spatially separated superconducting qubits. This same approach can be applied to couple spatially separated QDs in nanowires/nanotubes suspended across the cavity[211]. This is a scalable approach to manipulate and readout spin and charge qubits in NWs[212], and it can act as a quantum bus to connect spatially separated multi-QD NWs. Additionally, suspending NWs/NTs across cavities reduces the decohering effects of charge noise and can extend the relaxation time by reducing the spontaneous phonon emission rate[213].

### *2.2.1 Ge-Si core-shell NWs (CSNWs)*

Type-II band alignment in Ge-Si CSNWs results in a one-dimensional hole gas (1DHG) in the germanium core (Fig. 4a,b)[214]. This enables the formation of hole QDs that are, like their 2DHG counterparts, promising for hole spin qubits[215–217]. Strong, gate-tunable SOC is a primary benefit of the CSNW structure as compared to planar Ge[218]. The ultra-strong SOC allows for ultra-fast, all electrical spin manipulation. Dephasing times up to 180 ns have been demonstrated for hole spins in Ge-Si CSNWs, consistent with a combined mechanism of scattered nonzero-spin Ge nuclear isotopes and a SOC mechanism[219], suggesting improvement with isotopic purification. Recently, Froning *et al.* demonstrated ultrafast control of hole spins in Ge-Si CSNWs with gate-tunable Rabi frequencies (Fig. 4c), with a maximum $f_{Rabi}$ of 435 MHz, approximately 4 times faster than planar Ge[220]; this result in spin-flipping times of ~1 ns. Moreover, they showed a coherence time extended up to 250 ns using a Hahn echo pulse sequence; the driven coherence time ranged from 7 to 59 ns depending on the gate voltage, respectively. It is expected that charge noise or other sources of fluctuating electric fields are also a source of decoherence[153,219,220]. The "sweet spot" for minimal charge noise occurs when the direct Rashba spin-orbit coupling (DRSOC) is zero[158]. Coherence times may be improved by only switching "on" the DRSOC via applied gate voltage for logic gates. The impact of the Ge-Si interface has not been explored, to our knowledge, but given the high surface to volume ratio of the Ge core and the sensitivity to charge noise, it is possible that interfacial defects and trap states may contribute to decoherence.

Ge-Si CSNWs are typically grown by the vapor-liquid-solid (VLS) method using Au nanodroplets to catalyze the growth of vertical CSNW arrays. This process can lead to Au atoms diffusing to the surface of the germanium core, preventing the formation of clean interfaces with Si and degrading electronic properties[221]. One solution to this problem is growing a small layer of Si between the Au and Ge to act as a diffusion barrier[222]. Recently, more effort has been focused on self-seeded growth of Ge-based nanowires[223], but the various processes attempted so far still result in an entangled mesh. A self-seeded growth process for horizontal, aligned arrays of Ge-Si CSNWs or individual Ge CSNWs with deterministic placement would be an important step towards Ge-Si CSNW-based quantum devices. Overcoming the processing challenges and eliminating sources of decoherence would make Ge-Si CSNWs a viable platform for QD qubits.

*2.2.2 In-V NWs*

Indium antimonide and indium arsenide have attracted attention for quantum devices due to their high electron mobility[224], large Lande g-factor[225], and strong spin orbit coupling[226]. While In-V NWs have primarily been studied for topological systems, as we discuss in a later section, fast, all-electrical initialization[227] and manipulation of spin-orbit qubits make InSb and InAs NWs promising for future QD devices. Single[225,228], double[229,230], and multi-QD devices[231] have been realized in InSb/As nanowires, and a spin-orbit qubit in an InSb NW QD was demonstrated with fast Rabi oscillations (104 MHz) and a coherence time up to ~35 ns[226]. Coupling of In-V NWs to superconducting microcavities has been demonstrated[212,229,232]. Fig. 4d shows a InAs double quantum dot coupled to a microwave resonator. Hyperfine decoherence, as in planar III-V heterostructures, limits the potential promise of In-V NWs for pure QD qubits. Compared to group IV-based 1D materials, it is unlikely that In-V NWs will be useful for practical applications of QD spin qubits. In-V NWs remain promising for topological and Andreev qubits.

*2.2.3 Carbon nanotubes (CNTs)*

Carbon nanotubes, like other group IV semiconductors, have low hyperfine interactions between electronic and nuclear spins that can be further reduced with isotopic purification. CNTs additionally have a high SOC that results in a zero-field splitting, enabling all electrical control of spins[233]. Ultra-long coherence times have made CNTs another attractive nanomaterial platform for QD qubits.

Spin-valley qubits can be formed in bent CNT DQD devices[234,235]. Although there have been considerably fewer efforts in this area than other potential QD qubit platforms, coherence times up to 198 ns have been demonstrated[235]. Charge noise contributes to dephasing, but the dominant source is hyperfine interactions[236]. This suggests that isotopic purity is especially crucial for CNT spin qubits.

It has been theoretically predicted that spins in CNT QDs can be initialized and manipulated optically when coupled to a cQED system[237]. These CNT/microcavity structures can be fabricated using a dry stamping technique to transfer CNTs from a growth substrate to the microcavity[238]. Cubaynes *et al.*[213] recently demonstrated the coupling of a single electron spin in a DQD device to a microwave photon in a superconducting cQED setup. By contacting a CNT with ferromagnetic electrodes, an artificial spin-orbit interaction is introduced, several orders of magnitude larger than natural SOC. Coherence times on the order of several microseconds were demonstrated, even without isotopic purification.

CNT QD qubits are promising for various types of quantum sensing. A charge qubit in a CNT positioned on a scanning probe tip (Fig. 4e) demonstrated high sensitivity to electric and magnetic fields (Fig. 4g)[239]. CNTs have quantized flexural modes that can be coupled to spin[240] and charge[241] states in CNT QDs. Collectively, CNT nanomechanical-QD hybrid qubits could be utilized for next-generation scanning probe microscopy techniques.

Recent advances in CNT processing have enabled dense, wafer-scale arrays of aligned nanotubes[242,243], making CNTs a feasible material system for large-scale applications. The hyperfine interaction-driven decoherence can be minimized by using isotopically purified $^{12}$C sources for CNT growth.

CNT QDs are a highly versatile platform with potential for quantum computing and quantum sensing. CNT QD qubits coupled to superconducting microcavities exhibit long coherence times and may be controlled optically or electrically. Further, the ability to couple CNTs to cQED systems will enable future entanglement of many spatially separated CNT QDs. The sensitivity of CNT QD devices to force and electromagnetic fields and the ability to integrate the devices on a scanning probe tip have the potential to enable next-generation quantum scanning probe microscopy techniques.

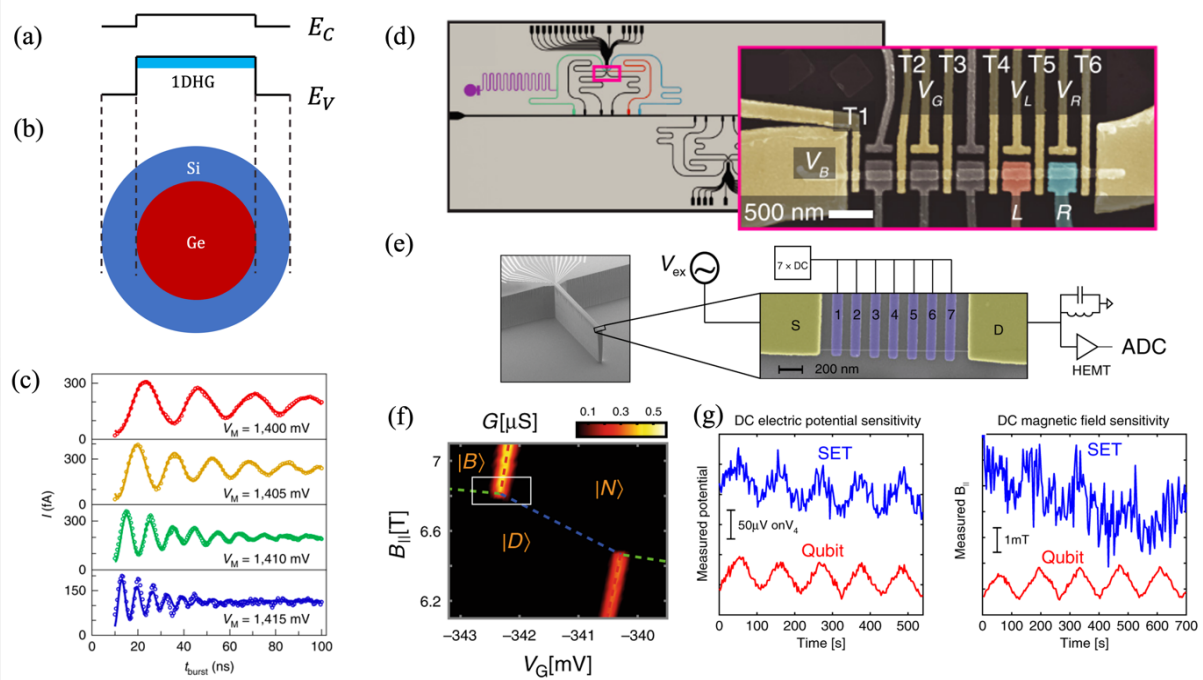

*Figure 4. Quantum dot qubits in 1-dimensional nanostructures.* (a.) Schematic of a 1DHG in a Ge-Si CSNW corresponding to a schematic of the CSNW cross-section in (b.). (c.) Gate-tunable coherent control of hole spin qubits in a Ge-Si CSNW. (d.) InAs multi-QD NW coupled to microwave resonator. (e.) A SEM image of a CNT QD charge qubit on a scanning probe tip (left) and a higher magnification, false-color SEM image of the device plus the control and readout circuitry (right). (f) A charge stability diagram for the device shown in panel (e.). The device is operated at a triple point between three charge states. (g.) Sensitivity of the CNT charge qubit scanning probe device shown in (e.) to a DC electric potential (left) and a DC magnetic field (right) as compared to a single electron transistor sensor. Panel (c.) adapted from ref. [220] Panel (d.) adapted from ref.[232]Panels (e, f, and g) reprinted from ref.[239]

## 3.0.0 Spin Qubits in Defects

Spin defects in wide-bandgap crystals are a promising system for qubits in quantum computation, quantum networks, and quantum sensing. Electronic spin defects can be rapidly initialized through optical pumping and then magnetically controlled via ESR, then readout of states can be performed optically. Defect center qubits offer two primary advantages over QD and superconducting qubits: room temperature operation and an intrinsic spin-photon interface. Coherence times of defect spin qubits are also generally longer than QD qubits. The optical properties and single photon emission from defect centers are discussed in more detail in the next section. The sensitivity of electronic defect states to magnetic field[244–246], strain, and temperature[247] makes them particularly attractive for quantum sensing applications. Weber, Koehl, and Varley[25] identify the criteria spin defects must meet to be used as qubits:

I. Long-lived paramagnetic defect states with a spin-dependent energy splitting.
II. Polarizable by optical pumping with spin-selective transitions with one or more nonradiative transitions between states with different spins.
III. Luminescence that differs between qubit sublevels by intensity, energy, etc.
IV. Optical readout without transitions that interfere with energy levels of host (i.e., the defect states must be within the bandgap of the host).
V. Energy level splitting must be large enough to avoid thermal excitation.

Like electron spins in QD qubits, spin-orbit interaction and hyperfine interactions with nuclei decohere spins. Hence, high-quality crystals with minimal non-zero nuclear spins are desirable. Group-IV, wide-bandgap materials are therefore ideal candidates to act as hosts.

The leading technology for defect-based quantum information processing is the nitrogen vacancy ($NV^-$) in diamond, an electronic spin defect[248]. Silicon (SiV)[249] and other defect centers[250] have been explored in diamond, but $NV^-$ centers remain the most mature technology. $NV^-$ centers can be manipulated and read out by optically detected magnetic resonance (ODMR), a double-resonance technique that combines ESR with optical readout. Electrical[251] and microwave cavity[252] readout schemes are also possible. $NV^-$ centers exhibit room temperature coherence times up to 2.4 ms, limited by electron-phonon interactions[253]. The coherence time is extended to >1s at cryogenic temperatures[254]. $NV^-$ centers can be coupled to other $NV^-$ centers or $^{13}C$ nuclear spins (as many as 27 $^{13}C$ nuclear spins have been addressed by a single $NV^-$ center) via magnetic dipole interactions[255]. $^{13}C$ nuclear spins exhibit coherence times >10s[256], but are not optically active. Optically addressable $NV^-$ centers coupled to dark nuclear spins with long storage times form a promising platform for quantum memories[256,257] and entanglement swapping procedures[258,259]. $NV^-$ centers coupled to $^{13}C$ quantum memories have also been shown to improve quantum sensing sensitivity[260]. Single gate fidelities of ~99.995% and two-qubit gate fidelities of ~99.2% have been demonstrated at room temperature[261], and basic QEC has been demonstrated for NV- centers in diamond[262]. The excellent properties of NV- centers and the ability to address $^{13}$C nuclear spins make diamond a strong candidate for quantum computing[25,263], quantum repeaters[259], and quantum sensors[244,246,247,252,260,264].

Silicon carbide (SiC) is another potential host for defect spin qubits. SiC is a wide bandgap semiconductor with many polytypes, the most common being hexagonal 4H-SiC. 4H-SiC is commonly used in microelectronic devices and is a more mature material than diamond from a processing standpoint. Because of the low naturally occurring percentage of non-zero nuclear spins for Si and C, SiC can be grown from isotopically purified precursors[265] to minimize hyperfine interaction. While isotopically pure SiC crystals have been produced, their use in spin-qubit engineering is still in its nascency yet remains a promising direction[266]. The negatively charged silicon vacancy ($V_{Si}^-$) and various neutral divacancies ($VV^0$) are the most studied defects in SiC. Like diamond color centers, SiC color centers are optically active electronic defects. Due to spin-locking, SiV centers in SiC have shown coherence times up to 20 ms[267]. Additionally, entanglement between optically active electronic defects and nuclear spins was demonstrated for divacancy centers[266]. The same group exhibited control over single qubits with a fidelity exceeding 99.98%. Due to the high temperature growth of SiC crystals, many impurities and intrinsic defects are present, leading to fluctuating charge and spin states in unintended defects that can lead to decoherence[268].

Kane proposed using the nuclear spins of shallow donor atoms in Si as qubits, with rotations performed by gate voltages and spin readout via a spin-polarized current[269]. While Kane's exact proposal has not yet been realized, dopant spins in Si have been explored with success. Qubits encoded in the nuclear spins of $^{31}$P dopant atoms in $^{28}$Si have been demonstrated with coherence times >30s and single qubit gate fidelities >99.99%[270]. Until recently, donor spins have been controlled using ESR, which - as mentioned early - is slow and requires bulky striplines. Electrical control of a high-spin (S = 7/2) $^{123}$Sb donor in Si was recently demonstrated with a dephasing time, $T_2^*$, of ~0.1s[271]. The electron bound to the Sb donor in its un-ionized state could be used to form a 'flip-flop' qubit out of the electron spin and nuclear spin states of the donor[272]. Acceptor dopant nuclear spins and acceptor-bound hole spins have also been explored. Like QD-based qubits, acceptor-bound hole spins can be manipulated via EDSR[273]. Rabi frequencies on the order of GHz are attainable for acceptor-bound light hole states in group IV quantum wells, with strong manipulability near charge noise sweet spots[274]. This approach leverages the minimal hyperfine and spin-orbit interaction present in Si and the maturity of CMOS fabrication techniques. These defect spins can be introduced into gate-defined Si/SiGe quantum dots for additional functionality[275]. One downside of Si dopant qubits is that they are not optically active. Without optical addressability and readout, the applications of a quantum memory formed from the long-lasting nuclear states are limited. Moreover, the location of the dopant near the end of the conduction (valence) band for a donor (acceptor) requires operation at cryogenic temperatures, unlike other defect qubits.

Rare earth ions (REIs) are another type of interesting extrinsic defect and exhibit highly coherent states in the 4f orbital that are shielded from the environment by a full 5d shell. 4f optical transitions normally parity-forbidden become allowed in the presence of an inversion symmetry-breaking crystal field[276]. The interaction between the 4f electron states and the high-spin nuclei leads to a high density of states with long coherence times as high long as several milliseconds[277]. The coherence properties, density of states, and ability to write and read out states with high fidelity makes single REI and ensembles of REI dopants particularly attractive for quantum memories[278,279]. Unlike many of the *s*- and *p*- hybridized defect states commonly found in wide-gap semiconductors (NV⁻:diamond, $V_{Si}$-$V_C$:SiC, etc.), *f* states have weak oscillator strengths and, thus, are significantly less optically active[276]. The light-matter coupling can be enhanced by incorporating the dopants in nanophotonic cavities[278,280].

### *3.1.0 2D Host: Hexagonal boron nitride (hBN)*

hBN is a vdW wide-bandgap semiconductor with an indirect bandgap of ~6 eV[281]. Negatively charged boron vacancies ($V_B^-$) are stable up to 600K[282] and can be deterministically created by electron beam irradiation[283], a focused ion beam (FIB)[284], or femtosecond laser pulses[285] - as seen in Fig. 5a. The van der Waals layered structure of hBN allows better spatial control of defect placement than in bulk crystals, and having defects at or near the surface offers the potential for easier integration with photonic nanostructures to enhance light matter coupling. The 2D structure also may allow hBN spin defects to be integrated in vdW heterostructures with other 2D materials, offering optical readout of QD spins or other qubits[286]. hBN additionally has low SOC due to the small atomic masses of B and N.

Fig 5b schematically shows the energetic structure of the $V_B^-$ spin defect. The spin-triplet ground state, $^3A$, is split into $m_s = 0$ and $\pm 1$ substates. At zero magnetic field, the $m_s = 0$ substate is split from the $m_s = \pm 1$ substates by the ZFS energy $\delta_{GS}$. The degeneracy of the $m_s = \pm 1$ substates is lifted by a magnetic field. The system can be optically pumped (green arrows) to the excited state $^3B$, from which there are two decay pathways: radiative decay to the bright $m_s = 0$ state (red arrow) or nonradiative decay via intersystem crossing into the metastable $^1A$, singlet state to the ground state (black dashed lines).

Recently, Gottschall *et al.* used microwave pulses at low magnetic fields to demonstrate coherent manipulation of spins in $V_B^-$ defects in hBN at room temperature[287]. ODMR contrast results as a function of microwave frequency (Fig. 5c) demonstrate transitions from the $m_s = 0$ ground state to the nondegenerate $m_s = \pm 1$ states. ODMR contrast results as a function of pulse length (Fig. 5d) show Rabi oscillations on the order of 10 MHz. They further measured T₁ to be 18 $\mu s$ at 300K and increase to 12 ms at 20K. This first demonstration of coherent control in hBN spin defects is an important step towards the potential realization of $V_B^-$ spin qubits.

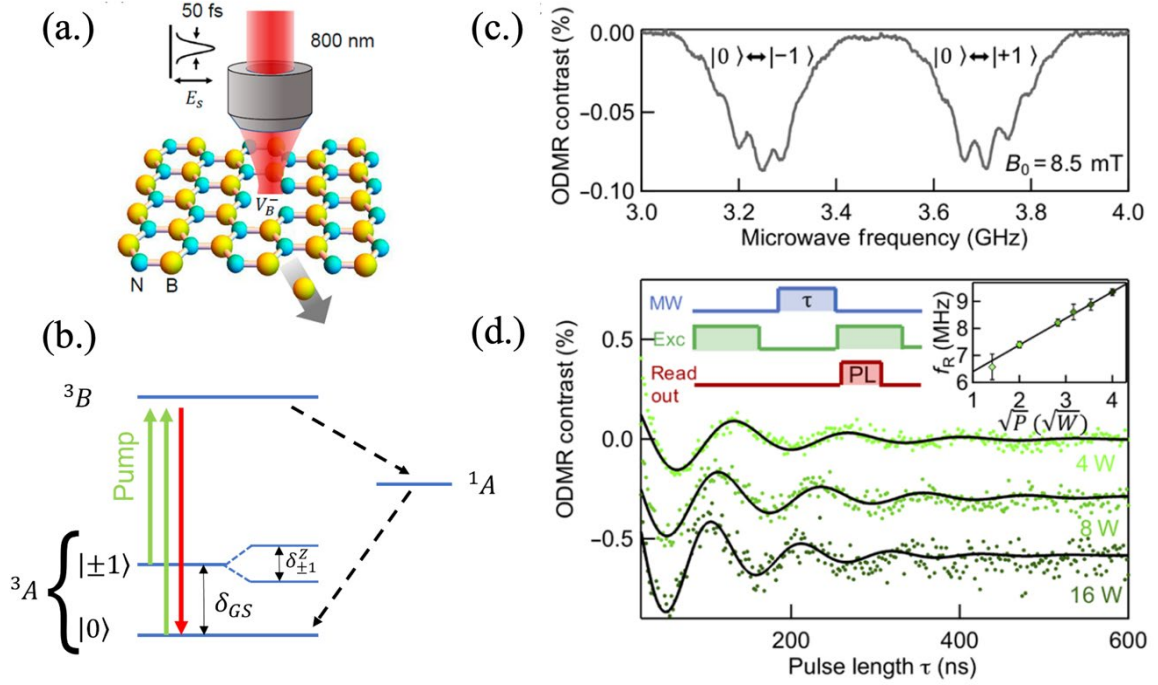

*Figure 5. Creation and control of boron vacancy spins in hBN.* (a) Creation a $V_B^-$ defect using a femtosecond laser pulse. (b) Energy structure of the $V_B^-$ electronic defect. A triplet ground state, $^3A$ is composed of $m_s = 0, \pm 1$ states. There exists a zero-field splitting, $\delta_{GS}$, between the $|0\rangle$ and $|\pm 1\rangle$ states. Under an applied magnetic field, the $|\pm 1\rangle$ state experiences a Zeeman splitting, $\delta_{\pm 1}^Z$. Optical pumping of the system (green arrows) causes a transition to the triplet excited state, $^3B$. From the excited state, there exists a radiative decay pathway (red arrow) and a nonradiative pathway via an intersystem crossing (black dashed arrows) through a metastable state, $^1A$. (c) ODMR contrast vs frequency, demonstrating the $m_s = 0$ to $m_s = -1$ and $m_s = 0$ to $m_s = +1$ transitions of a $V_B^-$ center electron spin. (d) Rabi oscillations shown via ODMR. Panel a is adapted with permission from ref.[285]. Copyright 2021 American Chemical Society. Panels c and d are adapted from ref.[287]

It was recently observed via ODMR measurements that the $V_B^-$ ground state ZFS varies with temperature and lattice strain. The change in the ground state ZFS due to temperature, $\Delta \delta_{GS}^T$, for hBN has a significantly higher coupling constant (-623 kHz/K) than NV$^-$:diamond (-74 kHz/K) and defects in SiC (-1.1 kHz/K)[288]. However, diamond has temperature and magnetic field resolutions of 4 orders of magnitude better than hBN, likely due to stronger PL emission and improved measurement protocols and contrast. Notably, while the dominant mechanism for the $\Delta \delta_{GS}^T$ in diamond is believed to be electron-phonon interactions[289], it has been suggested that the $\Delta \delta_{GS}^T$ in hBN is likely due to thermal expansion-related strain[282]. Once again, this suggests that the 2D structure of hBN could be leveraged to enable a next-generation strain sensor integrated in novel geometries.

Current diamond and SiC devices offer coherence times orders of magnitude longer than hBN. It should be noted, however, that control of $V_B^-$ defects as spin qubits is largely unexplored, and further research is needed to determine both fundamental and practical limitations of the coherence times in hBN. One clear issue is the non-zero spins of B and N nuclei, but spectral hole burning may mitigate this problem[287]. If coherence times in hBN cannot be extended to times comparable with diamond or SiC, spin defects in hBN may still be useful for quantum sensing and single photon emission, as discussed in the next section.

### 3.2.0 0D Hosts

*3.2.1 Nanodiamond*

Diamond nanocrystals, like bulk diamond, are an excellent host for defect centers. We discussed the phenomenal properties of $NV^-$ centers in diamond in this section's introduction. Diamond has high chemical stability and is biocompatible, so $NV^-$ centers in nanodiamonds have been explored primarily for quantum biosensing[290,291]. $NV^-$ centers are highly sensitive to magnetic fields, allowing single-molecule nuclear magnetic resonance and sensing of biomagnetic signals[292]. Obtaining $NV^-$: nanodiamond coherence times comparable to $NV^-$ centers in bulk diamond has been a challenge due to surface spin states and a greater density of N impurities in nanodiamonds. Nanodiamonds for quantum applications are typically synthesized by milling of bulk high-pressure-high-temperature (HPHT) diamond or by plasma-assisted CVD[290]. HPHT diamond tends to have a large amount of naturally occurring N impurities that can lead to decoherence. Using dynamic decoupling, a coherence time of 67 $\mu$s was demonstrated in a milled nanodiamond[293], still significantly shorter than in the bulk. Synthesis of nanodiamonds with controlled N doping and passivated surfaces is necessary to improve the coherence times – and therefore the sensitivity – of color centers in nanodiamond, making this a materials challenge. Achieving coherence times comparable to those $NV^-$ centers in bulk diamond would not only improve the sensitivity of $NV^-$: nanodiamond for biosensing, but also potentially allow nanodiamond to replace bulk diamond altogether. This would be highly advantageous from a cost standpoint and would also allow easier integration for sensing applications beyond biological systems.

*3.2.2 Magnetic Dopants in II-VI QDs*

Individual magnetic transition metal (TM) dopants in semiconducting II-VI quantum dots are another promising technology for spin qubits and qudits (multi-level quantum states). The hyperfine interaction of d electrons with the nuclei results in a zero-field splitting into a multilevel quantum system with a Hilbert space of more than 2 dimensions, enabling greater computational and memory capabilities. Moreover, *d* electrons are decoupled from the environment, potentially offering long coherence times for nuclear and electronic spins. Like REI dopants, TM dopants have weak oscillator strengths. It is therefore desirable to enhance the light-matter coupling of single TM dopants in an environment that does not decohere spin states.

Semiconducting II-VI QDs are highly optically active and have zero-spin nuclei that can be isotopically purified. TM dopants like $Mn^{2+}$, $Fe^{2+}$, and $Cr^+$ can be easily incorporated into II-VI cQDs at low dopant concentrations[294]. Dopant spins in II-VI QDs can be controlled electrically[295,296], optically[297,298], or via pulsed electron paramagnetic resonance (EPR)[299]. Because of the confinement, the wavefunction overlap of the spin defect and an exciton localized in the QD enables spin-photon coupling through the engineered interactions of the exciton with the defect[298]. Enhancement of the optical properties can also be done by incorporating the doped cQDs in optical nanocavities. With improvements in the positioning of individual nanoparticles[294,300,301], this has become a viable approach.

$Mn^{2+}$ dopants have an S = 1/2 electron and an I = 5/2 nucleus. The hyperfine interactions between these spins splits the $m_s = \pm\frac{1}{2}$ states into two separate six-level systems, between which there are six allowed transitions. This state space can be exploited for quantum computing and quantum memory. NOT and $\sqrt{SWAP}$ universal quantum gates have been experimentally realized for qudits in $Mn^{2+}$-doped cQDs. Using electron double resonance-detected nuclear magnetic resonance (EDNMR), coherent manipulation of $Mn^{2+}$ qudits between various states was demonstrated. Coherence times of approximately 8 $\mu s$ were shown, with $\pi/2$ pulse times of approximately 24 ns[302].

The multidimensional Hilbert space and optical addressability make TM:cQD systems interesting for optical quantum memories. Understanding and eliminating the sources of decoherence will be necessary to make these systems viable for practical applications. In addition to hosting TM dopant qubits, II-VI cQDs are also promising as single photon sources, as we discuss in the next section. This may present a unique platform to explore a spin-photon interface in which the mechanism for photon emission is excitonic emission from the host and not transitions of the defect qubit.

*3.2.3 Organic Molecular and Radical Spins*

Organic molecules have emerged as an unconventional potential qubit system. This includes the spins of molecular nanomagnets (MNMs), photogenerated states, and radical spins. The chemical tunability of these complexes is a unique feature that allows greater room for optimization and molecular design.

MNMs are metalorganic molecules with one[303,304] or multiple[305] TM/RE ions, surrounded by organic ligands. Fig. 7a shows the molecular structures of the single ion magnet (SIM), $[V(C_8S_8)_3]^{2-}$ and the MNM $Cr_7Ni$. MNMs can be processed in solution and then sublimated as ordered monolayers[306]. As discussed above for TM and REI dopants in solids, hyperfine interaction of an electron with a high-spin nuclei produces a high-dimensional Hilbert space. Leuenberger and Loss[307] proposed the use of MNMs as qubits. Since, coherent control via EPR has been demonstrated for various types of MNM qubits[304,308–310].

A promising application of MNMs is quantum memories with embedded QEC[305,311,312]. Multiple ancilla qubits can be encoded in the multilevel system and can be used to encode one logical qubit[312]. Additionally, molecules can be designed to have multiple coupled qubits or qudits, one of which serves as a memory while the rest act as a processor or ancilla bit for QEC. Lockyer *et al.*[305] proposed and synthesized a molecule with a $S = 1/2$ $Cr_7Ni$ ring coupled via an exchange interaction to a Cu ion, composed of an electronic spin coupled by hyperfine interaction to the $I = 3/2$ nuclear spin. In this scheme, the $Cr_7Ni$ qubit serves as a processor, the electron spin is an ancilla qubit, and the nuclear qudit acts as the quantum memory, respectively. Alternatively, Macaluso *et al.*[311] proposed implementing a 3-qubit phase-flip repetition code using a molecule containing a Er-Ce-Er complex. This scheme uses the center Ce ion as a quantum memory and the two Er ions as ancilla bits. The ability to design single molecules with a quantum memory and built-in error correction makes the pursuit of MNM qubits and qudits worthwhile.

MNM qubits are limited by relatively short coherence times. Coherence times are greatly influenced by the presence of nuclear spins in both the organic ligands and the surrounding environment, and, with careful tuning of molecular structure and choice of a solvent with minimal nuclear spins, coherence times up to 0.7 ms have been demonstrated in a $[V(C_8S_8)_3]^{2-}$ system in a $CS_2$ solvent [313]. However, coherence times for other complexes have been limited to several microseconds or less[304,310]. Given the particular interest in these systems for quantum memories, it is crucial to improve $T_2$.

Photogenerated states in organic molecules are another emerging molecular platform for qudits. Coherent control of photogenerated quartet states in organic molecules was recently demonstrated with coherence times up to 1.8 μs[314]. Recently, Wang *et al.* utilized the photoexcited triplet state of a $C_{70}$ fullerene molecule as a qutrit (three level analogue of a qubit). The quantum state interference a function of time while applying pulsed electron paramagnetic resonance was interpreted as coherent evolution of the qutrit states[315].

## 4.0.0 Superconducting Qubits

Superconducting circuits (SCs) are one of the most explored solid-state technologies for quantum computing. In this platform, qubits are encoded in the energy eigenstates of SCs. Compared to other platforms for quantum computing, SCs are macroscopic. Supercurrents are carried by Cooper pairs - paired electrons with zero net spin - sharing one macroscopic wavefunction with some phase, $\phi$[316]. The single wavefunction and the quantization of magnetic flux, $\Phi$, in a superconducting ring result in SCs behaving as microscopic quantum systems despite a macroscopic size. In this way, SCs are like artificial atoms with quantum properties that can be designed[317].

The Josephson Junction (JJ) is the building block of SCs. A JJ is a nonlinear device in which two superconductors are joined by a thin (< 5-10 nanometers) non-superconducting material – the weak link (aka tunnel barrier) – through which Cooper pairs can coherently tunnel. The current, $I$, and voltage, $V$, are related to the difference, $\delta$, in the superconducting phase across the junction:

$$I = I_C \sin\delta \qquad V = \frac{\Phi_0}{2\pi}\dot\delta$$

Where $\Phi_0$ is the flux quantum, $I_C$ the critical current, and $\dot\delta$ the time derivative of the phase difference, respectively. The Hamiltonian for this system can be found from the above Josephson relations to be[318]:

$$H = 4E_C n^2 - E_J \cos\phi$$

Where $E_C = e^2/(2C_J)$ is the charging energy with $C_J$ the total junction capacitance; $n = Q/(2e) = C\dot\Phi/(2e)$ is the excess number of Cooper pairs; $E_J = I_C \Phi_0/(2\pi)$ is the Josephson energy; and $\phi = 2\pi\Phi/\Phi_0$ is the reduced flux. One can therefore see that this system is essentially a perturbed quantum harmonic oscillator (QHO). Unlike a QHO, however, the anharmonicity causes unequal energy spacings between the energy eigenstates of the oscillator, enabling transitions between energy levels to be selectively addressed. Qubit operations are conducted by applying pulses of microwave radiation.

JJs are typically made using superconducting metal thin films (typically Al or Nb) with a thermal oxide (AlO$_x$) as the weak link. Figure 7b shows a SEM image of a typical JJ. Different types of superconducting qubits can be formed from different circuit configurations. Common qubit types include the transmon[319], X-mon[320], flux qubit[321], and fluxonium[322]. These different types of qubits arise from differences of how JJs are shunted in a circuit. Circuit-level analysis of superconducting qubits is covered in a recent review[318]. Larsen *et al.* introduced a new type of qubit – the gatemon – using an InAs-Al core shell NW[323]. In this device, the insulating "normal" material in a superconductor-normal-superconductor (SNS) JJ is replaced with a semiconductor. In a JJ with the N material being an insulator, $E_J$ is fixed. However, if the N material is a semiconductor, $E_J$ can be tuned by a gate voltage. A qubit based on this kind of JJ is called a gatemon. Electrostatic control of qubits instead of flux-based control reduces resistive dissipation.

Readout of superconducting qubits is most commonly done by dispersive readout using superconducting microcavity waveguides[324]. In these systems, superconducting qubits are entangled with cavity photons, enabling long distance coupling, qubit control, and operations on multiple logical qubits[325]. State-of-the-art superconducting quantum computers have qubits with $T_1$ and $T_2$ on the order of $10^2$-$10^3$ μs, with two-qubit gate times $10^3$ orders of magnitude shorter[324,326]. Two-qubit gate fidelities of 99% have been demonstrated on logical qubits[327]. Superconducting NISQ processors have been realized as well, including the computer that was used to demonstrate quantum supremacy[13]. Scalability and technological maturity make SCs very promising for quantum computation.

Two level systems (TLSs) are the dominant source of noise and decoherence in superconducting quantum devices[328]. These parasitic systems have an electric dipole that couples to the superconducting resonator, leading to dissipation and dephasing. TLSs can arise from the spins of molecular adsorbates on the surface, contaminants, and structural damage[329]. The dominant source of decoherence, however, is dielectric loss[330]. These sources arise from defects in the dielectric layer, oxidation, and impurities/damage induced during the fabrication process. It is therefore clear that solving decoherence issues in superconducting qubits is largely a materials science challenge.

**4.1.0** *2D Materials for Josephson Junctions*

2D vdW materials are excellent candidates to replace conventional materials for both the weak link and the superconducting material. vdW heterostructures can be formed by stacking individual 2D materials, either by dry transfer or direct growth. These heterostructures have clean, atomically precise interfaces and high crystallinity. The crystalline, atomically flat and dangling-bond free surfaces are in stark contrast to the inherent roughness and nano-crystallinity of any sputtered or evaporated metal thin film. Moreover, a highly crystalline weak link without dangling bonds, such as graphene or hBN, will have fewer TLSs than an amorphous thermal oxide. 2D JJs are therefore extremely promising alternatives to conventional devices.

*4.1.1 2D Materials as the Weak Link*

JJs based on 2D materials as a weak link between bulk superconductors have been explored in recent years. vdW materials are an ideal tunnel barrier for vertically stacked JJs and should be able to overcome issues with $AlO_x$ thermal oxides. vdW materials lack dangling bonds and therefore eliminate TLSs arising from these defects. Additionally, atomically sharp interfaces can be created. Finally, atomically-thin tunnelling barriers make this interface transparent to superconductors.

Devices using graphene as the weak link between bulk superconducting metals have been demonstrated. A gatemon made from Al electrodes contacting a gated graphene channel (Fig. 6a) was shown with a coherence time of 55 ns[331] (Fig. 6b). Lee *et al.*[332] demonstrated a qubit with a vertically stacked JJ composed of Al superconducting leads and few-layer MOCVD-grown $MoS_2$. The coherence time was limited to 12 ns but could likely be improved by optimizing the junction dimensions, eliminating polymer residue on the MoS2 from the transfer process, and by improving the quality of the growth material. MoS2 (and other TMD semiconductors) could also be used for gatemon qubits or other superconductor-semiconductor hybrid devices[333]. hBN is another potential weak link material due to its low dielectric loss, lack of dangling bonds, and robustness as a tunnel barrier[334].

Heterostructures of 2D materials have also been explored as the weak link material. It was demonstrated Gr/WS2 heterostructures exhibited greater critical magnetic fields due to the enhancement of the spin-orbit interaction in graphene by the TMD[335]. Signatures of quasi-ballistic edge states in graphene stabilized by spin-orbit interaction were seen for longer lengths of the normal material. This suggests that heterostructures of 2D materials as the weak link in a lateral JJ could be used to study topological effects.

*4.1.2 2D Superconductors*

Superconductivity has been discovered in various TMD materials, at interfaces, in topological systems, and in twisted heterostructures. Superconductivity at the 2D limit was first demonstrated in monolayers of Pb atoms on a Si surface[336].

Superconductivity in 2D TMDs has been demonstrated and obeys conventional Bardeen-Cooper-Schrieffer (BCS) theory[337]. NbSe2[338], TaS2, IrTe2[339], etc. are intrinsic superconductors, and superconductivity can be induced in other TMDs by ion-gating[340], doping[341,342], or the proximity effect[343]. Large Ising spin-orbit coupling (ISOC) in TMDs allows large in-plane critical magnetic fields due to Ising spin-orbit protection[344]. Superconductivity down to the monolayer limit has been demonstrated in NbSe2 and TaS2. The effect of thickness on the critical temperature, $T_C$, differs for these materials: $T_C$ increases with decreasing thickness for TaS2, while the opposite occurs for NbSe2[345] (Fig. 6e). While group V-b TMDs oxidize readily, they can be encapsulated with hBN in an inert environment to protect from oxidation and create a surrounding dielectric environment with low loss due to hBN's high crystallinity and lack of dangling bonds. Group V-b TMDs tend to exhibit competition/interplay between the charge density wave (CDW) phases and superconducting phases[346,347]. In both NbSe2 and TaS2, it is found that applying compressive pressure favors the formation of the superconducting phase[348,349]. Understanding the interplay between these two phases in these materials may elucidate the unconventional superconductivity in these systems.

Superconductivity has been shown in other classes of 2D materials as well. 2D transition metal carbides (TMCs) such as Mo2C have been fabricated and exhibit superconductivity in few-layer films[350]. Monolayer FeSe grown on SrTiO3 (STO)[351] and other oxide substrates[352,353] has emerged as a high-$T_C$ superconductor, with a $T_C$ exceeding 100 K in some experiments. Unconventional superconductivity was discovered in Moiré superlattices of graphene (and graphene with hBN) arising from flat minibands[354–356]. tBLG has a low density of carriers compared to most superconductors, indicating ultra-strong correlation between paired electrons. Interestingly, tBLG and twisted trilayer graphene (tTLG) enable electrostatic tuning of the superconducting phase[354,355]. This makes twisted graphene structures an exciting system for studying novel superconducting physics and highly correlated electron states. Tian *et al.* fabricated hybrid Josephson Junctions superconducting films and 2D vdW superconducting materials by implementing a Nb/Au/NbSe2 hybrid Josephson junction[357] (Figure 7e). This device demonstrated supercurrent inhomogeneity, likely due to the roughness of the Nb/Au surface. By using an all-2D structure, surface roughness should be negligible.

*4.1.3 All-2D JJs*

All-2D JJs have been fabricated, combining some of the above 2D superconductors with 2D weak link materials. With large-area CVD synthesis of 2D superconductors[358,359], all-2D superconducting circuits are feasible. Strong proximity coupling in all-2D $NbSe_2$/Gr/$NbSe_2$ JJs provides a system with high transparency[360,361], displaying promise for coherent qubits to be formed from these devices. Fig. 6f shows a vertical vdW JJ constructed with a graphene weak link between several-layer $NbSe_2$ superconducting leads. Similarly, Xu *et al.* made an all-2D JJ using graphene as the weak link between 2D superconducting $Mo_2C$, again displaying strong coupling and transparent interfaces[358]. Ai *et al.* recently fabricated vdW JJs using $NbSe_2$ as the superconductor and $Cr_2Ge_2Te_6$ (CGT), a 2D ferromagnetic insulator, as the weak link[362]. These devices exhibited a hysteretic response of the supercurrent and resistance in an in-plane applied magnetic field and also showed evidence of π-phase coupling. This result hints at the promise of using the vast and ever-expanding catalog of (anti)ferromagnetic and multiferroic vdW insulators and semiconductors in vdW JJs. To our knowledge, neither coherence time measurements nor coherent control of qubits based on all-2D JJs have been demonstrated. Despite this, all-2D JJs are an intriguing platform for future devices with high coherence and for the studies of exotic quantum transport due to the unconventional superconductivity and other potential topological effects.

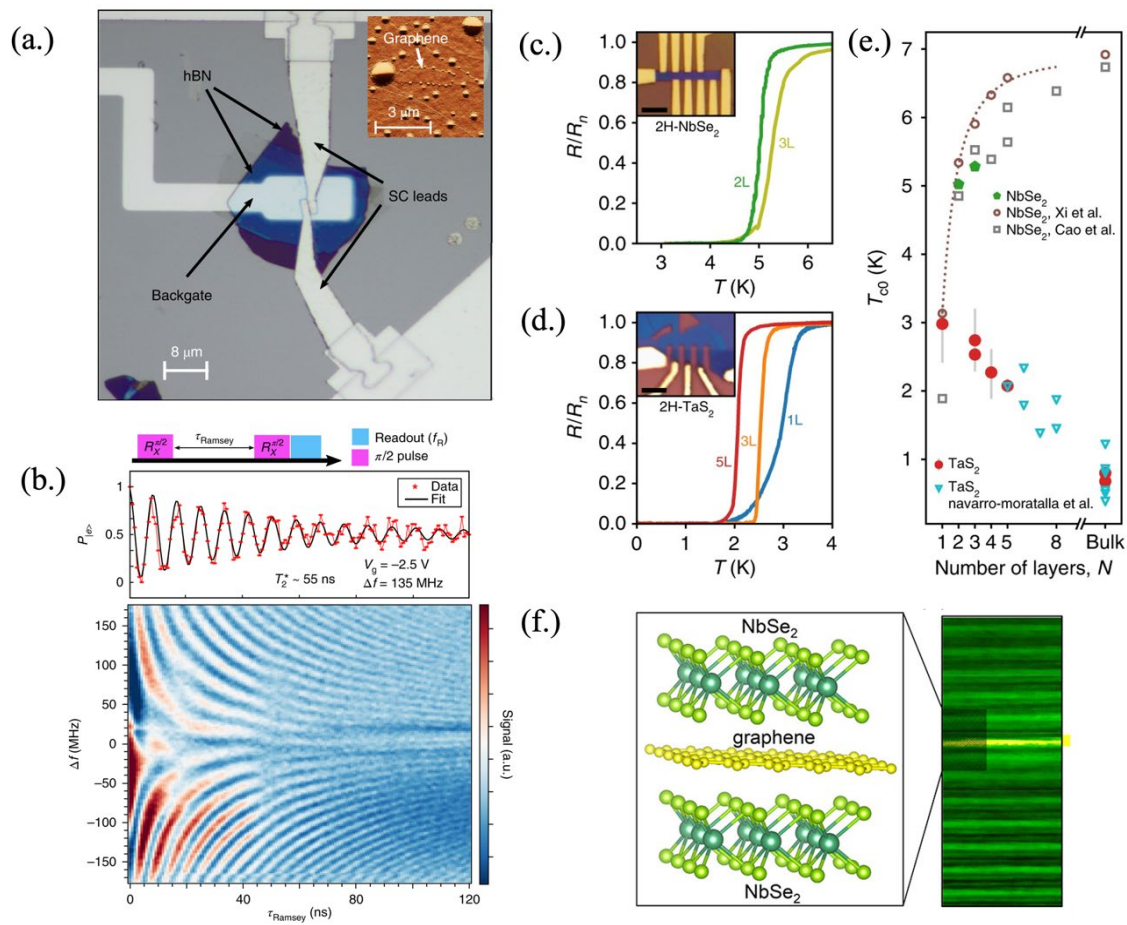

*Figure 6. 2D Materials for Superconducting Qubits.* **(**a.) Optical image of an Al-Gr-Al JJ. (b.) Ramsey fringing at a finite detuning (top) and as a function of detuning (bottom) indicating coherent control of the device in panel (a.). Layer-dependent resistivity vs. temperature curves for (c.) 2H-NbSe$_2$ and (d.) 2H-TaS$_2$. (e.) Layer-dependent critical temperature for NbSe$_2$ and TaS$_2$.(f.) A schematic and false-color TEM image of a NbSe$_2$-Gr-NbSe$_2$ JJ. Panels (a. and b.) adapted from ref.[331] Panels (c, d, e.) adapted from ref.[345] Panel (f.) adapted from ref.[361]

## 5.0.0 Topological Qubits and Superconductor-Semiconductor Hybrid Devices

Non-Abelian anyons are topological states of matter that are neither Fermionic nor Bosonic and obey non-Abelian braiding statistics. Kitaev theoretically proposed that non-Abelian anyons could be utilized for inherently fault-tolerant computation, using anyon braiding to protect against decoherence[363]. Topological quantum computation with anyons has since become a long-term goal and ideal approach for fault tolerant quantum computation. A physical implementation of a topological quantum computer requires that non-Abelian anyons be found, manipulated/braided, and read out[27]. Abelian anyons – anyons that obey Abelian statistics – have been experimentally demonstrated in $\nu = 1/3$ fractional quantum Hall (FQH) states in 2DEGs[364], and Nakamura *et al.* recently observed anyonic braiding[365].

Discovery of non-Abelian anyons is the current challenge. Degeneracies of Majorana bound states (MBSs) have been proposed for discovering non-Abelian anyons. Majorana fermions are particles that are their own antiparticle. Majorana bound states have been predicted in vortices of *p*-wave superconductors[366], at topological insulator-superconductor interfaces[367], and at the end of superconductor-proximitized semiconductor NWs (Josephson NWs)[368]. The $\nu = 5/2$ FQH state was also predicted to be non-Abelian [369]. However, experimental reports of Majorana bound states and non-Abelian anyons have been controversial[370]. In this section, we briefly introduce MBSs and then consider novel topological materials that may potentially be used to realize these states.

### 5.1.0 Hybrid Semiconductor-Superconductor Devices

Hybrid semiconductor-superconductor devices are one potential platform for the realization of a type of MBS called Majorana zero modes (MZMs). In general, hybrid semiconductor-superconductor devices have been primarily explored using nanowires, but 2D materials can also be used. We introduced the gatemon qubit in section 4. This device can be modified by electrostatically defining quantum dots in the semiconductor weak link of the JJ. The resulting confinement results in the emergence of hybrid states.

Andreev reflection occurs when an electron traveling in a normal metal is incident on a superconductor. In this situation, a Cooper pair is injected into the superconductor and a hole is reflected back into the normal material. This enables conductance through a JJ below the superconducting gap. Under confinement, this process yields discrete Andreev bound states (ABSs). Despite sometimes exhibiting a zero-bias conductance peak, these states are topologically trivial[371]. Chtchelkatchev and Nazarov predicted that quasiparticles in ABSs could be brought into long-lasting spin-1/2 states that, with spin-orbit interaction, have a spin-dependence coupling to the supercurrent carried by the Andreev levels[372]. Coherent control of an Andreev spin qubit (ASQ) in an InAs Josephson NW was recently demonstrated[373].

It is predicted that narrow-bandgap semiconductor nanowires with strong Rashba SOC and a hard superconducting gap proximity-induced by superconducting leads can undergo a topological phase transition resulting in the existence of Majorana fermions at the ends of the NW[368]. Experimental realization of this requires defect-free, transparent superconductor-semiconductor interfaces[132]. InAs and InSb NWs have strong Rashba SOC and superconducting metals can be epitaxially grown on these materials[374]. Therefore, these materials are the leading candidate for devices in the search for Majorana fermions[375,376].

Majorana fermions have also been theoretically predicted in Ge-Si core-shell nanowires (CSNW) with superconducting leads[377]. On the experimental side, helical states – a prerequisite for Majorana fermions – have also been demonstrated in Ge-Si CSNWs[378]. Moreover, Ge-Si CSNW Josephson junctions have been demonstrated with high-quality interfaces and coherent transport, indicating promise for Andreev qubits and Majorana devices[379,380]. As discussed in section 2.2.1, however, improved quality of Ge-Si CSNWs and growth without Au diffusion is necessary for potential topological devices, likely more so than for spin qubits.

Similarly, CNTs exhibit promise for topological superconductor-semiconductor hybrid devices. It has been theoretically predicted that proximity-induced superconductivity and gapless topological states can be induced in CNTs, enabling Majorana fermions/quasiparticles[381–384]. Recently, Desjardins *et al.* used a proximal ferromagnetic multilayer to enhance spin-orbit coupling in a CNT contacted by superconducting electrodes; they observed oscillations of sub-gap states with applied magnetic field and a zero-bias peak[385]. Bäuml *et al.* have since demonstrated supercurrent and phase slips in a bundle of CNTs encapsulated by h-BN and 2D $NbSe_2$[386]. These recent results indicate promise for CNTs in superconductor-semiconductor hybrid devices and – potentially – also for confirming presence of Majorana states in devices.

### 5.2.0 2D Topological Materials

#### 5.2.1 2D Topological Insulators (2DTIs)

A seminal work by Kane and Mele predicted that spin-orbit coupling in graphene would open a topological gap in the bulk and convert graphene into a quantum spin Hall insulator (QSHI) with metallic edge states[387]. The Kane-Mele model can be generalized to other 2D systems with a similar honeycomb lattice. 2D materials exhibiting a spin-orbit topological gap and metallic edge states are referred to as 2D topological insulators (2DTIs). Unlike 3D topological insulators, which have a 2D metallic surface and insulating bulk, 2DTIs have 1D metallic states that are protected against weak disorder[122].

The topological gap predicted in graphene is on the order of $\sim 10^{-3}$ meV due to the minimal intrinsic SOC[387]. Instead, other 2D materials with a similar honeycomb lattice but greater SOC have been suggested. Silicene, germanene, and stanene are elemental group-IV 2D materials with a honeycomb lattice like graphene, but the slightly buckled structure and greater atomic mass lead to greater SOC; these materials have been predicted as QSHIs with greater spin-orbit gaps than graphene[388–390]. Various other emerging 2D materials have been experimentally shown to be QSHIs: bilayer bismuth[391]; antimonene[392]; MBE-grown monolayer 1T'-$WSe_2$ [393]; a 2D In-Sb compound grown on Si[394]; $ZrTe_5$[395]; etc. Various other 2D QHSIs have been theoretically predicted[396–399]. Notably, phosphorene was theoretically shown to have an electric-field induced topological insulator phase[400]. Similarly, graphene nanoribbons were predicted to have an electric-field-tunable topological phase[401]. Work remains to synthesize the theoretically predicted QHSIs and to further characterize the topological states of the QHSIs that have been experimentally realized.

*5.2.2 2D Topological Superconductivity*

*p*-wave superconductivity is desirable for the realization of Majorana fermions, but materials exhibiting p-wave superconductivity are rare in nature. Alternatively, heterostructures of topological insulators (or other materials exhibiting strong Rashba SOC) with *s*-wave superconductors can lead to *p*-wave-like topological superconducting states by the proximity effect[367]. 2D materials are an excellent candidate for this approach for several reasons: (i.) 2D materials can be easily stacked together via dry transfer to form ultra-clean interfaces, eliminating the need for direct growth of dissimilar materials on one another; (ii.) many 2D materials, particularly transition metal compounds or x-enes with a buckled structure, exhibit high Rashba SOC; and (iii.) because proximity induced superconductivity occurs at interfaces, 2D materials inherently lack any undesirable contribution of bulk states.

Proximity-induced topological superconductivity has been demonstrated in heterostructures of 1T'-$WTe_2$/$NbSe_2$[402], ultra-thin Bi/$NbSe_2$[403], $Bi_2Te_3$/$NbSe_2$[404], etc. Continuing to explore heterostructures of QHSIs and *s*-wave superconductors will most likely yield more candidates for topological superconductivity.

FeSe/STO, which we discussed previously as a high temperature superconductor, was shown to have an antiferromagnetic quantum spin hall insulating gap of ~40 meV with topological edge states[405]. The discovery of topological states and high temperature superconductivity in one material make this system interesting for exotic devices and studies of fundamental physics.

As mentioned in section 4.1.1, using a graphene/TMD heterostructure as the weak link in a JJ resulted in spin-orbit stabilized quasi-ballistic edge states in graphene. Exploring heterostructures of 2D materials with exhibiting a spin-orbit proximity effect as the weak link in a JJ may result in the discovery of new topological modes.

**6.0.0 Nanomaterials for Quantum Communications and Photonic Quantum Computing**

*6.1.0 Single photon sources*

Travelling qubits encoded in the polarization of single photons are an important resource for quantum information processing and can be utilized for all-optical quantum computation[30], quantum simulation[406], photonic quantum sensing[407], and quantum communications[32]. Single photons can also be used to interface with stationary qubits such as spins in quantum dots, spin-valley qubits, and spins in defects. Thus, a reliable single photon source is crucial for the realization of many quantum technologies.

A quantum emitter (QE) must demonstrate bright, deterministic emission of high-purity, indistinguishable single photons with high extraction efficiencies. Photon brightness is characterized by the emission rate; $10^9$ photon/sec brightness is generally desired[408]. Deterministic operation requires coherent absorption and emission and is crucial for read/write based protocols. Perfect coherence requires that the optical coherence time $T_2 = 2T_1$ where $T_1$ is the radiative lifetime; this is demonstrated by Fourier transform (FT) - limited linewidths: the spectral linewidth of the emitter is limited by the FT of its PL lifetime[409].

Photon indistinguishability is directly related to coherence and is how identical (emission energy and polarization) the individual photons are to each other. Hong-Ou-Mandel (HOM) interference measurements can be used to determine indistinguishability, $I$. For perfectly indistinguishable photons, $I = 1$. Indistinguishability is not a crucial metric for QKD, but $I > 0.99$ is required for photonic quantum computing and all-optical quantum repeaters[408]. Indistinguishability is related to the linewidth, spectral broadening, and dephasing of an emitter. The linewidth of an emitter ($\Gamma$) depends on the radiative decay rate ($\gamma$), dephasing rate ($\gamma^*$), and spectral broadening ($\Delta\delta$): $\Gamma = \gamma + \gamma^* + \Delta\delta$. $I$ is approximately $\gamma/\Gamma$. To maximize $I$, $\gamma^* + \Delta\delta$ must be negligibly small compared to $\gamma$. Emission is coherent and photons are indistinguishable when the linewidth is approximately equal to the radiative decay rate, therefore, demonstration of Fourier transform (FT)-limited linewidths demonstrates coherence and spectral indistinguishability. At room temperature, $\gamma^*$ is typically between $10^3\gamma$ to $10^6\gamma$ for most emitters[410]. Therefore, $\Gamma \gg \gamma$ and $I$ approaches zero. By operating at lower temperatures, $\gamma^*$ is reduced. Indistinguishability can also be significantly improved by integrating the emitter with an optical cavity[410,411]. Use of optical cavities also enhances the extraction efficiency and increases the emission rate by Purcell enhancement. For integrated photonics, this is done by incorporating the emitter in a dielectric cavity, photonic crystal, or plasmonic microcavity/nanoantenna[412–414]. Choosing materials platforms for QEs that can easily be integrated with cavities is therefore crucial for practical applications.

Narrow linewidths can be achieved in solid-state sources by emission into zero-phonon lines (ZPLs)[248]. The percent of emission into the ZPL is called the Debye-Waller (DW) factor and should be maximized. For applications requiring transmission over long distances through fiber optic cables, it is ideal to have photon emission at a wavelength in one of the telecom bands, so it is ideal to find single photon sources with a ZPL in the 1.3-1.55 μm range.

The purity of a photon source is determined from Hanbury-Brown-Twiss (HBT) interferometry and characterized by the resulting second-order autocorrelation function value at a time 0, $g^{(2)}(0)$. The threshold for single photon emission is $g^{(2)}(0) < 0.5$. Quantum key distribution requires $g^{(2)}(0) < 0.1$, but other photonic quantum applications require a $g^{(2)}(0) < 0.001$[408]. Finally, extraction efficiencies should be > 99%.

Defect-based single photon emission (SPE) is one of the most common mechanisms. Various defect centers in large bandgap materials (diamond, SiC, etc.) exhibit SPE up to room temperature[415]. $NV^-$ centers in diamond, which we mentioned in the context of defect spin qubits in section 3, are among the most widely explored single photon sources. $NV^-$ centers have a relatively high quantum efficiency of ~70%, but a Debye-Waller factor of 0.04 limits the linewidths that are achievable closer to room temperature[416]. The electron-phonon coupling-based change in the ZFS energy[289] and electric dipole[408] cause the emission energy of $NV^-$ centers to be highly sensitive to strain and temperature. As discussed in the previous section, this is useful for quantum sensing, but it is not ideal for most applications of QEs. SiV centers are an alternative with a large DW factor of 0.8 and preserved inversion symmetry. Silicon vacancies also have a shorter radiative lifetime (~1 ns) than $NV^-$ centers, but they have a low quantum efficiency that limits the brightness[408]. SiC color centers also are single photon sources. SiC color centers emit closer to telecom wavelengths: the ZPL for the $V_{Si}^-$ center ranges from 861-916 nm and the ZPL for the $V_C V_{Si}^0$ center ranges from 1037-1149 nm[268].

### *6.2.0 EPR pair sources*

The emission of polarization-entangled photon pairs is the building block of quantum communications, as discussed above in the introduction. Spontaneous parametric down-conversion (SPDC)[417,418] is among the oldest and most commonly used EPR pair sources. In SPDC, a pump photon with frequency $\omega_p$ incident on a crystal with a susceptibility of quadratic order may spontaneously split into signal and idler photons. Energy and momentum conservation introduce two phase-matching conditions that must be satisfied for SPDC to occur[419]:

$$\omega_{sig} + \omega_{idle} = \omega_p$$

$$k_{sig} + k_{idle} = k_p$$

The signal and idler photons are produced in an entangled state. SPDC is categorized by the polarizations of the two output photons. If the signal and idler photons share the same polarization to each other, it is Type I SPDC. If the signal and idler photons have perpendicular polarizations, it is deemed Type II SPDC. While this process is simple and produces EPR pairs with high-fidelity entanglement, it is poissonian and therefore not reliable for practical applications. Greatly improved SPDC efficiency with compact, integrable sources or another EPR pair source is clearly needed for a large-scale quantum internet.

Biexciton decay in semiconductor quantum dots is the leading technology for integrated EPR pair emission. Biexcitons localized in the QD (Fig. 7a) decay radiatively from the $XX$ state to one of two $X$ states, then from $X$ to ground. Depending on the spin of the intermediate $X$ state, the photons are emitted as $|L\rangle_{xx}|R\rangle_x$, denoting a left-hand-circularly polarized (LHCP) photon emitted from the $XX \to X$ transition and RHCP photon emitted via the $X \to G$, or $|R\rangle_{xx}|L\rangle_x$, denoting the inverse pathway. Assuming that the two intermediate X states are degenerate, the state of the emitted photons can be re-written in the basis of horizontal (H) and vertical (V) polarization as the maximally entangled Bell state $|\Phi^+\rangle = \frac{1}{\sqrt{2}}(|HH\rangle + |VV\rangle)$. In reality, strain, asymmetry, and concentration gradients in the dots cause the two $X$ states to hybridize into bonding and antibonding states with a fine-structure splitting, $S$[420]. This introduces a phase shift between the two decay pathways, making them no longer equivalent and therefore degrading the entanglement fidelity (Fig. 7b). $S$ can be tuned and minimized by applying strain or an electric or magnetic field[421–423].

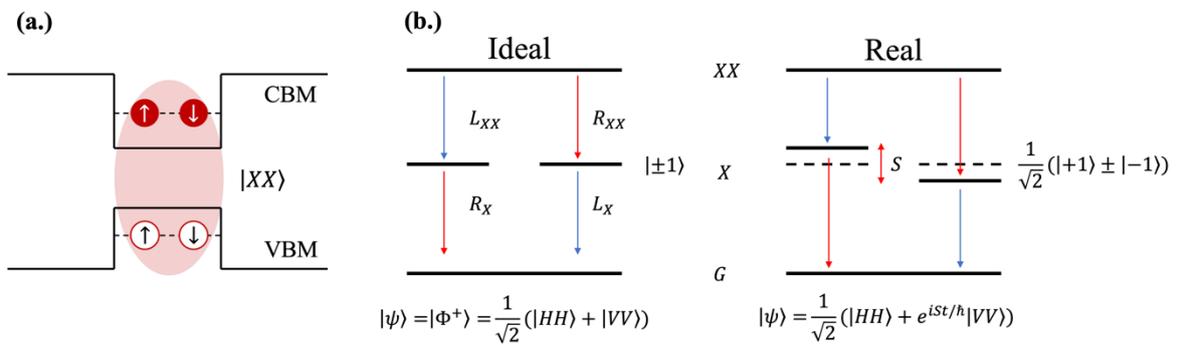

*Figure 7. EPR pair generation via biexciton decay in semiconductor QDs.* (a.) Band alignment and quantum confinement enable the formation of a biexciton. Filled circles in the conduction band denote electrons while unfilled circles in the valence band denote holes. The spins are shown by the arrows. A bright biexciton has a net

spin of 0. (b.) The process of biexciton decay in ideal and real quantum dots. In a real quantum dot, the fine structure splitting $S$ introduces a phase factor in the wavefunction that degrades the fidelity of the state to the Bell state $|\Phi^+\rangle$.

### 6.3.0 2D Material-based QEs

### 6.3.1 QEs in hBN

Defects in hBN were discussed as spin qubits in the previous section. Here we focus on their optical properties and their potential as a room-temperature single photon source. hBN-based defects emit polarized single photons with a high quantum efficiency (87±7%)[424], high purity, and narrow linewidths. Additionally, the emission energy is highly tunable by the Stark effect[425], surface acoustic waves[426], and strain[427].

Defects can be created with near-deterministic placement in hBN flakes using femtosecond laser pulses[285], focused ion beam (FIB) radiation[284,428], electron beam radiation[429,430], transferring of flakes onto nanopillars[431], direct growth of hBN by CVD on nanopillars[432], and annealing of arrays of hBN nanocrystals[433]. Most of these techniques result in large spectral inhomogeneities. Recently, Fournier *et al.*[430] used E-beam radiation to fabricate arrays of emitters in hBN with high spectral homogeneity (Fig. 8d and e). Continued improvement over the deterministic creation and placement of emitters is crucial.

Identifying the nature of defect emitters in hBN is a current challenge that is complicated by the many potential defect types. A variety of defect emitters in hBN have been reported with different emission wavelengths[283,434–436]. Fig. 8a shows common emission peaks. The $V_B^-$ defect investigated primarily as a spin defect has a broad emission peak at ~1.5 eV[437], and the origins of the ZPLs at 2.0 eV, 2.1 eV, and 4.1 eV have been proposed to be boron dangling bonds[438,439], a $V_B C_N^-$ defect[440], and a carbon dimer, $C_B C_N$[441], respectively. The atomic structure of the boron dangling bond defect is shown in Fig. 8b. Its proposed electronic structure is a singlet ground ($^1A$) and excited state ($^1B$) with an intersystem crossing to a metastable triplet state ($^3B$) (Fig. 8c). More studies are needed to solidify the understanding of the active defects and their electronic structures. Additionally, it is crucial to understand how to deterministically create specific defects with control over position.

Defects in hBN have a high DW factor[435] and a low Huang-Rhys factor, enabling narrow linewidths even at room temperature. Resonant excitation has been used to demonstrate room-temperature emission at ~635.5 nm with linewidths at the FT limit[409]. This result suggests the potential for hBN color centers to be sources of highly indistinguishable single photons at room temperature.

The 2D structure of hBN allows it to be easily integrated with various structures and on arbitrary substrates. Additionally, it causes emitters to be located at – or close to – the surface. These two factors allow hBN QEs to be integrated into optical cavities with relative ease, which enhance the properties of the emitters. Emitters in a DBR cavity exhibited a reduction in the $g^{(2)}(0)$ from 0.051 to 0.018, a reduction in the PL lifetime from 837 ps to 366 ps, and a reduction in linewidth by a factor > 25[442], as compared to bare emitters. Metal-dielectric antennas coupled to hBN emitters increased the extraction efficiency to 98%[443]. The $V_B^-$ spin defect considered for spin qubits has a relatively weak, broadband emission peak on its own. However, Fröch *et al.*[444] recently demonstrated that coupling a $V_B^-$ spin qubit to a bullseye cavity (Fig. 8f) can enhance emission intensity and linewidth with negligible effects on the spin coherence time. As spectral homogeneity improves, coupling hBN emitters to cavities will continue to improve the properties of hBN QEs even more.

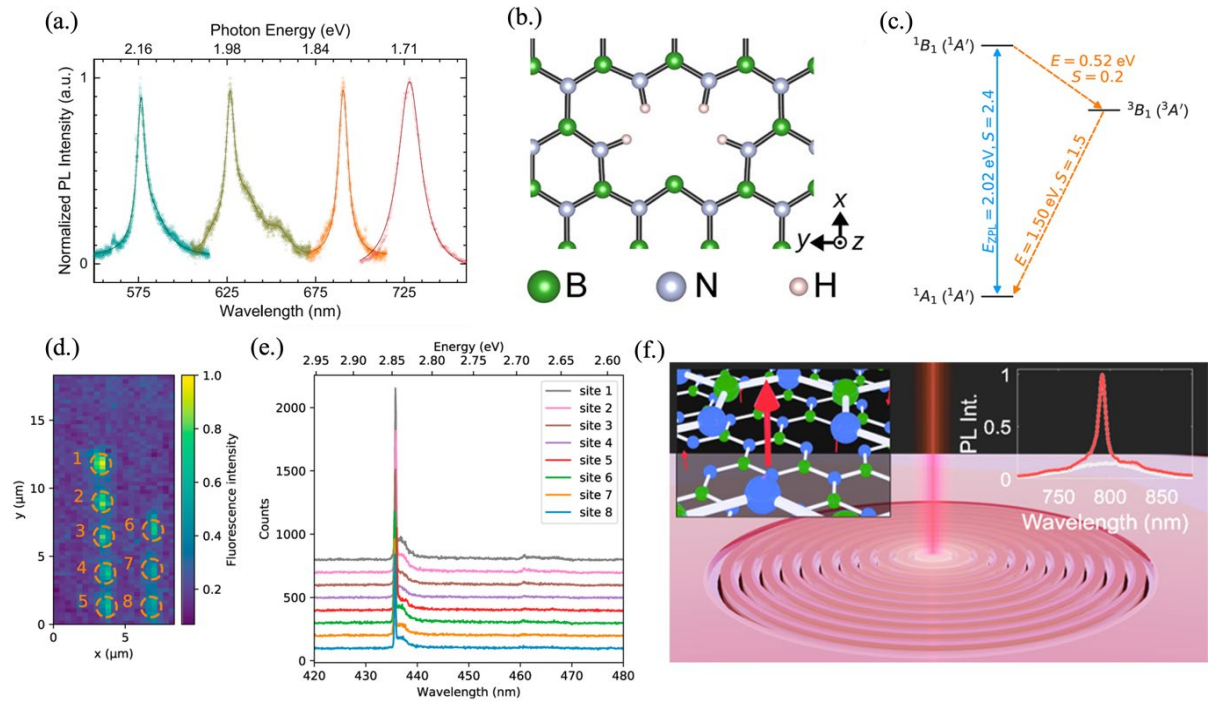

*Figure 8. Single photon sources in hBN.* (a.) Common ZPLs from defects in hBN. (b.) The atomic structure and (c.) the calculated electronic structure of the boron dangling bond defect in hBN, the defect proposed to be responsible for the ~2 eV ZPL. (d.) PL map and (e.) PL spectra of an array of position-controlled emitters in hBN. All peaks are within a 0.7 nm window. (f.) A schematic of a $V_B^-$ defect (upper left inset) integrated with a bullseye resonator and the enhancement of the emission due to the resonator (upper right inset). Panel (a.) is adapted from ref.[445] Panels (b. and c.) are adapted from ref.[439] Panels (d. and e.) are adapted from ref.[430] Panel (f.) reprinted from ref.[444]

One advantage of hBN over bulk hosts for defect QEs is the ability to transfer hBN layers to arbitrary substrates and integrate them in vdW heterostructures. The incorporation of hBN QEs in heterostructures with graphene QD spin qubits or other 2D spin qubits could enable a spin-photon interface and allow for optical readout of spins[286].

Understanding of the origin and structure of QEs in hBN is currently well behind that of QEs in more mature hosts like diamond and SiC. Improved understanding of SPE defect types and their electronic structures will allow improved spectral homogeneity, deterministic defect placement, and improved control of defect qubits and QEs in hBN. Improving the spectral homogeneity will also make it easier to design integrated nanophotonic cavities for Purcell enhancement of the emitters. Despite the current limits of understanding, FT-limited linewidths[409], sub-ns radiative lifetimes[443], and near-unity extraction efficiencies[442,443] have already been demonstrated in hBN. As the understanding of the physics, chemistry and materials science related to the defects improves, hBN is expected to become competitive with diamond and SiC for practical applications.

*6.3.2 QEs in TMDs*

2D TMDs are another promising 2D host for QEs. Group-VI-b TMDs have direct bandgaps in the monolayer limit and are highly excitonic due to quantum confinement and reduced dielectric screening[446]. The two-dimensionality of TMDs allows QEs to be more easily integrated with nanophotonic cavities to enhance emission rates. SPE has been demonstrated or theoretically predicted using defects[447–453], strain[454–457], moiré potentials[458,459], or combinations of the above[460–462]. Fig. 9a shows the different effects of defects and localized strain on the electronic band-structure. Defect sites create bright states in the bandgap of the TMD into which a carrier in the conduction band continuum can relax before recombining. Meanwhile, localized strain creates a spatially small well in the conduction band into which one or few electrons can relax, depending on the size of the quantum-confined area.

Like hBN, arrays of defect-based QEs can be deterministically created in TMDs using a Helium ion FIB[450,452,463] or e-beam irradiation[460]. The optically active defects have been shown to be chalcogen vacancies[451]. Barthelmi et al. used a He-ion beam to create sulfur vacancies in hBN-encapsulated $MoS_2$ (Fig. 9d)[453]. Ultra-sharp emission was shown from these defects with relatively high purity (Fig. 9e), indicating S vacancies in $MoS_2$ are a promising new color center for SPE. The ability to deterministically position defect emitters with sub-10 nm accuracy is conducive to embedding emitters in photonic nanostructures[463]. Hötger et al. showed SPE from an array of defect QEs in hBN-encapsulated monolayer $MoS_2$ created with a He-ion source could be electrically tuned[452]. By applying a gate voltage, the emission could be switched between defect-related single photons, the neutral exciton, and trions.

Localized strain is typically created by transferring TMD flakes onto arrays of patterned nanostructures[454,464,465]. Gold nanostars (AuNSs) with tip radii <5 nm dispersed on the surface of 1L-$WSe_2$ were used by Peng et al. to create strain-localized emitters[455]. In addition to creating multiple localized QEs per site through applying compressive strain to the TMD, the AuNS plasmonic resonance coupled to the emitter and reduced the radiative lifetime as compared to using dielectric nanopillars from 11.2±1.67 ns to 5.5±0.66 ns via Purcell enhancement, improving brightness. Recently, Zhao et al. showed site controlled, tunable telecom-wavelength SPE from a $MoTe_2$ monolayer transferred onto a nanopillar array at temperatures up to 77K[466]. A $g^2(0)$ value of 0.058±0.03 (at 11K) indicates high purity, but a radiative lifetime of ~20 ns must be improved upon. Nevertheless, this is a promising step towards high purity, telecom wavelength SPE from TMDs.

Recently, Parto et. al (2021)[460] used a combination of defect and strain engineering to demonstrate excitonic and biexcitonic SPE in hBN-encapsulated $WSe_2$ at temperatures up to 150K. They transferred hBN-encapsulated $WSe_2$ onto $SiO_2$ nanopillars to create strain, then irradiated the strained regions with an e-beam to create a defect level inside the bandgap. Fig. 9b shows this structure schematically, and the PL maps show emission not corresponding to the bulk exciton is localized at the strain/defect locations. The exciton, X, and biexciton, XX, emission peaks and the fine structure splitting between the horizontally (H) and vertically (V) polarized emission peaks are seen in Fig. 9c. The second order correlation plots in Fig. 9c and extracted $g^2(0)$ values of 0.05±0.04 for exciton emission and 0.09±0.05 for the biexciton emission confirm high purity single photons. The biexciton decay emission in a TMD is an exciting result as it could allow for the emission of entangled photon pairs. For this, more studies are needed to investigate the origins of the fine structure splitting, S, in the device and how S can be tuned. The ability to tune S could allow the ability to switch between SPE or high-fidelity entangled pair emission in TMD QEs.

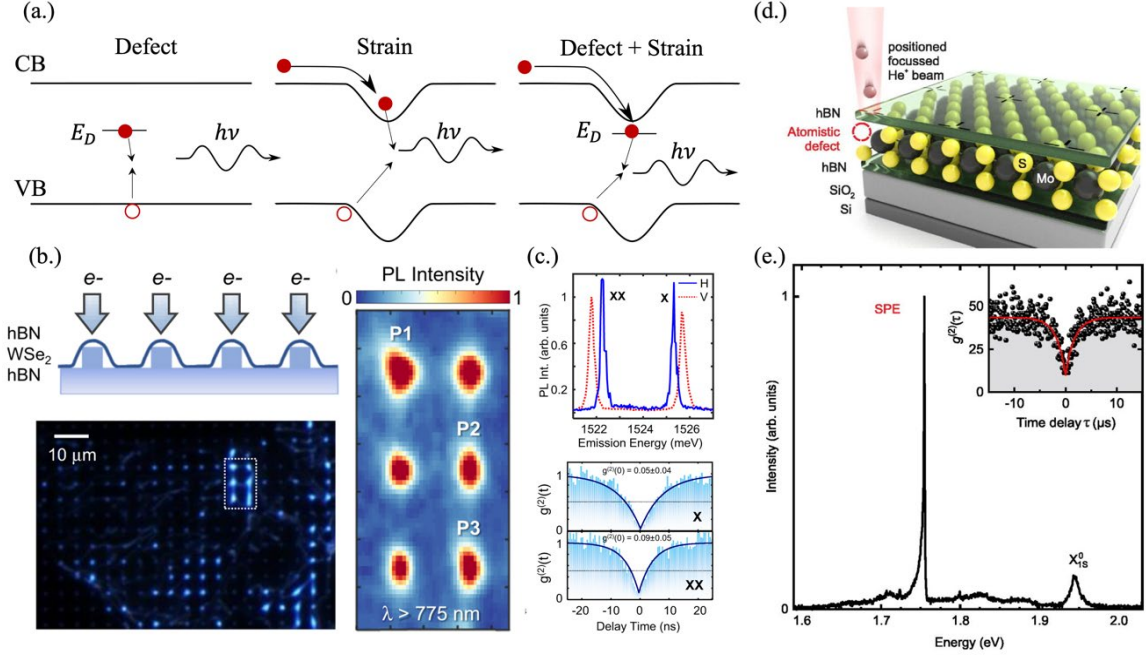

*Figure 10. **Quantum emitters in TMDs.*** (a.) Mechanism for single photon emission from defects, strain, and a combination of the two. (b.) Formation of strain/defect emitters in hBN-encapsulated WSe$_2$ via transfer onto nanopillars and e-beam irradiation (upper left). PL maps showing site-controlled emission (bottom and right). (c.) Polarization-resolved emission spectra (top) and antibunching results for exciton (X) and biexciton (XX) SPE from the device in (b.). Fig. (b, c.) adapted from ref.[460] (d.) Site-controlled creation of sulfur vacancy defects in hBN-encapsulated MoS$_2$ using a He ion FIB and (e.) the emission spectra (inset: antibunching results). Figures (d, e.) are reprinted from ref.[453], with the permission of AIP Publishing.

SPE has also recently been demonstrated from interlayer excitons (IXs) in twisted type-II TMD heterobilayers[459]. Type-II band alignment in heterobilayers leads to the formation of IXs. Depending on the in-plane twist angle between the two layers, carriers are spatially localized by the confining moiré potential, forming arrays of bound excitons[467]. Because of the permanent dipole, IX energy can be tuned by the DC stark effect by as much as 40 meV[459]. Moiré superlattices are therefore an intriguing route to tunable SPE.

Realizing reliable single photon emission from TMDs currently faces several challenges. First, linewidths are 1-2 orders of magnitude higher than FT-limited linewidths in the TMD emitters that have been demonstrated so far; this can potentially be improved with resonant excitation and Purcell enhancement[468]. Additionally, brightness must be improved. For defect-based emission, greater understanding of the defect photophysics is necessary for this novel system.

The emission of biexcitons in WSe$_2$[446] and MoSe$_2$[469] makes these materials candidates for entangled photon emission. Price *et al.* (2019) proposed that lateral heterostructures of smaller-gap TMD QDs <20 nm in diameter embedded in a matrix of a larger-gap TMD could be an optimal platform for the room-temperature confinement of massive Dirac fermions[470]. By forming these QDs in type-I heterostructures with optimal size and shape, the room-temperature quantum confinement of biexcitons could be experimentally realized, potentially enabling the emission of entangled photon pairs via biexciton decay. This is, first and foremost, a synthesis challenge, and an appropriate example of how advances in materials synthesis could enable next-generation quantum technologies.

While most efforts on QEs in 2D semiconductors have focused on group VI-b TMDs, other 2D semiconductors have been explored in a limited capacity. Tonndorf *et al.* demonstrated SPE from layered GaSe thin films in which Se nanoclusters create local strain[471]. A following paper used GaSe thin films on a rib/slot waveguides to as waveguide-coupled, strain-induced QEs[472]. A $g^2(0)$ under 0.1 has yet to be demonstrated for this system. Despite the various classes of 2D metal-chalcogenide semiconductors, studies of quantum emitters in 2D semiconductors beyond group Mo and W chalcogenides are scarce.

### *6.4.0 1D Material-Based QEs*

### *6.4.1 QEs in CNTs*

CNTs are a promising platform for room-temperature quantum emission. CNTs have a high exciton binding energy due to limited coulomb screening, enabling exciton stability at room temperature[473]. In pristine CNTs, rapid exciton diffusion results in multiphoton emission, and a collection of dark states lower in energy than the single bright exciton state leads to many nonradiative decay processes, resulting in low quantum efficiency (Fig. 10a). Localizing excitons within CNTs allows for room temperature SPE with improved quantum yield. Excitons can be localized by functionalizing nanotubes with organic color centers (OCCs)[474] or by non-covalently decorating nanotubes with molecular adsorbents[475]. Both OCCs and molecular adsorbents localize excitons, but the mechanisms differ.

OCCs are *sp³* defects that behave as two-level systems within the host. The excited state, $E_{11}^-$, is lower in energy than the dark exciton states, preventing the non-radiative decay processes that plague pristine CNTs (Fig. 10a). The electronegativity of the functional molecule determines the depth of the quantum well and also adjusts the energies of the LUMO and HOMO states that form the two-level system[474]. SPE in the telecom range has been demonstrated from OCCs in nanotubes with purities exceeding 99% at room temperature (Fig. 10c)[476]. The emission wavelength is tunable by CNT diameter and functionalization chemistry (Fig. 10b). OCC QEs demonstrate $g^2(0)$ values of ~0.01 (Fig. 10d) and radiative lifetimes of ~100 ps[476] (Fig. 10e), comparable to III-V QDs (Fig. 11).

Unlike OCCs, molecular adsorbents do not create distinct two-level defect states. There is also no sp³ bonding involved between molecular adsorbents and CNTs. Instead, SPE from this platform relies on the process of exciton-exciton annihilation (EEA): mobile excitons undergo an Auger recombination process until a single exciton remains, which can finally recombine radiatively. Mobile excitons in a pristine CNT can emit single photons at room temperature via EEA[477], but the efficiency is still limited by non-radiative processes. Molecular adsorbents create a well lower in energy than the dark states. Excitons diffusing along the nanotube enter this well, undergo EEA, and then recombine radiatively. CNTs decorated with pentacene were recently shown to produce an improvement in PL emission over pristine nanotubes for room-temperature SPE[475], but $g^2(0)$ approaching 0.5 indicate poor indistinguishability relative to OCCs and even pristine nanotubes. By using molecules that create deeper wells, the indistinguishability may be improved.

Optical cavities may enable new applications for CNT-based SPE. Hybrid exciton-polariton formation resulting from the coupling of pristine CNTs to optical cavities enhances PLQY from CNTs due to the emission from a lower polariton (LP) branch at a lower energy than the dark exciton states[478,479]. Interestingly, the polariton branch structure is not changed when the CNTs are functionalized with OCCs[479]. Further, strong coupling of photons in an optical cavity with CNTs to bright excitons and phonon-brightened dark excitons hybridizes these states, potentially enabling the addressability and manipulation of dark states in a CNT optical cavity via cavity photons[480]. It may be possible to apply a similar scheme to functionalized CNTs.

Quantum emitters in CNTs are very promising due to high purity, chemical tunability, SWIR emission, and room temperature SPE with short radiative lifetimes. However, much work remains to be done with regards to QEs in nanotubes since they need to have improved brightness and reduced linewidths to be useful for quantum information applications. Current CNT brightness ($10^5$-$10^7$ photons/sec)[476] is comparable to diamond but is still a couple orders of magnitude less than the desirable $10^9$ photons/sec brightness. The radiative lifetimes of nanotubes are comparable to those of III-V QDs (Figure 13), so $10^9$ photons/sec brightness is possible with improved quantum efficiency. Coupling to an optical cavity may improve the extraction efficiency and brightness, but the inability to deterministically place molecular defects may make it difficult to effectively integrate OCC QEs in photonic nanostructures such as cavities or waveguides. Deterministic creation and placement of single OCCs in CNTs with high yield will be a significant step towards realizing functionalized CNTs as a practical platform for room temperature SPE. Photochemical functionalization[481] may enable a process similar to lithographic patterning for generation of OCCs, offering a potential route to overcoming this issue.

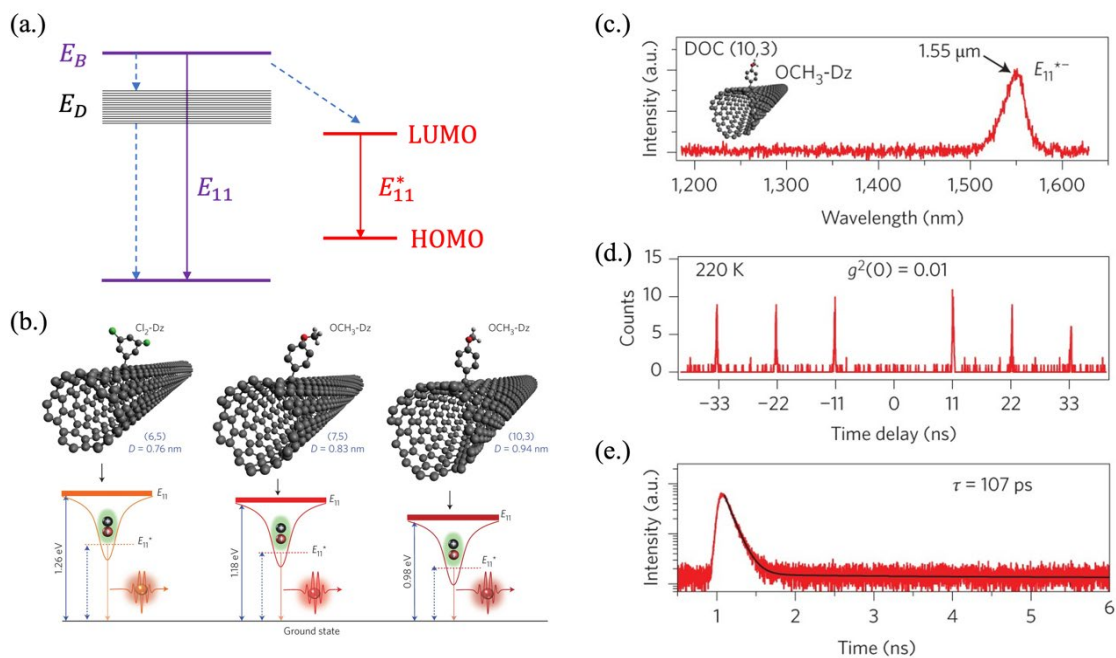

*Figure 11. Carbon nanotubes as a platform for single photon emission.* (a.) In a pristine nanotube, the single bright exciton state, $E_B$, is higher in energy than the dark states, $E_D$, resulting in relaxation and nonradiative decay, in turn causing the bright transition, $E_{11}$, to have low quantum yield. The addition of an OCC creates a two-level system and a bright state lower in energy than the band of dark excitonic states. The emission from the LUMO to the HOMO of the OCC, $E_{11}^*$, is therefore significantly more efficient. (b.) Structure of $sp^3$ OCC defects. Altering the functionalization chemistry and the nanotube chirality changes the emission energy. (c.) Emission telecom c-band from the OCH$_3$-Dz defect on a (10,3) chiral CNT. (d.) HBT measurements results yielding a $g^2(0)$ of 0.01, indicating high-purity SPE. (e.) Time resolved PL demonstrating ultra-fast radiative lifetimes of 107 ps. Panels b-e adapted from ref.[476]

## 6.5.0 0D QEs

0D platforms for light emission are attractive due to the inherent quantum confinement, yielding atomic-like transitions. These structures generally can support individual excitons and biexcitons, making them excellent candidates for both SPE and EPR pair emission. QD QEs generally have high quantum efficiencies, high brightness, and emission that is tunable by varying size and composition.

### 6.5.1 III-V Epitaxial QDs

III-V eQDs are one of the most promising technologies for SPE and EPR pair emission. Processing techniques for III-V self-assembled eQDs and optoelectronics in general are well defined, and the ability to form different heterostructures and alloys within the III-V system yields tunable properties. Emission of entangled photon pairs via biexciton decay in III-V eQDs compares favorably to the probabilistic process of SPDC.

Self-assembled InAs QDs at a buried GaAs interface have long been a leading technology for single photon or entangled photon pair emission in the telecom range. Emission can be triggered electrically, and single photon LEDs[482] and entangled-LEDs[483] were demonstrated over a decade ago. Despite the ability to electrically trigger emission, pulsed resonant two-photon excitation is a better approach to prepare biexcitons for highly entangled, indistinguishable photons.

Imperfections due to strain and concentration gradients in InAs/GaAs QDs, however, result in a fine structure splitting that leads to a degradation of entanglement fidelity. Strain and concentration gradients resulting from the Stranski-Krastanov (SK) growth method and the large nuclear spin of In, which interacts with electrons in the QD, create the Overhauser field responsible for the fine structure splitting[420]. As discussed in 6.0.2, it is ideal to minimize this fine structure splitting.

The SK method has been replaced with alternate synthetic approaches in recent years that aim to reduce strain. High-quality InAs eQDs emitting in the telecom C-band can be grown by metal organic vapor phase epitaxy (MOVPE) on a metamorphic buffer layer on a GaAs substrate[484]. Phonon-assisted two-photon resonant excitation of dots grown by this method was shown to trigger entangled photon pairs with an indistinguishability of ~0.91 and a fidelity of ~0.95[485] – a significant improvement over SK-grown QDs.

Another growth method – droplet epitaxy (DE)[486] – improves QD quality by reducing strain and improving the symmetry of the dots[487]. Unlike QDs grown by the SK method, which is commonly done on a (100) surface having $C_{2V}$ symmetry, DE is performed on a (111) surface, which has triangular $C_{3V}$ symmetry and therefore yields symmetric hexagonal eQDs[488].

GaAs eQDs grown on AlGaAs by droplet epitaxy method have also shown promise, with the highest entanglement fidelities and indistinguishability values shown for any III-V eQD system so far. Droplet epitaxy produces QDs with negligible strain or concentration gradients, and the lower nuclear spins of Ga and Al relative to In result in smaller Overhauser fields and, therefore, a smaller intrinsic $S$[489]. Consequently, GaAs QDs produce single photons and entangled photon pairs with ultra-high purity, high indistinguishability, and high entanglement fidelity. With piezoelectric strain tuning and the application of a magnetic field to eliminate S, indistinguishability >0.97 and an entanglement fidelity of 0.978 were demonstrated from GaAs eQDs[422]. GaAs QDs were recently used to demonstrate quantum teleportation with a fidelity of 0.842[490], demonstrating their potential for deployment in a quantum network. $g^{(2)}(0)$ values on the order of $10^{-5}$ have been observed for this system, demonstrating ultra-high purity SPE[491].

GaAs QDs currently face two challenges for scalability. First, emission typically occurs in the 750-800 nm range, not in a low-loss telecom band. This emission, however, can be tuned to transitions in $^{87}$Rb atoms; $^{87}$Rb atomic vapors are commonly used as quantum memories[492–494]. Second, the devices are operated at ~5 K, potentially limiting scalability. By reducing dephasing rates and careful integration in nanophotonic cavities, the operating temperature could theoretically be increased[411].

III-Nitride epitaxial quantum dots have also attracted attention as single photon sources at room temperature[495–498]. Like III-As heterostructure eQDs, III-N heterostructure eQDs have a type-I band alignment that localizes both holes and electrons, enabling the formation of single biexcitons under certain conditions. III-N emitters are highly versatile due to their room temperature operation, ability to grow on Si and SiC substrates, and high emission energy tunability[499]. However, purity and linewidths are behind those of their III-As relatives.

Achieving $g^{(2)}(0) < 0.1$ in III-N QDs has been a challenge, but $g^{(2)}(0)$ values of < 0.05 were recently achieved in InGaN/GaN QDs using near-resonant, pulsed, low-power excitation[500]. III-N QDs have excitons with relatively large dipole moments and are consequently sensitive to charge fluctuations in the surrounding environment, leading to dephasing and therefore broadened linewidths and reduced indistinguishability[501]. Currently, emission rates are on the order of several $10^6$ photons/sec[500,502], but sub-ns emission lifetimes indicate the potential for larger emission rates[500]. Still, this has yet to be realized and the current performance is well behind III-As QEs.

While III-As QDs have biexciton states with spin $S_{xx} = 0$, allowing radiative decay into $S_x = \pm 1$ states, Hönig et al. proposed that electrons in III-N QDs experience piezo- and pyroelectric fields that result in the formation of a hybrid-biexciton state with a spin of $S_{xx} = \pm 3$[503]. Conservation of angular momentum then requires the existence of dark exciton states with $S_x^D = \pm 2$ in addition to the bright exciton with $S_x^B = \pm 1$. Dark excitons then require a phonon-assisted spin-flip process to transition into the bright state and emit light[498]. This complicated and inefficient process would make III-N eQDs poor candidates for entangled light emission. We note, however, that Arita et al. demonstrated GaN/AlGaN eQDs with a $g^{(2)}(0)$ of ~0.02 and did not see evidence of this complicated biexcitonic decay mechanism[497]. However, they did observe a large FSS that must be reduced for EPR pair emission.

To summarize the different III-V eQD materials systems, In(Ga)As/GaAs QDs emit entangled photons in the telecom range, and emerging growth techniques are enabling improved indistinguishability and fidelity. GaAs/AlGaAs QDs grown by droplet epitaxy produce entangled photon pairs with the highest indistinguishability, fidelity, and purity, but they do not emit in the telecom range. III-N systems have a larger bandgap and therefore are usable up to room temperature, but emission quality is not yet up to the standards of III-As QDs due to inherent, system-specific challenges. More research into the biexciton decay physics in III-N eQDs is needed.

III-V quantum dots embedded in site-controlled III-V nanowires have emerged as an interesting single photon and EPR pair source. The nanowire behaves as a cavity/waveguide that enhances the extraction efficiency of the emitters[504]. Moreoever, narrow linewidths have been demonstrated at temperatures as high as 140K[505]. $g^{(2)}(0)$ values <0.005 have been demonstrated for InAsP QDs in InP NWs[506]. Haffouz *et al.* demonstrated InAsP QD emitters embedded in an InP nanowire waveguide with tunable wavelength throughout the telecom range[507]. In dot-in-nanowire systems, multiple quantum dots can be vertically stacked and – by tuning the composition of the dot – can be a source of frequency-multiplexed single photons[508]. So far, entanglement fidelities on par with eQDs at buried interfaces have not been demonstrated, but enhanced extraction efficiencies and frequency multiplexing nonetheless make this system promising as a single photon source.

*6.5.2 Colloidal nanocrystal emitters*

Solution-processed colloidal quantum dots (cQDs) are attractive 0D materials for room temperature quantum emission due to high quantum efficiency. Solution processing is a scalable, inexpensive means of fabrication, and recent advancements in the deterministic placement of cQDs make this a practical platform for integrated quantum nanophotonics. Fluorescent blinking – intermittent periods of high PL brightness (on) and low brightness (off) – is a challenge for both II-VI and perovskite cQDs.

*6.5.2.1 II-VI cQDs*

II-VI cQDs were one of the first systems for which SPE was reported[509]. SPE in this system is the result of quantum confinement. While SPE has been explored in this system due to high quantum efficiency and room temperature operation, II-VI cQDs face three challenges: (i.) a large dephasing rate, $\gamma^*$, (ii.) long radiative lifetimes, and (iii.) blinking.

For bare (i.e., not coupled to a cavity) cQDs, $\gamma^* \approx 10^5 \gamma$ at room temperature[510], resulting in $I \sim 10^{-5}$: unusable for practical applications. Excitonic radiative lifetimes in bare II-VI cQDs are on the order of 10-100 ns[511,512], far longer than other systems. It is therefore clear that Purcell enhancement is necessary to improve both the brightness and indistinguishability of II-VI cQDs. Coupling of CdSe/ZnS core-shell QDs to Ag nanocube plasmonic nanocavities on an Au surface was shown to drastically enhance emission rate via Purcell enhancement, achieving a PL lifetime of ~10 ps at room temperature[513]. However, this approach had the adverse effect of increasing $g^{(2)}(0)$ from 0.17 to 0.32 as compared to the same particles on glass. Another approach to enhance single photon emission rates is to highly charge the QDs using an electrochemical cell[511]. This was shown to enhance the PL decay rate by a factor of 140, but it reduced the quantum yield by a factor of 12 and worsened $g^{(2)}(0)$. It is imperative to achieve a means of enhancing emission rate without increasing $g^{(2)}(0)$. Lin *et al.* recently demonstrated that electroluminescence enables shorter lifetimes for SPE in CdSe/CdS core-shell cQDs due to the shorter lifetimes of trions than excitons[514]. This approach enabled the lifetime to be reduced while maintaining a $g^{(2)}(0)$ value of < 0.05 at room temperature. While electrically triggered SPE may enhance brightness, it will still be necessary to couple emitters to a cavity to reduce dephasing.

II-VI cQDs have yet to demonstrate sufficient indistinguishability; bright but pure emission; and blinking-free performance. Further, precise placing of individual cQDs in arrays or at desired locations with uniform emission characteristics has not been realized. Until these challenges have been overcome, it is difficult to envision II-VI cQDs being reliable single photon sources for any practical quantum application.

*6.5.2.2 0D Perovskites*

Lead halide perovskites with chemical formula $APbX_3$ (X= Cl, Br, I) are solution-processable compounds for SPE applications. The A-site element can be either organic (methylammonium, MA or formamidium, FA) or inorganic atoms (Cs) to form organic-inorganic or fully inorganic perovskite, respectively. Further, the dimension of the perovskites can be widely varied from 3D to 2D to 1D and 0D[71,515–517]. For SPE applications, 0D cubes and 2D nanosheets have been extensively studied[518,519]. The ground state exciton emission from PQDs was confirmed to be a bright triplet state, while the singlet state is dark[520]. Further, the photoluminescence (PL) emission from PQDs at low temperatures shows multiple emission peaks with polarization dependence below the exciton emission[518–520]. PL recorded by varying PQDs size and excitation fluence shows that multiple emissive states originate from exciton, biexciton, trion, and two longitudinal-optical modes[519–521]. PQDs possessing rich PL features are promising for single-photon emission applications only after meeting the stringent demands such as optical coherence and reduced fluorescent blinking as discussed below. For discussion on compatibility with electrical injection, the readers are referred to prior review articles[522,523].

Single-particle spectroscopy studies at room temperature have shown strong antibunching with $g^2(0) = 0.05$ from PQDs confirming their potential single-photon emission; however, fluorescence blinking (ON/OFF) due to charging/discharging of the PQDs was observed[524]. Charging/discharging of PQDs originates from exciton-trion formation, which is called type-A blinking. Blinking is reduced upon cooling to 6 K[525]. From several studies, the origin for fluorescence blinking was ascribed to Auger non-radiative recombination[524], memory effect due to ion migration[526–528], or combination of Auger recombination, trap states, and hot-carrier relaxation[529]. Similar blinking mechanism applies to 2D perovskites nanosheets[530], thereby concluding that fluorescence blinking is present irrespective of the semiconductor dimensionality. Blinking was suppressed in PQDs by halide ion and organic ligand vacancy filling, which resulted in a long exciton lifetime and high PL intensity[531]. This study shows that achieving stoichiometry in PQDs synthesis to obtain phase purity compounds, thereby suppressing blinking, is critical for SPE applications. Alternatively, encapsulating PQDs with alumina[532] or polystyrene[533] has been shown to reduce blinking and enhance air stability. Collectively, these strategies are promising to realize room temperature, air-stable QEs in PQDs with suppressed fluorescent blinking.

Beyond fluorescence blinking, optical coherence is an important parameter to leverage the application of PQDs for QIP applications demanding single photons or entangled photon pairs. For these applications, the optical coherence of the emitter must achieve twice the spontaneous exciton decay time ($T_2 = 2T_1$). Phonon scattering and spin-noise often limit $T_2 \ll 2T_1$[534,535]. Further, the charge density perturbation in the environment results in shifting the emission or spectral diffusion, which causes decoherence[525,536]. Using photon-correlation Fourier spectroscopy, the $T_2$ was ~80 ps, and $T_1$ was ~210 ps[519], a ratio of ~0.19 which approaches the transform limit and is two orders of magnitude higher than other low-dimensional semiconductor systems and is comparable to epitaxial QDs. Placing PQDs in nanocavities is also a viable strategy to enhance indistinguishable photons for SPE applications[537,538]. From the above discussion, the origin of multiple emissive states from PQDs is clear at this point. However, the optical coherence in PQDs needs to be addressed to implement them in quantum optics and remains an area for future research exploration.

*6.5.2.3 Other Colloidal Emitters*

Silicon is well known to have an indirect bandgap in bulk form, resulting in poor emission. However, Si nanoparticles become emissive due to quantum confinement and passivation with organic ligands. Si NPs have been shown to have high quantum efficiency[539]. The mechanism for emission is a charge transfer state between the ligand and the Si NP surface[540]. This charge transfer emission has been shown to produce non-classical light, as photon antibunching measurements have shown $g^{(2)}(0)$ values as low as 0.05 with high dependence on the substrate and ligand[541].

Zhao *et al.* have shown room temperature SPE from graphene quantum dots[542]. Unlike the GQDs discussed in section 2.1.1, these GQDs are carbon nanodots that are 0D in structure and were chemically synthesized. Chemical functionalization with Cl instead of alkyl groups enabled the tuning of the emission peak by 100 nm. The $g^{(2)}(0)$ for these emitters was found to be ~0.05, indicating high purity, and the emission lifetime was on the order of a few nanoseconds. These results suggest that colloidal GQDs could be promising for chemically tunable, bright emission with high single photon purity.

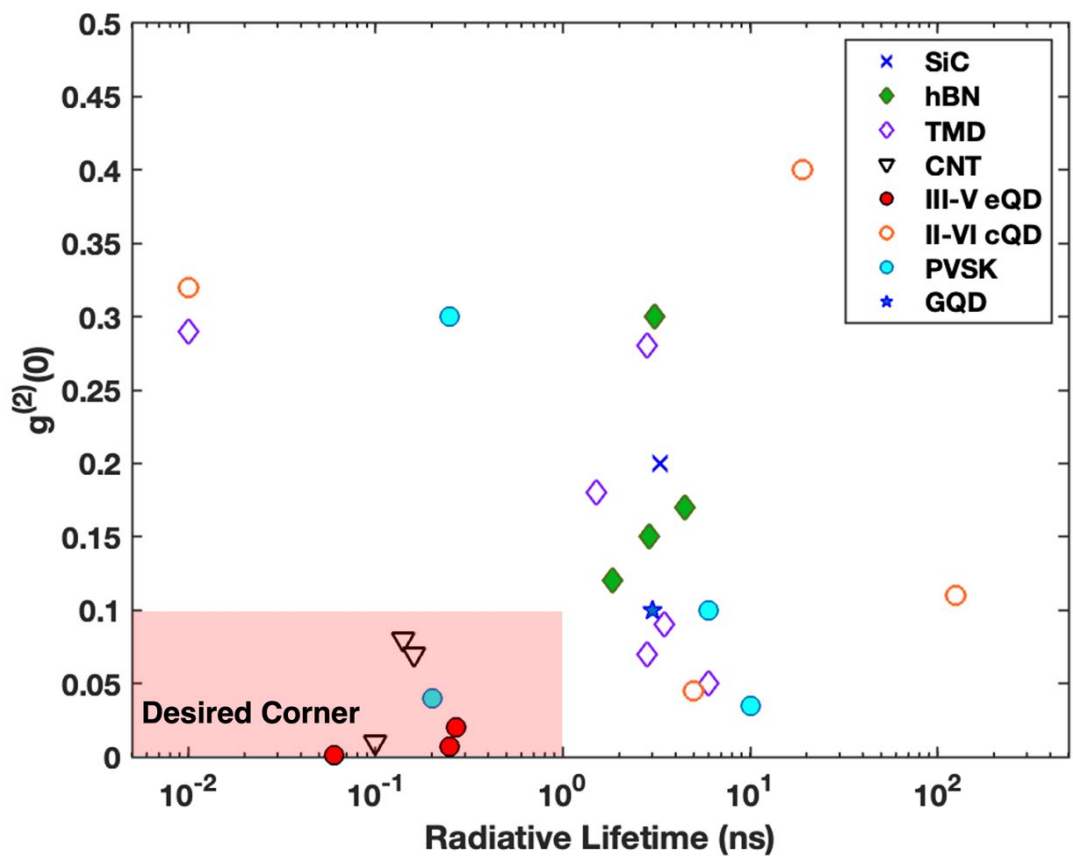

*Figure 12. Comparison of quantum light emission platforms.* Scatterplot of the radiative lifetime in ns vs. the second order autocorrelation function for quantum emitters based on silicon carbide (blue 'X', ref.[543]), hBN (green diamonds, refs.[283,430,435,443]), TMDs (violet diamonds, refs.[448,454,456,459,460,544]), carbon nanotubes (black triangles, refs.[476,545,546]), , III-V epitaxial QDs (red circles, refs.[489,547,548]), II-VI colloidal QDs (orange circles, refs.[511–514]), perovskites (labeled PVSK, cyan circles, refs.[518,525,533,549]), and graphene quantum dots (blue star,

ref.[542]). $g^{(2)}(0) < 0.5$ indicates quantum emission, but $g^{(2)}(0) < 0.1$ is required for QKD. Ideally, both the lifetime and $g^{(2)}(0)$ would be minimized. The desired corner ($g^{(2)}(0) < 0.1$, lifetime < 1 ns) for quantum communications applications is highlighted in pink.

### *6.6.0 Nanoscale Nonlinear Materials for SPDC*

In the introduction of this section, we briefly introduced spontaneous parametric down-conversion (SPDC); we discuss it in more detail here. SPDC was first demonstrated in 1967[550,551] and first shown to produce non-classical light in the late 1980s[552]. Simplicity, high indistinguishability, stability, high entanglement fidelity, and room-temperature operation make SPDC sources advantageous for many quantum optics experiments and applications in quantum cryptography, quantum simulation, quantum metrology, as well as for testing fundamental laws of physics in quantum materials[419,553,554].

$\chi^{(2)}$, the quadratic order optical susceptibility, is typically small (on the order of $10^{-12}$ m/V), limiting the efficiency of the SPDC process: the highest efficiency obtained for SPDC is $4 \times 10^{-6}$ in periodically poled lithium niobate waveguides[555]. The lack of efficiency makes the process stochastic and limits the brightness. Therefore, enhancement of efficiency is necessary. Efficiency can be improved by choosing materials with a larger $\chi^{(2)}$ or higher optical density of states or by Purcell enhancement to enhance emission and modify the phase-matching conditions[553]. The latter can be achieved using metamaterials[553,556,557]. Davoyan *et al.* have analytically shown that hyperbolic metamaterials can enable non-resonant, broadband, phase-mismatch-free Purcell enhancement of spontaneous nonlinear light emission[553].

Low dimensional materials open an opportunity for compact optical cavities for Purcell enhancement of SPDC. Tokman *et al.* theoretically describe the the process of parametric down-conversion coupled to a subwavelength 2D cavity made from 2D nonlinear materials[556]. They predicted a significant reduction in the parametric instability threshold can be achieved in this structure made from $MoS_2$. While the above theoretical efforts have indeed shown that engineering of materials on the nanoscale to enhance light-matter interactions can help improve SPDC efficiencies, nanostructuring non-linear optical materials is still in its infancy. Therefore, this provides an important opportunity for materials and nanophotonic engineers to design and experimentally demonstrate compact and efficient SPDC sources.

### 7.0.0 Single Photon Detectors

Photodetectors sensitive to single photons are a key component in quantum communications and optical quantum computing. In general, single photon detectors (SPDs) should have a low dark current, high signal-to-noise ratio, and low timing jitter[558]. For quantum information, it is also ideal to be able to resolve the number of incident photons.

The single photon avalanche diode (SPAD) one of the most common solid-state SPD technologies. Avalanche photodiodes (APDs) are heavily doped p-n junctions operating under a large reverse bias. Electron-hole pairs (EHPs) created by a single incident photon in the depletion region experience a large electric field, causing them to accelerate and generate additional carriers, which then further excite more carriers. However, SPADs have detection efficiencies well below unity, high timing jitter, and high dark count rates[559–561].

Superconducting SPDs are another general class of detectors that generally exhibit high efficiency at telecom wavelengths and low timing jitter[562]. This includes superconducting nanowire SPDs and transition edge superconducting (TES) detectors. Superconducting nanowire SPDs (SNSPDs) are the most studied type of SPD based on low-dimensional materials. In this device, an incident photon creates a normal "hotspot" region within the superconductor. This region quickly grows to block supercurrent in the superconductor, resulting in an increase in resistance that is detected as a voltage pulse. The transition from normal to superconductor then occurs on the order of picoseconds[563]. Despite the name, these devices are not true nanowires but rather ~100 nm wide strips patterned in a thin film of a superconducting material such as NbN, MoSi, or WSi[562]. TES detectors operate as bolometers: an incident photon interacting with the material introduces a small amount of heat that takes the superconductor above its critical temperature, introducing a finite resistance that can be read as a voltage pulse[564]. This method enables photon number resolution. While superconducting SPDs exhibit high efficiencies, low timing jitter, and fast operation, cryogenic operation is not ideal.

Another type of SPD can be made utilizing the photogating effect, which is unique to low-dimensional semiconductors and semimetals. The photogating effect involves the local gating of a channel by photoinduced filling of trap states[565]. This increases the carrier lifetime and increases the conductance of the channel, offering high gain.

### 7.1.0 2D Materials for SPDs

The ability to form vdW heterostructures with various different band alignments, ultra-sharp interfaces, and thicknesses smaller than the mean-free path of carriers make 2D materials promising for APDs[566]. Gao *et al.* demonstrated APDs based on InSe/BP exhibiting ballistic avalanche breakdown at low voltages enabled by the thin device structure and band alignment engineering[567]. One issue, however, is interlayer recombination of carriers in 2D APDs. By inserting a graphene layer between InSe and BP, Wang *et al.* showed a suppression of interlayer carrier recombination while maintaining a relatively high sensitivity[568]. Neither of the two aforementioned devices demonstrated single-photon-level sensitivity, but these results are nonetheless important starting points in this infant field.

2D materials exhibit strong photogating effects due to the strong sensitivity to interfacial effects[569–571]. Roy *et al.* developed a number-resolved BLG/MoS$_2$ SPD (Fig. 12b) operated at ~100K with simultaneously high gain, low noise, and low dark count, outperforming common, established technologies in these areas[572]. One challenge, however, for this device is a quantum efficiency of < 4%, limited by absorption.

Despite the high absorption coefficient and strong light-matter interaction in 2D materials, absorption is limited due to the ultra-thin structure. The absorption must be maximized to achieve an improvement in quantum efficiency. Various approaches can be taken to improve the absorption of 2D materials, including using back reflectors, dielectric spacer layers, dielectric/plasmonic resonators, or a combination of the above[99]. Another approach is the formation of superlattices that maintain the low-dimensional electronic structure but enhance absorption[573].

As discussed in section 4, various 2D materials have been shown to exhibit superconductivity or to be high-performance weak-link layers in a JJ. Atomically thin superconductors will have a lower heat capacity than the bulk and therefore could be more sensitive to single photons[561]. Recently, Walsh *et al.* utilized the localized surface plasmon modes of graphene to couple light to a graphene-based JJ[574]. Incident single photons couple to plasmons in the graphene and break Cooper pairs in the superconductor, forming quasiparticles that, in turn, induce a diffusion current across the junction, enabled by quasi-ballistic transport through graphene. This enabled detection of single telecom wavelength photons. We expect this paper to inspire more research on 2D materials for superconducting SPDs, a largely unexplored field.

### *7.2.0 1D Materials for SPDs*

The narrow channel and high surface-to-volume ratio of 1D semiconductor nanowires makes these materials particularly interesting for photogating (PG) SPDs. 1D materials have anisotropic optical properties that result in polarization-dependent absorption[575]. Polarization-selective photodetection is potentially useful for Bell state measurements. Bottom-up synthesized 1D materials have been used for PG SPDs and 1D-APDs. Luo *et al.* fabricated a CdS core-shell nanowire (CdS core with a self-assembled photogate shell) photogate SPD demonstrating sensitivity to single photons at room temperature[576]. This device displayed a very low dark count rate and high gain, with polarization-selective performance. Farrell *et al.* recently demonstrated 1D-APDs based on vertical InGaAs-GaAs nanowire arrays with low dark count rates, low timing jitter, and photon count rates higher than commercial InP-InGaAs SPADs[577]. In this structure, each nanowire is an individual photodiode, this opens the possibility of using each photodiode as a pixel with high spatial density, enabling increased information density via spatial multiplexing. While this device was not sensitive to single photons, it is a step towards 1D-SPADs. Like for 2D materials, single-photon detection based on bottom-up synthesized 1D materials is still an emerging, largely unexplored field.

### *7.3.0 0D Materials for SPDs*

Quantum dot field effect transistors (QDFETs) are FETs gated by an array of semiconductor quantum dots – typically III-Vs. Shields *et al.* demonstrated in 2000 that QDFETs can detect single photons: the QDs trap photoexcited carriers, modulating the conductance of a 2DEG channel[578]. Kardynał *et al.* showed that the change in conductance in the channel was proportional to the number of photons incident on the device, allowing the photon number to be resolved[579]. Another QD-based device, the QD resonant tunneling diode, was introduced and showed an improved dark count rate and higher quantum efficiency[580]. However, this device still suffers from high timing jitter. QD devices, in general, possess excellent sensitivity but struggle with timing jitter and must be operated at cryogenic temperatures to have low dark counts[561].

**Conclusions and Outlook**

We have discussed various aspects of quantum information science and engineering applications with a focus on low-dimensional nanomaterials as the key ingredients or platforms for next generation quantum devices. QIP requires long coherence times and fast, high-fidelity gate operations, and minimizing sources of decoherence requires high quality materials with clean interfaces. This places enormous emphasis on materials synthesis and interface engineering and creates an opportunity for the introduction of novel materials. In addition, high-resolution structural and electronic characterization of buried interfaces is equally important. To conclude, we highlight some of the critical needs for future research and developments in terms of materials in various QISE applications.

Spin qubits in gate-defined semiconducting quantum dots are one of the more mature quantum computing platforms, with few-qubit processors already demonstrated[157]. For QD based qubits, Ge/Si core-shell NWs are a potential nanomaterial candidate to further explore since they inherit the advantages of group-IV materials while also having strong Rashba spin orbit coupling, enabling ultra-fast all-electrical control[220]. Understanding all sources of decoherence and improving synthetic methods would make this system potentially competitive for all-electrical QD-based qubits. While In-V NWs similarly exhibit the potential for electrical control over spins[581], coherence times of spins are inherently limited by hyperfine interactions with the non-zero nuclear spins of group III and group V elements. Therefore, pure spin qubits in In-V NWs are impractical for scalable computing applications. Despite having been explored minimally, carbon nanotube-based QD qubits also present an opportunity for spin and spin-valley qubits[234]. Optical control over spins in CNT QDs coupled to superconducting microwave cavities is a potentially scalable approach to QD computing with long coherence times[213], especially with isotopic purification of the carbon precursor. Additionally, nanomechanical-charge/spin hybrid qubits in CNT QDs exhibit promise for a quantum sensing analogue of scanning probe microscopy[239–241]. 2D materials are still in their infancy in terms of QD-based qubits. Graphene is currently the most promising 2D material for quantum computation based on spin qubits in QDs due to its ultra-high mobility, minimal hyperfine interaction, and minimal intrinsic spin-orbit coupling[160]. Valley qubits may also be realized in graphene due to the large valley $g$-factor[162]. Spin-valley qubits in graphene quantum dots could theoretically be viable, but this will require significant enhancement of the spin-orbit coupling. Recent experiments have demonstrated the prerequisites for the use of graphene quantum dots as qubits[166,169,170], but coherent control of spins and the measurement of coherence times remains the next task. While recent experiments have focused on bulk bilayer graphene, graphene nanoribbons grown by bottom-up methods are also worth exploring for spin/valley qubits due to the inherent lateral confinement, lifting of the valley degeneracy, and bandgap[160]. 2D transition metal chalcogenides are interesting for spin-valley QD qubits due to strong spin-valley coupling and valley-dependent optical selection rules[188,189], potentially enabling spin-valley qubits with an intrinsic spin-photon interface. Despite single-particle level transport having been demonstrated in gate-defined TMD QDs[190,191,582], the readout and coherence times of spin/valley states in these devices has not been demonstrated yet, likely due to high contact resistance at low temperatures. Moreover, mobilities lower than graphene and conventional QD materials makes it difficult to confine individual carriers electrostatically; other means of confinement may be explored.

Defect spin qubits have shown some of the longest coherence times of any solid state system – particularly in diamond[253,254,256] – and exhibit promise for quantum sensing[246,247,264]. Research on defect spins in hexagonal boron nitride is in its infancy. Despite deterministic defect placement[285] and coherent manipulation of defect spins in hBN[287], the non-zero nuclear spins of boron and nitrogen make achieving coherence times as long as those in diamond and SiC speculative at best. The high surface to volume ratio also may lead to greater noise and limit coherence times. Defect centers in low dimensional materials are highly attractive for quantum sensing as qubits exist at or near the surface, compared to color centers in bulk crystals, which are buried. This enables qubits to be placed closer to the system of interest. Despite the challenges of hBN for quantum computing, we expect h-BN to be useful for quantum sensing. h-BN can be easily integrated in vdW heterostructures and transferred to arbitrary substrates, and defects in hBN exhibit strong sensitivity to external parameters[288]. hBN can also be synthesized as nanoplatelets[433], expanding the potential for sensing. $NV^-$ centers in nanodiamond can maintain the strong sensing capabilities of $NV^-$ centers in bulk diamond while overcoming the geometric limitation of the bulk. Nanodiamonds are particularly exciting for quantum biosensing due to the nanoscale dimensions and biochemical stability[291,583]. However, synthesizing nanodiamonds with controlled shape, size, and nitrogen concentration is necessary to realize the potential of this technology. Rare earth element (REE) and transition metal (TM) magnetic dopants are another form of defect qubits and are promising for quantum memory application[584]. II-VI cQDs are a promising host for magnetic TM dopant spins, as they can be doped with single TM dopants and provide an inherent optical interface via excitonic *sp-d* interaction[585]. This possibly will allow for optically addressable TM quantum memories with a multidimensional Hilbert space. Likewise, halide perovskites are also amenable to TM and REE doping[586–589]. There is also an opportunity to explore TM and REE dopants in 2D semiconductors.

Superconducting (SC) Josephson junction qubits are the most mature of all solid-state qubit technologies and have been integrated into fully functional NISQ processors which can run primitive quantum algorithms[13,324]. Two-level systems in the dielectric tunnel junction and surrounding dielectric environment are the primary source of noise and decoherence in SC qubits[328]. Materials and interface engineering of both the superconductor and weak link layer are therefore crucial for increasing coherence times in SC qubits. In this regard, 2D materials provide a notable advantage and research opportunity due to their atomically controlled thicknesses and dangling bond free surfaces. 2D superconductors demonstrate the potential for eventual use in superconducting qubits: TMD superconductors such as $NbSe_2$ have a large critical field due to Ising SOC[345,360,590], and FeSe monolayers epitaxially grown on oxide substrates is a novel high critical temperature superconductor[352,353,591]. All-2D JJs with highly transparent interfaces have been demonstrated[358,361] and remain promising. However, wafer-scale growth of highly crystalline 2D materials with minimal defects remains challenging in comparison with Al, Nb and Al oxide deposition processes, which have been optimized over wafer scales for SC qubits.

Despite limited success with QD-based spin qubits, 1D In-V NWs are highly attractive for topological qubits and are an excellent platform to explore superconductor-semiconductor hybrid devices such as gatemon qubits[323] and Andreev spin qubits[373]. 2D semiconductors and graphene are similarly interesting for gatemon qubit types[333,360].

Finally, ideas, concepts, and preliminary experimental demonstrations in terms of topological qubits and the search for non-abelian anyons are almost completely reliant on low-dimensional materials such as InAs and InSb nanowires[375]. Similarly, Ge-Si core-shell NWs[218,377,378] and CNTs[384,385] have exhibited potential for use in topological superconductor-semiconductor devices. Emerging 2D quantum spin-Hall insulators[122,392,393] and/or topological superconductors[337,405] may eventually be ideal for topological quantum computation. Theoretical predictions and experimental demonstrations of novel 2D materials exhibiting topological edge states, the fractional quantum Hall effect, and topological superconductivity will continue to be important discoveries as the search for non-Abelian anyons continues.

III-V eQDs are – by far – the most advanced solid-state source for quantum emission. Purity ($g^2(0) < 10^{-4}$)[491], indistinguishability (>0.97)[422], and entanglement fidelity (0.978)[422] in this system are currently unparalleled. Among III-V eQDs, GaAs QDs grown by droplet epitaxy and InAs QDs grown on a metamorphic buffer layer currently exhibit greater promise than III-N and III-P systems. III-V eQDs are currently limited to cryogenic operating temperatures due to rapid dephasing rates at higher temperatures. With integration in well-designed, cascaded cavities, increased operation temperature without reducing indistinguishability is feasible[411]. Otherwise, it is worth identifying single photon and entangled photon pair sources that can perform at higher temperature for future applications. While room temperature SPE has been shown from II-VI cQDs[513,514], we are not optimistic about II-VI cQDs as quantum emitters due to difficulties with indistinguishability, blinking, and demonstrating the coexistence of purity and brightness. Use of II-VI cQDs as QEs requires that these issues be resolved. Perovskite QDs have exhibited some of the same issues with blinking, but short (some < 1 ns) radiative lifetimes, relatively high coherence[519], and high purity make PQDs worth exploring for room temperature quantum emission[524]. CNTs functionalized with organic color centers are a viable platform for high-purity SPE at room temperature with high chemical tunability and short (~100 ps) radiative lifetimes[476]. With controlled defect placement and enhanced brightness and indistinguishability via Purcell enhancement, OCCs in CNT could realistically become a practical single photon source. We note that CNTs have displayed promise for QD qubits, quantum sensing, hybrid and topological quantum devices, and SPE. Despite research interest in CNTs having waned significantly over the past decade, prior decades of research in controlled growth[592–594], purification[595–597], placement[238,243], and chemistry[598,599] of semiconducting CNTs is expected to pay off as CNTs with their highly confined 1D electronic structure and seamless atomic structure exhibit the potential to be highly useful in QISE. We therefore expect a CNT renaissance, of sorts, in the nanomaterials and QISE research communities. 2D materials are promising hosts for quantum emitters as the atomically thin structure enables easier integration with nanophotonic cavities and deterministic creation/placement of emitters via defect, strain, or both[600]. Defects in hBN have a large Debye-Waller factor and low electron-phonon coupling[438] and are therefore promising for room temperature single photon emission with transform-limited linewidths[409], sub-ns radiative lifetimes[443], and near-unity extraction efficiencies[442,443]. Emitters in 2D TMDs are interesting in this application given the ease of tunability, control (via strain and electric fields), and integration with nanophotonic elements. However, the purity ($0.05 < g^2(0) < 0.3$) and radiative lifetimes (typically >1 ns) are currently behind III-V eQDs[448,454,456,459,460,544]. For both hBN and TMDs, it is crucial to better understand the physics and chemistry of emissive defect states. An interesting direction to pursue for 2D chalcogenides in this application is to engineer quantum emitters via composition confinement induced quantum dot formation (like III-V eQDs) instead of defect/strain engineering[470]. Additionally, while a plethora of moderate to wide-gap magnetic 2D chalcogenides and chalcophosphides are available, quantum emitters in such intrinsically magnetic hosts have been scarcely explored and therefore present a ripe opportunity.

Overall, while some aspects of QISE have attained maturity, the overall technology performance still fails to challenge their classical counterparts in useful applications. While that may change in the future, advances at all levels – materials, devices, circuits/architecture, and algorithms – are desired to attain quantum advantage in multiple domains of information and computing technology. We have identified that low-dimensional and nanoscale materials offer potential for realizing qubits with large decoherence times, quantum emitters with better purity and brightness, and as serve as conduits for novel quantum sensing and imaging modalities by virtue of their large surface area and electronic structure. It is therefore conceivable that the next big advance in QISE may emerge from novel materials and interface engineering.

Acknowledgements:

D.J. and A.A. acknowledge primary support for this work by the Air Force Office of Scientific Research (AFOSR) FA2386-20-1-4074 and FA2386-21-1-4063. D.J. and H.Z. acknowledge partial support from the U.S. Army Research Office under contract number W911NF-19-1-0109. H.Z. acknowledges partial support from by Vagelos Institute of Energy Science and Technology graduate fellowship. S.B.A. gratefully acknowledges partial funding received from the Swiss National Science Foundation (SNSF) under the Early Postdoc Mobility program (grant 187977) to conduct this work.

**References:**

[1]   C. H. Bennett, D. P. DiVincenzo, *J. Phys. A. Math. Gen.* **2001**, *34*, 6723.

[2]   Y. Wang, Z. Hu, B. C. Sanders, S. Kais, *Front. Phys.* **2020**, *8*, 479.

[3]   R. Horodecki, P. Horodecki, M. Horodecki, K. Horodecki, *Rev. Mod. Phys.* **2009**, *81*, 865.

[4]   R. Jozsa, N. Linden, *Proc. R. Soc. London. Ser. A Math. Phys. Eng. Sci.* **2003**, *459*, 2011.

[5]   C. H. Bennett, G. Brassard, C. Crepeau, R. Jozsa, A. Peres, W. K. Wootters, *Phys. Rev. Lett.* **1993**, *70*, 1895.

[6]   C. Lupo, S. Pirandola, *Phys. Rev. Lett.* **2016**, *117*.

[7]   C. L. Degen, F. Reinhard, P. Cappellaro, *Rev. Mod. Phys.* **2017**, *89*, 035002.

[8]   R. P. Feynman, *Int. J. Theor. Phys.* **1982**, *21*.

[9]   P. W. Shor, In *Proceedings 35th Annual Symposium on Foundations of Computer Science*, **1994**, pp. 124–134.

[10]  P. W. Shor, *SIAM J. Comput.* **1997**, *26*, 1484.

[11]  D. S. Abrams, S. Lloyd, *Phys. Rev. Lett.* **1999**, *83*, 5162.

[12]  J. Preskill, *Quantum* **2018**, *2*, 79.

[13]  F. Arute, K. Arya, R. Babbush, D. Bacon, J. C. Bardin, R. Barends, R. Biswas, S. Boixo, F. G. S. L. Brandao, D. A. Buell, B. Burkett, Y. Chen, Z. Chen, B. Chiaro, R. Collins, W. Courtney, A. Dunsworth, E. Farhi, B. Foxen, A. Fowler, C. Gidney, M. Giustina, R. Graff, K. Guerin, S. Habegger, M. P. Harrigan, M. J. Hartmann, A. Ho, M. Hoffmann, T. Huang, T. S. Humble, S. V. Isakov, E. Jeffrey, Z. Jiang, D. Kafri, K. Kechedzhi, J. Kelly, P. V. Klimov, S. Knysh, A. Korotkov, F. Kostritsa, D. Landhuis, M. Lindmark, E. Lucero, D. Lyakh, S. Mandrà, J. R. McClean, M. McEwen, A. Megrant, X. Mi, K. Michielsen, M. Mohseni, J. Mutus, O. Naaman, M. Neeley, C. Neill, M. Y. Niu, E. Ostby, A. Petukhov, J. C. Platt, C. Quintana, E. G. Rieffel, P. Roushan, N. C. Rubin, D. Sank, K. J. Satzinger, V. Smelyanskiy, K. J. Sung, M. D. Trevithick, A. Vainsencher, B. Villalonga, T. White, Z. J. Yao, P. Yeh, A. Zalcman, H. Neven, J. M. Martinis, *Nature* **2019**, *574*, 505.

[14]  D. P. Divincenzo, *Fortschr. Phys.* **2000**, *48*, 771.

[15]  E. Knill, R. Laflamme, W. H. Zurek, *Science (80-. ).* **1998**, *279*, 342.

[16]   P. W. Shor, *Phys. Rev. A* **1995**, *52*.

[17]   A. Steane, In *Proc. R. Soc. Lond. A.*, The Royal SocietyLondon, **1996**, pp. 2551–2577.

[18]   R. Laflamme, C. Miquel, J. P. Paz, W. H. Zurek, *Phys. Rev. Lett.* **1996**, *77*.

[19]   M. Takita, A. W. Cross, A. D. Córcoles, J. M. Chow, J. M. Gambetta, *Phys. Rev. Lett.* **2017**, *119*, 180501.

[20]   N. H. Nguyen, M. Li, A. M. Green, C. H. Alderete, Y. Zhu, D. Zhu, K. R. Brown, N. M. Linke, *Phys. Rev. Appl.* **2021**, *10*, 24057.

[21]   S. J. Devitt, W. J. Munro, K. Nemoto, *Reports Prog. Phys.* **2013**, *76*, 076001.

[22]   M. Blencowe, *Nature* **2010**, *468*, 44.

[23]   P. Rabl, D. Demille, J. M. Doyle, M. D. Lukin, R. J. Schoelkopf, P. Zoller, *Phys. Rev. Lett.* **2006**, *97*, 033003.

[24]   C. Kloeffel, D. Loss, *Annu. Rev. Condens. Matter Phys.* **2013**, *4*, 51.

[25]   J. R. Weber, W. F. Koehl, J. B. Varley, A. Janotti, B. B. Buckley, C. G. Van De Walle, D. D. Awschalom, *PNAS* **2010**, *107*, 8513.

[26]   J. Clarke, F. K. Wilhelm, *Nature* **2008**, *453*, 1031.

[27]   C. Nayak, S. H. Simon, A. Stern, M. Freedman, S. Das Sarma, *Rev. Mod. Phys.* **2008**, *80*, 1083.

[28]   M. Saffman, T. G. Walker, K. Mølmer, *Rev. Mod. Phys.* **2010**, *82*, 2313.

[29]   C. D. Bruzewicz, J. Chiaverini, R. Mcconnell, *Appl. Phys. Rev* **2019**, *6*, 21314.

[30]   P. Kok, W. J. Munro, K. Nemoto, T. C. Ralph, J. P. Dowling, G. J. Milburn, *Rev. Mod. Phys.* **2007**, *79*, 135.

[31]   H. J. Kimble, *Nature* **2008**, *453*, 1023.

[32]   T. E. Northup, R. Blatt, *Nat. Photonics* **2014**, *8*, 356.

[33]   C. H. Bennett, *Phys. Rev. Lett.* **1992**, *68*, 3121.

[34]   M. Müller, S. Bounouar, K. D. Jöns, M. Glässl, P. Michler, *Nat. Photonics* **2014**, *8*, 224.

[35]   G. Molina-Terriza, J. P. Torres, L. Torner, *Nat. Phys.* **2007**, *3*, 305.

[36]   A. Mair, A. Vaziri, G. Weihs, A. Zeilinger, *Opt. Angular Momentum* **2016**, *412*, 287.

[37]   J. D. Franson, *Phys. Rev. Lett.* **1989**, *62*, 2205.

[38]   N. Gisin, G. Goire Ribordy, W. Tittel, H. Zbinden, *Rev. Mod. Phys.* **2002**, *74*.

[39]   R. L. Rivest, A. Shamir, L. Adleman, *Commun. ACM* **1978**, *21*, 120.

[40]   C. H. Bennett, G. Brassard, In *International Conference on Computers, Systems, and Signal Processing*, **1984**, pp. 175–179.

[41]   C. Z. Peng, J. Zhang, D. Yang, W. B. Gao, H. X. Ma, H. Yin, H. P. Zeng, T. Yang, X. Bin Wang, J. W. Pan, *Phys. Rev. Lett.* **2007**, *98*, 010505.

[42]   H. L. Yin, T. Y. Chen, Z. W. Yu, H. Liu, L. X. You, Y. H. Zhou, S. J. Chen, Y. Mao, M. Q. Huang, W. J. Zhang, H. Chen, M. J. Li, D. Nolan, F. Zhou, X. Jiang, Z. Wang, Q. Zhang, X. Bin Wang, J. W. Pan, *Phys. Rev. Lett.* **2016**, *117*, 190501.

[43]   S. K. Liao, W. Q. Cai, W. Y. Liu, L. Zhang, Y. Li, J. G. Ren, J. Yin, Q. Shen, Y. Cao, Z. P. Li, F. Z. Li, X. W. Chen, L. H. Sun, J. J. Jia, J. C. Wu, X. J. Jiang, J. F. Wang, Y. M. Huang, Q. Wang, Y. L. Zhou, L. Deng, T. Xi, L. Ma, T. Hu, Q. Zhang, Y. A. Chen, N. Le Liu, X. Bin Wang, Z. C. Zhu, C. Y. Lu, R. Shu, C. Z. Peng, J. Y. Wang, J. W. Pan, *Nature* **2017**, *549*, 43.

[44]   H. K. Lo, X. Ma, K. Chen, *Phys. Rev. Lett.* **2005**, *94*, 230504.

[45]   X. Bin Wang, *Phys. Rev. Lett.* **2005**, *94*, 230503.

[46]   W. K. Wootters·, W. H. Zurek, *Nature* **1982**, *299*, 802.

[47]   D. Bouwmeester, J. W. Pan, K. Mattle, M. Eibl, H. Weinfurter, A. Zeilinger, *Nature* **1997**, *390*, 575.

[48]   X. S. Ma, T. Herbst, T. Scheidl, D. Wang, S. Kropatschek, W. Naylor, B. Wittmann, A. Mech, J. Kofler, E. Anisimova, V. Makarov, T. Jennewein, R. Ursin, A. Zeilinger, *Nature* **2012**, *489*, 269.

[49]   J. Yin, J.-G. Ren, H. Lu, Y. Cao, H.-L. Yong, Y.-P. Wu, C. Liu, S.-K. Liao, F. Zhou, Y. Jiang, X.-D. Cai, P. Xu, G.-S. Pan, J.-J. Jia, Y.-M. Huang, H. Yin, J.-Y. Wang, Y.-A. Chen, C.-Z. Peng, J.-W. Pan, *Nature* **2012**, *488*, 185.

[50]   J. Yin, Y. Cao, Y.-H. Li, S.-K. Liao, L. Zhang, J.-G. Ren, W.-Q. Cai, W.-Y. Liu, B. Li, H. Dai, G.-B. Li, Q.-M. Lu, Y.-H. Gong, Y. Xu, S.-L. Li, F.-Z. Li, Y.-Y. Yin, Z.-Q. Jiang, M. Li, J.-J. Jia, G. Ren, D. He, Y.-L. Zhou, X.-X. Zhang, N. Wang, X. Chang, Z.-C. Zhu, N.-L. Liu, Y.-A. Chen, C.-Y. Lu, R. Shu, C.-Z. Peng, J.-Y. Wang, J.-W. Pan, *Science (80-. ).* **2017**, *356*, 1140.

[51]   H.-J. Briegel, W. Dür, J. I. Cirac, P. Zoller, *Phys. Rev. Lett.* **1998**, *81*, 5932.

[52]   S. Muralidharan, L. Li, J. Kim, N. Lütkenhaus, M. D. Lukin, L. Jiang, *Sci. Rep.* **2016**, *6*, 20463.

[53]   D. Deutsch, A. Ekert, R. Jozsa, C. Macchiavello, S. Popescu, A. Sanpera, *Phys. Rev. Lett.* **1996**, *77*, 2818.

[54]   W. Dür, H.-J. Briegel, J. I. Cirac, P. Zoller, *Phys. Rev. A* **1999**, *59*.

[55]   W. J. Munro, A. M. Stephens, S. J. Devitt, K. A. Harrison, K. Nemoto, *Nat. Photonics* **2012**, *6*, 777.

[56]   J. Borregaard, H. Pichler, T. Schröder, M. D. Lukin, P. Lodahl, A. S. Sørensen, *Phys. Rev. X* **2020**, *10*, 021071.

[57]   I. K. Kominis, T. W. Kornack, J. C. Allred, M. V. Romalis, *Nature* **2003**, *422*, 596.

[58]   R. Maiwald, D. Leibfried, J. Britton, J. C. Bergquist, G. Leuchs, D. J. Wineland, *Nat. Phys.* **2009**, *5*, 551.

[59]   M. Bal, C. Deng, J.-L. Orgiazzi, F. R. Ong, & A. Lupascu, *Nat. Commun.* **2012**, *3*, 1324.

[60]   R. C. Jaklevic, J. Lambe, J. E. Mecereau, A. H. Silver, *Phys. Rev.* **1965**, *140*, 1628.

[61]   J. Michl, J. Steiner, A. Denisenko, A. Andrébü Lau, A. Andrézimmermann, K. Nakamura, H. Sumiya, S. Onoda, P. Neumann, J. Isoya, J. Rg Wrachtrup, *Nano Lett* **2019**, *19*, 4904.

[62]   G. S. Waters, P. D. Francis, *J. Sci. Instrum.* **1958**, *35*, 88.

[63]   P. Bhattacharya, S. Ghosh, A. D. Stiff-Roberts, *Annu. Rev. Mater. Res.* **2004**, *34*, 1.

[64]   Y. Arakawa, M. J. Holmes, *Appl. Phys. Rev.* **2020**, *7*.

[65]   H. Liu, T. Wang, Q. Jiang, R. Hogg, F. Tutu, F. Pozzi, A. Seeds, *Nat. Photonics* **2011**, *5*, 416.

[66]   J. Phillips, K. Kamath, P. Bhattacharya, *Appl. Phys. Lett.* **1998**, *72*, 2020.

[67]   J. Phillips, K. Kamath, T. Brock, P. Bhattacharya, *Appl. Phys. Lett.* **1998**, *72*, 3509.

[68]   I. Moreels, K. Lambert, D. Smeets, D. De Muynck, T. Nollet, J. C. Martins, F. Vanhaecke, A. Vantomme, C. Delerue, G. Allan, Z. Hens, *ACS Nano* **2009**, *3*, 3023.

[69]   L. Protesescu, S. Yakunin, M. I. Bodnarchuk, F. Krieg, R. Caputo, C. H. Hendon, R. X. Yang, A. Walsh, M. V Kovalenko, *Nano Lett* **2015**, *15*, 58.

[70]   S. Ithurria, M. D. Tessier, B. Mahler, R. P. S. M. Lobo, B. Dubertret, A. L. Efros, *Nat. Mater.* **2011**, *10*, 936.

[71]   G. Schileo, G. Grancini, *J. Phys. Energy* **2020**, *2*, 021005.

[72]   Y. Shirasaki, G. J. Supran, M. G. Bawendi, V. Bulović, *Nat. Photonics* **2013**, *7*, 13.

[73]  Y.-S. Park, J. Roh, B. T. Diroll, R. D. Schaller, V. I. Klimov, *Nat. Rev. Mater.* **2021**, *6*, 382.

[74]  B. S. Mashford, M. Stevenson, Z. Popovic, C. Hamilton, Z. Zhou, C. Breen, J. Steckel, V. Bulovic, M. Bawendi, S. Coe-Sullivan, P. T. Kazlas, *Nat. Photonics* **2013**, *7*, 407.

[75]  J. Tang, K. W. Kemp, S. Hoogland, K. S. Jeong, H. Liu, L. Levina, M. Furukawa, X. Wang, R. Debnath, D. Cha, K. Wei Chou, A. Fischer, A. Amassian, J. B. Asbury, E. H. Sargent, *Nat. Mater.* **2011**, *10*, 765.

[76]  D. O. Sigle, L. Zhang, S. Ithurria, B. Dubertret, J. J. Baumberg, *J. Phys. Chem. Lett.* **2015**, *6*, 1099.

[77]  G. Hills, C. Lau, A. Wright, S. Fuller, M. D. Bishop, T. Srimani, P. Kanhaiya, R. Ho, A. Amer, Y. Stein, D. Murphy, A. Chandrakasan, M. M. Shulaker, *Nature* **2019**, *572*, 596.

[78]  B. Song, F. Liu, H. Wang, J. Miao, Y. Chen, P. Kumar, H. Zhang, X. Liu, H. Gu, E. A. Stach, X. Liang, S. Liu, Z. Fakhraai, D. Jariwala, **2021**, *2*, 39.

[79]  M. Meyyappan, *Small* **2016**, *12*, 2118.

[80]  V. Schroeder, S. Savagatrup, M. He, S. Lin, T. M. Swager, *Chem. Rev.* **2019**, *119*, 599.

[81]  K. Tomioka, M. Yoshimura, T. Fukui, *Nature* **2012**, *488*, 189.

[82]  H. J. Joyce, Q. Gao, H. Hoe Tan, C. Jagadish, Y. Kim, J. Zou, L. M. Smith, H. E. Jackson, J. M. Yarrison-Rice, P. Parkinson, M. B. Johnston, *Prog. Quantum Electron.* **2011**, *35*, 23.

[83]  Y. Huang, X. Duan, C. M. Lieber, *Small* **2005**, *1*, 142.

[84]  J. Xiang, W. Lu, Y. Hu, Y. Wu, H. Yan, C. M. Lieber, *Nature* **2006**, *441*, 489.

[85]  J. Goldberger, A. I. Hochbaum, R. Fan, P. Yang, *Nano Lett.* **2006**, *6*, 973.

[86]  K. S. Novoselov, A. K. Geim, S. V. Morozov, D. Jiang, Y. Zhang, S. V. Dubonos, I. V. Grigorieva, A. A. Firsov, *Science (80-. ).* **2004**, *306*, 666.

[87]  A. H. Castro Neto, F. Guinea, N. M. R. Peres, K. S. Novoselov, A. K. Geim, *Rev. Mod. Phys.* **2009**, *81*, 109.

[88]  V. W. Brar, M. S. Jang, M. Sherrott, J. J. Lopez, H. A. Atwater, T. J. Watson, *Nano Lett.* **2021**, *13*, 2541.

[89]  A. Vakil, N. Engheta, *Science (80-. ).* **2011**, *332*, 1291.

[90]  F. Wang, Y. Zhang, C. Tian, C. Girit, A. Zettl, M. Crommie, Y. R. Shen, *Science (80-. ).* **2008**, *320*, 206.

[91]  X. Ling, H. Wang, S. Huang, F. Xia, M. S. Dresselhaus, *PNAS* **2015**, *112*, 4523.

[92]  F. Xia, H. Wang, Y. Jia, *Nat. Commun.* **2014**, *5*, 4458.

[93]  K. F. Mak, C. Lee, J. Hone, J. Shan, T. F. Heinz, *Phys. Rev. Lett.* **2010**, *105*, 136805.

[94]  H. Wang, L. Yu, Y.-H. Lee, Y. Shi, A. Hsu, M. L. Chin, L.-J. Li, M. Dubey, J. Kong, T. Palacios, *Nano Lett.* **2012**, *12*, 4674.

[95]  D. Lembke, S. Bertolazzi, A. Kis, *Acc. Chem. Res.* **2015**, *48*, 100.

[96]  K. Kang, S. Xie, L. Huang, Y. Han, P. Y. Huang, K. F. Mak, C.-J. Kim, D. Muller, J. Park, *Nature* **2015**, *520*, 656.

[97]  X. Duan, C. Wang, A. Pan, R. Yu, X. Duan, *Chem. Soc. Rev* **2015**, *44*, 8859.

[98]  H. Zhang, B. Abhiraman, Q. Zhang, J. Miao, K. Jo, S. Roccasecca, M. W. Knight, A. R. Davoyan, D. Jariwala, *Nat. Commun.* **2020**, *11*.

[99]  D. Jariwala, A. R. Davoyan, J. Wong, H. A. Atwater, *ACS Photonics* **2017**, *4*, 2962.

[100] D. Jariwala, V. K. Sangwan, L. J. Lauhon, T. J. Marks, M. C. Hersam, *Emerging device applications for*

*semiconducting two-dimensional transition metal dichalcogenides*, Vol. 8, **2014**, pp. 1102–1120.

[101] R. Cheng, D. Li, H. Zhou, C. Wang, A. Yin, S. Jiang, Y. Liu, Y. Chen, Y. Huang, X. Duan, *Nano Lett.* **2014**, *14*, 5590.

[102] Y. Xue, Y. Zhang, Y. Liu, H. Liu, J. Song, J. Sophia, J. Liu, Z. Xu, Q. Xu, Z. Wang, J. Zheng, Y. Liu, S. Li, Q. Bao, *ACS Nano* **2016**, *10*, 25.

[103] H. Tan, Y. Fan, Y. Zhou, Q. Chen, W. Xu, J. H. Warner, *ACS Nano* **2016**, *10*, 7866.

[104] D. Jariwala, A. R. Davoyan, G. Tagliabue, M. C. Sherrott, J. Wong, H. A. Atwater, *Nano Lett* **2016**, *16*, 8.

[105] L.-H. Zeng, S.-H. Lin, Z.-J. Li, Z.-X. Zhang, T.-F. Zhang, C. Xie, C.-H. Mak, Y. Chai, S. Ping Lau, L.-B. Luo, Y. Hong Tsang, L. Zeng, S. Lin, C. Mak, Y. Chai, S. P. Lau, Y. H. Tsang, Z. Li, Z. Zhang, T. Zhang, C. Xie, L. Luo, **2018**.

[106] X. Meng, Y. Shen, J. Liu, L. Lv, X. Yang, X. Gao, M. Zhou, X. Wang, Y. Zheng, Z. Zhou, *Appl. Catal. A Gen.* **2021**, *624*, 118332.

[107] X. Chia, Z. Sofer, J. Luxa, M. Pumera, *ACS Appl. Mater. Interfaces* **2017**, *9*, 25587.

[108] C. Song, X. Yuan, C. Huang, S. Huang, Q. Xing, C. Wang, C. Zhang, Y. Xie, Y. Lei, F. Wang, L. Mu, J. Zhang, F. Xiu, H. Yan, *Nat. Commun.* **2021**, *12*, 386.

[109] J. Wang, H. Zheng, G. Xu, L. Sun, D. Hu, Z. Lu, L. Liu, J. Zheng, C. Tao, L. Jiao, *J. Am. Chem. Soc* **2016**, *138*, 43.

[110] A. W. Tsen, R. Hovden, D. Wang, Y. Duck Kim, J. Okamoto, K. A. Spoth, Y. Liu, W. Lu, Y. Sun, J. C. Hone, L. F. Kourkoutis, P. Kim, A. N. Pasupathy, A. contributions, A. designed research, L. performed research, A. analyzed data, **2015**, *112*, 15054.

[111] H. Wang, X. Huang, J. Lin, J. Cui, Y. Chen, C. Zhu, F. Liu, Q. Zeng, J. Zhou, P. Yu, X. Wang, H. He, S. H. Tsang, W. Gao, K. Suenaga, F. Ma, C. Yang, L. Lu, T. Yu, E. Hang, T. Teo, G. Liu, Z. Liu, .

[112] M. Bonilla, S. Kolekar, Y. Ma, H. C. Diaz, V. Kalappattil, R. Das, T. Eggers, H. R. Gutierrez, M.-H. Phan, M. Batzill, *Nat. Nanotechnol.* **2018**, *13*, 289.

[113] Z. Fei, B. Huang, P. Malinowski, W. Wang, T. Song, J. Sanchez, W. Yao, D. Xiao, X. Zhu, A. F. May, W. Wu, D. H. Cobden, J.-H. Chu, X. Xu, *Nat. Mater.* **2018**, *17*, 778.

[114] J.-U. Lee, S. Lee, J. Hoon Ryoo, S. Kang, T. Yun Kim, P. Kim, C.-H. Park, J.-G. Park, H. Cheong, *Nano Lett* **2016**, *16*, 30.

[115] B. Huang, G. Clark, E. Navarro-Moratalla, D. R. Klein, R. Cheng, K. L. Seyler, D. Zhong, E. Schmidgall, M. A. Mcguire, D. H. Cobden, W. Yao, D. Xiao, P. Jarillo-Herrero, & X. Xu, *Nature* **2017**, *546*, 270.

[116] K. S. Burch, D. Mandrus, J.-G. Park, *Nature* **2018**, *563*, 47.

[117] C. Gong, L. Li, Z. Li, H. Ji, A. Stern, Y. Xia, T. Cao, W. Bao, C. Wang, Y. Wang, Z. Q. Qiu, R. J. Cava, S. G. Louie, J. Xia, & X. Zhang, *Nature* **2017**, *546*, 265.

[118] M. Osada, T. Sasaki, *APL Mater.* **2019**, *7*, 120902.

[119] W. Ding, J. Zhu, Z. Wang, Y. Gao, D. Xiao, Y. Gu, Z. Zhang, W. Zhu, *Nat. Commun.* **2017**, *8*, 14956.

[120] F. Liu, L. You, K. L. Seyler, X. Li, P. Yu, J. Lin, X. Wang, J. Zhou, H. Wang, H. He, S. T. Pantelides, W. Zhou, P. Sharma, X. Xu, P. M. Ajayan, J. Wang, Z. Liu, *Nat. Commun.* **2016**, *7*, 12357.

[121] J. Qi, H. Wang, X. Chen, *Appl. Phys. Lett* **2018**, *113*, 43102.

[122] L. Kou, Y. Ma, Z. Sun, T. Heine, C. Chen, *J. Phys. Chem. Lett.* **2017**, *8*, 1905.

[123] F. Sheng, C. Hua, M. Cheng, J. Hu, X. Sun, Q. Tao, H. Lu, Y. Lu, M. Zhong, K. Watanabe, T. Taniguchi, Q. Xia, Z.-A. Xu, Y. Zheng, *Nature* **2021**, *593*.

[124] Y.-C. Lin, B. Jariwala, B. M. Bersch, K. Xu, Y. Nie, B. Wang, S. M. Eichfeld, X. Zhang, T. H. Choudhury, Y. Pan, R. Addou, C. M. Smyth, J. Li, K. Zhang, M. A. Haque, S. Fö, R. M. Feenstra, R. M. Wallace, K. Cho, S. K. Fullerton-Shirey, J. M. Redwing, J. A. Robinson, *ACS Nano* **2018**, *12*, 965.

[125] D. Hee Lee, Y. Sim, J. Wang, *APL Mater.* **2020**, *8*, 30901.

[126] D. Wei, Y. Liu, Y. Wang, H. Zhang, L. Huang, G. Yu, *Nano Lett.* **2009**, *9*, 1752.

[127] R. Muñoz, C. Gómez-Aleixandre, *Chem. Vap. Depos.* **2013**, *19*, 297.

[128] Z. Q. Shi, H. Li, Q. Q. Yuan, Y. H. Song, Y. Y. Lv, W. Shi, Z. Y. Jia, L. Gao, Y. Bin Chen, W. Zhu, S. C. Li, *Adv. Mater.* **2019**, *31*, 1806130.

[129] P. Barthelemy, L. M. K. Vandersypen, *Rev. Artic. Ann. Phys.* **2013**, *525*, 808.

[130] T. Hensgens, T. Fujita, L. Janssen, X. Li, C. J. Van Diepen, C. Reichl, W. Wegscheider, D. sarma, L. M. K Vandersypen, *Nat. Publ. Gr.* **2017**, *548*.

[131] J. P. Dehollain, U. Mukhopadhyay, V. P. Michal, Y. Wang, B. Wunsch, C. Reichl, W. Wegscheider, E. Demler, L. M. K. Vandersypen, *Nature* **2021**, *579*.

[132] N. P. de Leon, K. M. Itoh, D. Kim, K. K. Mehta, T. E. Northup, H. Paik, B. S. Palmer, N. Samarth, S. Sangtawesin, D. W. Steuerman, *Science (80-. ).* **2021**, *372*, 253.

[133] A. Chatterjee, P. Stevenson, S. De Franceschi, A. Morello, N. P. de Leon, F. Kuemmeth, *Nat. Rev. Phys.* **2021**, *3*, 157.

[134] D. Loss, D. P. DiVincenzo, *Phys. Rev. A* **1997**, *57*, 120.

[135] K. C. Nowack, F. H. L. Koppens, Y. V. Nazarov, L. M. K. Vandersypen, *Science (80-. ).* **2007**, *318*, 1430.

[136] B. Trauzettel, D. V. Bulaev, D. Loss, G. Burkard, *Nat. Phys.* **2007**, *3*, 192.

[137] N. Rohling, G. Burkard, *New J. Phys.* **2012**, *14*, 083008.

[138] M. Russ, G. Burkard, *J. Phys. Condens. Matter* **2017**, *29*, 393001.

[139] J. Medford, J. Beil, J. M. Taylor, E. I. Rashba, H. Lu, A. C. Gossard, C. M. Marcus, *Phys. Rev. Lett.* **2013**, *111*, 050501.

[140] J. M. Taylor, V. Srinivasa, J. Medford, *Phys. Rev. Lett.* **2013**, *111*, 050502.

[141] A. Pan, T. E. Keating, M. F. Gyure, E. J. Pritchett, S. Quinn, R. S. Ross, T. D. Ladd, J. Kerckhoff, *Quantum Sci. Technol* **2020**, *5*, 34005.

[142] M. Russ, G. Burkard, *Phys. Rev. B* **2015**, *92*, 205412.

[143] F. K. Malinowski, F. Martins, P. D. Nissen, S. Fallahi, G. C. Gardner, M. J. Manfra, C. M. Marcus, F. Kuemmeth, *Phys. Rev. B* **2017**, *96*, 45443.

[144] D. Kim, Z. Shi, C. B. Simmons, D. R. Ward, J. R. Prance, T. S. Koh, J. King Gamble, D. E. Savage, M. G. Lagally, M. Friesen, S. N. Coppersmith, M. A. Eriksson, *Nature* **2014**, *511*, 70.

[145] J. R. Petta, A. C. Johnson, J. M. Taylor, E. A. Laird, A. Yacoby, M. D. Lukin, C. M. Marcus, M. P. Hanson, A. C. Gossard, *Science (80-. ).* **2005**, *309*, 2180.

[146] F. H. L. Koppens, C. Buizert, K. J. Tielrooij, I. T. Vink, K. C. Nowack, T. Meunier, L. P. Kouwenhoven, L. M. K. Vandersypen, *Nature* **2006**, *442*, 766.

[147] A. V. Khaetskii, D. Loss, L. Glazman, *Phys. Rev. Lett.* **2002**, *88*, 1868021.

[148] P.-A. Mortemousque, E. Chanrion, B. Jadot, H. Flentje, A. Ludwig, A. D. Wieck, M. Urdampilleta, C. Bäuerle, T. Meunier, *Nat. Nanotechnol.* **2021**, *16*, 296.

[149] K. M. Itoh, H. Watanabe, *MRS Commun.* **2014**, *4*, 143.

[150] M. Veldhorst, J. C. C. Hwang, C. H. Yang, A. W. Leenstra, B. de Ronde, J. P. Dehollain, J. T.

[150] Muhonen, F. E. Hudson, K. M. Itoh, A. Morello, A. S. Dzurak, *Nat. Nanotechnol. 2014 912* **2014**, *9*, 981.

[151] C. H. Yang, K. W. Chan, R. Harper, W. Huang, T. Evans, J. C. C. Hwang, B. Hensen, A. Laucht, T. Tanttu, F. E. Hudson, S. T. Flammia, K. M. Itoh, A. Morello, S. D. Bartlett, A. S. Dzurak, *Nat. Electron. 2019 24* **2019**, *2*, 151.

[152] Y.-C. Yang, S. N. Coppersmith, M. Friesen, *Phys. Rev. A* **2020**, *101*, 12338.

[153] Z. Wang, E. Marcellina, A. R. Hamilton, J. H. Cullen, S. Rogge, J. Salfi, D. Culcer, *npj Quantum Inf.* **2021**, *7*, 54.

[154] A. Dobbie, M. Myronov, R. J. H. Morris, A. H. A. Hassan, M. J. Prest, V. A. Shah, E. H. C. Parker, T. E. Whall, D. R. Leadley, *Appl. Phys. Lett* **2012**, *101*, 172108.

[155] N. W. Hendrickx, D. P. Franke, A. Sammak, G. Scappucci, & M. Veldhorst, *Nature* **2020**, *577*, 487.

[156] D. Jirovec, A. Hofmann, A. Ballabio, P. M. Mutter, G. Tavani, M. Botifoll, A. Crippa, J. Kukucka, O. Sagi, F. Martins, J. Saez-Mollejo, I. Prieto, M. Borovkov, J. Arbiol, D. Chrastina, G. Isella, G. Katsaros, *Nat. Mater.* **2021**, *20*, 1106.

[157] N. W. Hendrickx, W. I. L Lawrie, M. Russ, F. van Riggelen, S. L. de Snoo, R. N. Schouten, A. Sammak, G. Scappucci, M. Veldhorst, *Nature* **2021**, *591*, 580.

[158] G. Scappucci, C. Kloeffel, F. A. Zwanenburg, D. Loss, M. Myronov, J.-J. Zhang, S. De Franceschi, G. Katsaros, M. Veldhorst, *Nat. Rev. Mater.* **2020**, *6*, 926.

[159] A. V. Kretinin, Y. Cao, J. S. Tu, G. L. Yu, R. Jalil, K. S. Novoselov, S. J. Haigh, A. Gholinia, A. Mishchenko, M. Lozada, T. Georgiou, C. R. Woods, F. Withers, P. Blake, G. Eda, A. Wirsig, C. Hucho, K. Watanabe, T. Taniguchi, A. K. Geim, R. V. Gorbachev, *Nano Lett.* **2014**, *14*, 3270.

[160] P. Recher, B. Trauzettel, *Nanotechnology* **2010**, *21*, 302001.

[161] J. B. Oostinga, H. B. Heersche, X. Liu, A. F. Morpurgo, L. M. K. Vandersypen, *Nat. Mater.* **2007**, *7*, 151.

[162] C. Tong, R. Garreis, A. Knothe, M. Eich, A. Sacchi, K. Watanabe, T. Taniguchi, V. Fal'ko, T. Ihn, K. Ensslin, A. Kurzmann, *Nano Lett.* **2021**, *21*, 1068.

[163] A. Hollmann, T. Struck, V. Langrock, A. Schmidbauer, F. Schauer, T. Leonhardt, K. Sawano, H. Riemann, N. V Abrosimov, D. Bougeard, L. R. Schreiber, *Phys. Rev. Appl.* **2020**, *10*, 34068.

[164] L. Banszerus, S. Möller, C. Steiner, E. Icking, S. Trellenkamp, F. Lentz, K. Watanabe, T. Taniguchi, C. Volk, C. Stampfer, *Spin-valley coupling in single-electron bilayer graphene quantum dots*, **2021**.

[165] A. Avsar, J. Y. Tan, T. Taychatanapat, J. Balakrishnan, G. K. W. Koon, Y. Yeo, J. Lahiri, A. Carvalho, A. S. Rodin, E. C. T. O'Farrell, G. Eda, A. H. Castro Neto, B. Özyilmaz, *Nat. Commun.* **2014**, *5*, 4875.

[166] L. Banszerus, S. Möller, E. Icking, K. Watanabe, T. Taniguchi, C. Volk, C. Stampfer, *Nano Lett.* **2020**, *20*, 2005.

[167] L. Banszerus, A. Rothstein, E. Icking, S. Möller, K. Watanabe, T. Taniguchi, C. Stampfer, C. Volk, *Appl. Phys. Lett.* **2021**, *118*, 103101.

[168] A. Kurzmann, M. Eich, H. Overweg, M. Mangold, F. Herman, P. Rickhaus, R. Pisoni, Y. Lee, R. Garreis, C. Tong, K. Watanabe, T. Taniguchi, K. Ensslin, T. Ihn, *Phys. Rev. Lett.* **2019**, *123*, 026803.

[169] C. Tong, A. Kurzmann, R. Garreis, W. W. Huang, S. Jele, M. Eich, L. Ginzburg, C. Mittag, K. Watanabe, T. Taniguchi, K. Ensslin, T. Ihn, *Pauli Blockade of Tunable Two-Electron Spin and Valley States in Graphene Quantum Dots*, **2021**.

[170] L. Banszerus, S. Möller, E. Icking, C. Steiner, D. Neumaier, M. Otto, K. Watanabe, T. Taniguchi, C. Volk, C. Stampfer, *Appl. Phys. Lett* **2021**, *118*, 93104.

[171] M. Mirzakhani, F. M. Peeters, M. Zarenia, *Phys. Rev. B* **2020**, *101*, 75413.

[172] L. Sun, Z. Wang, Y. Wang, L. Zhao, Y. Li, B. Chen, S. Huang, S. Zhang, W. Wang, D. Pei, H. Fang, S.

Zhong, H. Liu, J. Zhang, L. Tong, Y. Chen, Z. Li, M. H. Rümmeli, K. S. Novoselov, H. Peng, L. Lin, Z. Liu, *Nat. Commun.* **2021**, *12*, 2391.

[173] G. Kim, H. Lim, K. Yeol Ma, A.-R. Jang, G. Hee Ryu, M. Jung, H.-J. Shin, Z. Lee, H. Suk Shin, *Nano Lett.* **2015**, *15*, 4769.

[174] G. Kim, S.-S. Kim, J. Jeon, S. I. Yoon, S. Hong, Y. J. Cho, A. Misra, S. Ozdemir, J. Yin, D. Ghazaryan, M. Holwill, A. Mishchenko, D. V Andreeva, Y.-J. Kim, H. Y. Jeong, A.-R. Jang, H.-J. Chung, A. K. Geim, K. S. Novoselov, B.-H. Sohn, H. S. Shin, *Nat. Commun.* **2019**, *10*, 230.

[175] D. Bischoff, A. Varlet, P. Simonet, M. Eich, H. C. Overweg, T. Ihn, K. Ensslin, *Appl. Phys. Rev* **2015**, *2*, 31301.

[176] N. Kim, S. Choi, S.-J. Yang, J. Park, J.-H. Park, N. Ngan Nguyen, K. Park, S. Ryu, K. Cho, C.-J. Kim, *Cite This ACS Appl. Mater. Interfaces* **2021**, *13*, 28593.

[177] J. Cai, P. Ruffieux, R. Jaafar, M. Bieri, T. Braun, S. Blankenburg, M. Muoth, A. P. Seitsonen, M. Saleh, X. Feng, K. Müllen, R. Fasel, *Nature* **2010**, *466*, 470.

[178] N. Kim, S. Choi, S.-J. Yang, J. Park, J.-H. Park, N. Ngan Nguyen, K. Park, S. Ryu, K. Cho, C.-J. Kim, *ACS Appl. Mater. Interfaces* **2021**, *13*, 28593.

[179] A. J. Way, R. M. Jacobberger, M. S. Arnold, *Nano Lett.* **2018**, *18*, 898.

[180] A. J. Way, E. A. Murray, F. Gö, V. Saraswat, R. M. Jacobberger, M. Mavrikakis, M. S. Arnold, *J. Phys. Chem. Lett.* **2019**, *10*, 4266.

[181] C. Volk, C. Neumann, S. Kazarski, S. Fringes, S. Engels, F. Haupt, A. Müller, C. Stampfer, *Nat. Commun.* **2013**, *4*, 1753.

[182] L. Banszerus, K. Hecker, E. Icking, S. Trellenkamp, F. Lentz, D. Neumaier, K. Watanabe, T. Taniguchi, C. Volk, C. Stampfer, *Phys. Rev. B* **2021**, *103*, 81404.

[183] A. Kormányos, V. Zólyomi, N. D. Drummond, G. Burkard, *Phys. Rev. X* **2014**, *4*, 1.

[184] M. Brooks, G. Burkard, *Phys. Rev. B* **2020**, *101*, 35204.

[185] A. Altıntaş, M. Bieniek, A. Dusko, M. Korkusiński, J. Pawłowski, P. Hawrylak, **2021**, 1.

[186] J. Pawłowski, M. Bieniek, T. W. Woźniak, *Phys. Rev. Appl.* **2021**, *15*, 54025.

[187] W. Yao, D. Xiao, Q. Niu, *Phys. Rev. B* **2008**, *77*, 235406.

[188] T. Cao, G. Wang, W. Han, H. Ye, C. Zhu, J. Shi, Q. Niu, P. Tan, E. Wang, B. Liu, J. Feng, *Nat. Commun.* **2012**, *3*, 887.

[189] D. Xiao, G.-B. Liu, W. Feng, X. Xu, W. Yao, *Phys. Rev. Lett.* **2012**, *108*, 196802.

[190] X.-X. Song, D. Liu, V. Mosallanejad, J. You, T.-Y. Han, D.-T. Chen, H.-O. Li, G. Cao, M. Xiao, G.-C. Guo, G.-P. Guo, *Nanoscale* **2015**, *7*, 16867.

[191] S. Davari, J. Stacy, A. M. Mercado, J. D. Tull, R. Basnet, K. Pandey, K. Watanabe, T. Taniguchi, J. Hu, H. O. H. Churchill, *Phys. Rev. Appl.* **2020**, *13*, 1.

[192] X.-X. Song, Z.-Z. Zhang, J. You, D. Liu, H.-O. Li, G. Cao, M. Xiao, G.-P. Guo, *Nat. Publ. Gr.* **2015**, *5*, 16113.

[193] Z.-Z. Zhang, X.-X. Song, G. Luo, G.-W. Deng, V. Mosallanejad, T. Taniguchi, K. Watanabe, H.-O. Li, G. Cao, G.-C. Guo, F. Nori, G.-P. Guo, *Sci. Adv.* **2017**, *3*, e1701699.

[194] K. Lee, G. Kulkarni, Z. Zhong, *Nanoscale* **2016**, *8*, 7755.

[195] K. Wang, K. De Greve, L. A. Jauregui, A. Sushko, A. High, Y. Zhou, G. Scuri, T. Taniguchi, K. Watanabe, M. D. Lukin, H. Park, P. Kim, *Nat. Nanotechnol.* **2018**, *13*, 128.

[196] S. Reinhardt, L. Pirker, C. Bäuml, M. Remškar, A. K. Hüttel, S. Reinhardt, C. Bäuml, A. K. Hüttel, L. Pirker, M. Remškar, *Phys. Status Solidi* **2019**, *13*, 1900251.

[197] P. Kumar, J. P. Horwath, A. C. Foucher, C. C. Price, N. Acero, V. B. Shenoy, E. A. Stach, D. Jariwala, *npj 2D Mater. Appl.* **2020**, *4*, 16.

[198] Z. Wang, R. Luo, I. Johnson, H. Kashani, M. Chen, *ACS Nano* **2020**, *14*, 899.

[199] K. Momma, F. Izumi, IUCr, *J. Appl. Crystallogr.* **2011**, *44*, 1272.

[200] T. Hu, G. Zhao, H. Gao, Y. Wu, J. Hong, A. Stroppa, W. Ren, *Phys. Rev. B* **2020**, *101*, 1.

[201] I. Khan, B. Marfoua, J. Hong, *npj 2D Mater. Appl.* **2021**, *5*, 1.

[202] P. Zhao, Y. Ma, C. Lei, H. Wang, B. Huang, Y. Dai, *Appl. Phys. Lett* **2019**, *115*, 261605.

[203] R. Peng, Y. Ma, X. Xu, Z. He, B. Huang, Y. Dai, *Phys. Rev. B* **2020**, *102*, 035412.

[204] C. Lei, X. Xu, T. Zhang, B. Huang, Y. Dai, Y. Ma, *J. Phys. Chem. C* **2021**, *125*, 2802.

[205] X. Liu, M. C. Hersam, *Nat. Rev. Mater.* **2019**, *4*, 669.

[206] C.-C. Liu, H. Jiang, Y. Yao, *Phys. Rev. B* **2011**, *84*, 195430.

[207] B. Szafran, D. Żebrowski, A. Mreńca-Kolasińska, *Sci. Rep.* **2018**, *8*, 7166.

[208] B. Szafran, A. Mreńca-Kolasińska, B. Rzeszotarski, D. Zebrowski, *Phys. Rev. B* **2018**, *97*, 165303.

[209] B. Feng, Z. Ding, S. Meng, Y. Yao, X. He, P. Cheng, L. Chen, K. Wu, *Nano Lett.* **2012**, *12*, 3507.

[210] B. Lalmi, H. Oughaddou, H. Enriquez, A. Kara, S. Vizzini, B. Ealet, B. Aufray, *Appl. Phys. Lett* **2010**, *97*, 223109.

[211] K. D. Petersson, L. W. Mcfaul, M. D. Schroer, M. Jung, J. M. Taylor, A. A. Houck, J. R. Petta, *Nature* **2012**, *490*, 380.

[212] D. De Jong, J. Van Veen, L. Binci, A. Singh, P. Krogstrup, L. P. Kouwenhoven, W. Pfaff, J. D. Watson, *Phys. Rev. Appl.* **2019**, *11*, 044061.

[213] T. Cubaynes, M. R. Delbecq, M. C. Dartiailh, R. Assouly, M. M. Desjardins, L. C. Contamin, L. E. Bruhat, Z. Leghtas, F. Mallet, A. Cottet, T. Kontos, *npj Quantum Inf.* **2019**, *5*, 47.

[214] W. Lu, J. Xiang, B. P. Timko, Y. Wu, C. M. Lieber, *PNAS* **2005**, *102*, 10046.

[215] Y. Hu, H. O. H. Churchill, D. J. Reilly, J. Xiang, C. M. Lieber, † And, C. M. Marcus, *Nat. Nanotechnol.* **2007**, *2*, 622.

[216] M. Brauns, J. Ridderbos, A. Li, W. G. Van Der Wiel, E. P. A. M. Bakkers, F. A. Zwanenburg, *Appl. Phys. Lett* **2016**, *109*, 143113.

[217] F. N. M. Froning, M. K. Rehmann, J. Ridderbos, M. Brauns, F. A. Zwanenburg, A. Li, E. P. A. M. Bakkers, D. M. Zumbühl, F. R. Braakman, *Appl. Phys. Lett.* **2018**, *113*, 073102.

[218] C. Kloeffel, M. Trif, D. Loss, *Phys. Rev. B* **2011**, *84*, 195314.

[219] A. P. Higginbotham, T. W. Larsen, J. Yao, H. Yan, C. M. Lieber, C. M. Marcus, F. Kuemmeth, *Nano Lett* **2014**, *14*, 3582.

[220] F. N. M. Froning, L. C. Camenzind, O. A. H van der Molen, A. Li, E. P. A M Bakkers, D. M. Zumbühl, F. R. Braakman, *Nat. Nanotechnol.* **2021**, *16*, 308.

[221] I. A. Goldthorpe, A. F. Marshall, P. C. McIntyre, *Nano Lett.* **2008**, *8*, 4081.

[222] S. A. Dayeh, N. H. Mack, J. Y. Huang, S. T. Picraux, *Appl. Phys. Lett* **2011**, *99*, 023102.

[223] A. Garcia-Gil, S. Biswas, J. D. Holmes, R. Ramos-Barrado, *Nanomaterials* **2021**, *11*, 2002.

[224] Ö. Gül, D. J. van Woerkom, I. van Weperen, D. Car, S. R. Plissard, E. P. A M Bakkers, L. P. Kouwenhoven, *Nanotechnology* **2015**, *26*, 215202.

[225] H. A. Nilsson, P. Caroff, C. Thelander, M. Larsson, J. B. Wagner, L.-E. Wernersson, L. Samuelson, H. Q. Xu, *Nano Lett.* **2009**, *9*, 3151.

[226] J. W. G. Van Den Berg, S. Nadj-Perge, V. S. Pribiag, S. R. Plissard, E. P. A. M. Bakkers, S. M. Frolov, L. P. Kouwenhoven, *Phys. Rev. Lett.* **2012**, *110*.

[227] S. Bednarek, J. Pawłowski, M. Górski, G. Skowron, *Phys. Rev. Appl.* **2019**, *10*, 34012.

[228] D. Fan, S. Li, N. Kang, P. Caroff, L. B. Wang, Y. Q. Huang, M. T. Deng, C. L. Yu, H. Q. Xu, *Nanoscale* **2015**, *7*, 14822.

[229] R. Wang, R. S. Deacon, D. Car, E. P. A. M. Bakkers, K. Ishibashi, *Appl. Phys. Lett* **2016**, *108*, 203502.

[230] J.-Y. Wang, G.-Y. Huang, S. Huang, J. Xue, D. Pan, J. Zhao, H. Xu, *Nano Lett* **2018**, *18*, 4741.

[231] J. Mu, S. Huang, Z.-H. Liu, W. Li, J.-Y. Wang, D. Pan, G.-Y. Huang, Y. Chen, J. Zhao, H. Q. Xu, **2021**, *13*, 3983.

[232] D. De Jong, C. G. Prosko, D. M. A. Waardenburg, L. Han, F. K. Malinowski, P. Krogstrup, L. P. Kouwenhoven, J. V Koski, W. Pfaff, *Phys. Rev. Appl.* **2021**, *10*, 14007.

[233] F. Kuemmeth, S. Ilani, D. C. Ralph, P. L. McEuen, *Nature* **2008**, *452*, 448.

[234] E. A. Laird, F. Pei, L. P. Kouwenhoven, *Nat. Nanotechnol.* **2013**, *8*, 565.

[235] T. Pei, A. Pályi, M. Mergenthaler, N. Ares, A. Mavalankar, J. H. Warner, G. A. D. Briggs, E. A. Laird, *Phys. Rev. Lett.* **2017**, *118*, 1.

[236] H. O. H. Churchill, F. Kuemmeth, J. W. Harlow, A. J. Bestwick, E. I. Rashba, K. Flensberg, C. H. Stwertka, T. Taychatanapat, S. K. Watson, C. M. Marcus, *Phys. Rev. Lett.* **2009**, *102*, 166802.

[237] C. Galland, A. Imamog, *Phys. Rev. Lett* **2008**, *101*, 157404.

[238] T. Cubaynes, T; Contamin, L C; Dartiailh, M C; Desjardins M M; Cottet, A; Delbecq, M R; Kontos, *Appl. Phys. Lett.* **2020**, *117*, 114001.

[239] I. Khivrich, & S. Ilani, *Nat. Commun.* **2020**, *11*, 2299.

[240] A. Pályi, P. R. Struck, M. Rudner, K. Flensberg, G. Burkard, *Phys. Rev. Lett.* **2012**, *63*, 206811.

[241] F. Pistolesi, A. N. Cleland, A. Bachtold, *Phys. Rev. X* **2021**, *11*, 031027.

[242] L. Liu, J. Han, L. Xu, J. Zhou, C. Zhao, S. Ding, H. Shi, M. Xiao, L. Ding, Z. Ma, C. Jin, Z. Zhang, L. M. Peng, *Science (80-. ).* **2020**, *368*, 850.

[243] K. R. Jinkins, S. M. Foradori, V. Saraswat, R. M. Jacobberger, J. H. Dwyer, P. Gopalan, A. Berson, M. S. Arnold, *Sci. Adv.* **2021**, *7*.

[244] Z. Qiu, U. Vool, A. Hamo, A. Yacoby, *npj Quantum Inf.* **2021**, *7*, 39.

[245] J. R. Maze, P. L. Stanwix, J. S. Hodges, S. Hong, J. M. Taylor, P. Cappellaro, L. Jiang, M. V. G. Dutt, E. Togan, A. S. Zibrov, A. Yacoby, R. L. Walsworth, M. D. Lukin, *Nature* **2008**, *455*, 644.

[246] J. M. Taylor, P. Cappellaro, L. Childress, L. Jiang, D. Budker, P. R. Hemmer, A. Yacoby, R. Walsworth, M. D. Lukin, *Nat. Phys.* **2008**, *4*, 810.

[247] P. Neumann, I. Jakobi, F. Dolde, C. Burk, R. Reuter, G. Waldherr, J. Honert, T. Wolf, A. Brunner, J. H. Shim, D. Suter, H. Sumiya, J. Isoya, J. Wrachtrup, *Nano Lett.* **2013**, *13*, 2738.

[248] G. Zhang, Y. Cheng, J.-P. Chou, A. Gali, *Appl. Phys. Rev* **2020**, *7*, 31308.

[249] D. D. Sukachev, A. Sipahigil, C. T. Nguyen, M. K. Bhaskar, R. E. Evans, F. Jelezko, M. D. Lukin, *Phys. Rev. Lett.* **2017**, *119*, 223602.

[250] A. Pershin, G. Barcza, Ö. Legeza, A. Gali, *npj Quantum Inf.* **2021**, *7*, 99.

[251] F. M. Hrubesch, G. Braunbeck, M. Stutzmann, F. Reinhard, M. S. Brandt, *Phys. Rev. Lett.* **2017**, *118*, 037601.

[252] E. R. Eisenach, J. F. Barry, M. F. O'keeffe, J. M. Schloss, M. H. Steinecker, D. R. Englund, D. A. Braje, *Nat. Commun.* **2021**, *12*, 1357.

[253] E. D. Herbschleb, H. Kato, Y. Maruyama, T. Danjo, T. Makino, S. Yamasaki, I. Ohki, K. Hayashi, H. Morishita, M. Fujiwara, & N. Mizuochi, *Nat. Commun.* **2019**, *10*, 3766.

[254] M. H. Abobeih, J. Cramer, M. A. Bakker, N. Kalb, M. Markham, D. J. Twitchen, T. H. Taminiau, *Nat. Commun.* **2018**, *9*, 2552.

[255] M. H. Abobeih, J. Randall, C. E. Bradley, H. P. Bartling, M. A. Bakker, M. J. Degen, M. Markham, D. J. Twitchen, T. H. Taminiau, *Nature* **2019**, *576*, 411.

[256] C. E. Bradley, J. Randall, M. H. Abobeih, R. C. Berrevoets, M. J. Degen, M. A. Bakker, M. Markham, D. J. Twitchen, T. H. Taminiau, *Phys. Rev. X* **2019**, *9*, 031045.

[257] P. C. Maurer, G. Kucsko, C. Latta, L. Jiang, N. Y. Yao, S. D. Bennett, F. Pastawski, D. Hunger, N. Chisholm, M. Markham, D. J. Twitchen, J. I. Cirac, M. D. Lukin, *Science (80-. ).* **2012**, *336*, 1283.

[258] F. Dolde, I. Jakobi, B. Naydenov, N. Zhao, S. Pezzagna, C. Trautmann, J. Meijer, P. Neumann, F. Jelezko, J. Wrachtrup, *Nat. Phys.* **2013**, *9*, 139.

[259] N. Kalb, A. A. Reiserer, P. C. Humphreys, J. J. W. Bakermans, S. J. Kamerling, N. H. Nickerson, S. C. Benjamin, D. J. Twitchen, M. Markham, R. Hanson, *Science (80-. ).* **2017**, *356*, 928.

[260] S. Zaiser, T. Rendler, I. Jakobi, T. Wolf, S. Y. Lee, S. Wagner, V. Bergholm, T. Schulte-Herbrüggen, P. Neumann, J. Wrachtrup, *Nat. Commun.* **2016**, *7*, 12279.

[261] X. Rong, J. Geng, F. Shi, Y. Liu, K. Xu, W. Ma, F. Kong, Z. Jiang, Y. Wu, J. Du, *Nat. Commun.* **2015**, *6*, 8748.

[262] G. Waldherr, Y. Wang, S. Zaiser, M. Jamali, T. Schulte-Herbrüggen, H. Abe, T. Ohshima, J. Isoya, J. F. Du, P. Neumann, & J. Wrachtrup, *Nature* **2014**, *506*, 204.

[263] S. Pezzagna, J. Meijer, *Appl. Phys. Rev* **2021**, *8*, 11308.

[264] F. Dolde, H. Fedder, M. W. Doherty, T. Nöbauer, F. Rempp, G. Balasubramanian, T. Wolf, F. Reinhard, L. C. L. Hollenberg, F. Jelezko, J. Wrachtrup, *Nat. Phys.* **2011**, *7*, 459.

[265] B. Lundqvist, P. Raad, M. Yazdanfar, P. Stenberg, R. Liljedahl, P. Komarov, N. Rorsman, J. Ager, O. Kordina, I. Ivanov, E. Janzen, *19th Int. Work. Therm. Investig. ICs Syst.* **2013**, 58.

[266] A. Bourassa, C. P. Anderson, K. C. Miao, M. Onizhuk, H. Ma, A. L. Crook, H. Abe, J. Ul-Hassan, T. Ohshima, N. T. Son, G. Galli, D. D. Awschalom, *Nat. Mater.* **2020**, *19*, 1319.

[267] D. Simin, H. Kraus, A. Sperlich, T. Ohshima, G. V Astakhov, V. Dyakonov, *Phys. Rev. B* **2017**, *95*, 161201.

[268] N. T. Son, C. P. Anderson, A. Bourassa, K. C. Miao, C. Babin, M. Widmann, M. Niethammer, J. U. Hassan, N. Morioka, I. G. Ivanov, F. Kaiser, J. Wrachtrup, D. D. Awschalom, *Appl. Phys. Lett* **2020**, *116*, 190501.

[269] B. E. Kane, *Nature* **1998**, *393*, 133.

[270] J. T. Muhonen, J. P. Dehollain, A. Laucht, F. E. Hudson, R. Kalra, T. Sekiguchi, K. M. Itoh, D. N. Jamieson, J. C. Mccallum, A. S. Dzurak, A. Morello, *Nat. Nanotechnol.* **2014**, *9*, 987.

[271] S. Asaad, V. Mourik, B. Joecker, M. A. I. Johnson, A. D. Baczewski, H. R. Firgau, M. T. Mądzik, V. Schmitt, J. J. Pla, F. E. Hudson, K. M. Itoh, J. C. Mccallum, A. S. Dzurak, A. Laucht, A. Morello, *Nature* **2020**, *579*, 205.

[272] G. Tosi, F. A. Mohiyaddin, V. Schmitt, S. Tenberg, R. Rahman, G. Klimeck, A. Morello, *Nat. Commun. 8*, 450.

[273] J. Salfi, M. Tong, S. Rogge, D. Culcer, *Nanotechnology* **2016**, *27*, 244001.

[274] J. Salfi, M. Tong, S. Rogge, al -, M. Russ, G. Burkard, J. C. Abadillo-Uriel, M. J. Calderón, *New J. Phys* **2017**, *19*, 43027.

[275] B. Hensen, W. Wei Huang, C.-H. Yang, K. Wai Chan, J. Yoneda, T. Tanttu, F. E. Hudson, A. Laucht, K. M. Itoh, T. D. Ladd, A. Morello, A. S. Dzurak, *Nat. Nanotechnol.* **2020**, *15*, 13.

[276]  B. R. Judd, *Phys. Rev.* **1962**, *127*, 750.

[277]  P. Siyushev, K. Xia, R. Reuter, M. Jamali, N. Zhao, N. Yang, C. Duan, N. Kukharchyk, A. D. Wieck, R. Kolesov, & J. Wrachtrup, *Nat. Commun.* **2014**, *5*, 3895.

[278]  T. Zhong, J. M. Kindem, J. Rochman, A. Faraon, *Nat. Commun.* **2017**, *8*, 14107.

[279]  T. Zhong, J. M. Kindem, J. G. Bartholomew, J. Rochman, I. Craiciu, E. Miyazono, M. Bettinelli, E. Cavalli, V. Verma, S. W. Nam, F. Marsili, M. D. Shaw, A. D. Beyer, A. Faraon, *Science (80-. ).* **2017**, *357*, 1392.

[280]  T. Zhong, J. M. Kindem, E. Miyazono, A. Faraon, *Nat. Commun.* **2015**, *6*, 8206.

[281]  G. Cassabois, P. Valvin, B. Gil, *Nat. Photonics* **2016**, *10*, 262.

[282]  W. Liu, Z.-P. Li, Y.-Z. Yang, S. Yu, Y. Meng, Z.-A. Wang, Z.-C. Li, N.-J. Guo, F.-F. Yan, Q. Li, J.-F. Wang, J.-S. Xu, Y.-T. Wang, J.-S. Tang, C.-F. Li, G.-C. Guo, *ACS Photonics* **2021**, *8*, 1889.

[283]  T. T. Tran, C. Elbadawi, D. Totonjian, C. J. Lobo, G. Grosso, H. Moon, D. R. Englund, M. J. Ford, I. Aharonovich, M. Toth, *ACS Nano* **2016**, *10*, 45.

[284]  M. Kianinia, S. White, J. E. Frö, C. Bradac, I. Aharonovich, *ACS Photonics* **2021**, *7*, 2147.

[285]  X. Gao, S. Pandey, M. Kianinia, J. Ahn, P. Ju, I. Aharonovich, N. Shivaram, T. Li, *ACS Photonics* **2021**, *8*, 994.

[286]  A. L. Exarhos, D. A. Hopper, R. N. Patel, M. W. Doherty, L. C. Bassett, *Nat. Commun.* **2019**, *10*, 222.

[287]  A. Gottscholl, M. Diez, V. Soltamov, C. Kasper, A. Sperlich, M. Kianinia, C. Bradac, I. Aharonovich, V. Dyakonov, *Sci. Adv.* **2021**, *7*, 1.

[288]  A. Gottscholl, M. Diez, V. Soltamov, C. Kasper, D. Krauße, A. Sperlich, M. Kianinia, C. Bradac, I. Aharonovich, V. Dyakonov, *Nat. Commun.* **2021**, *12*, 4480.

[289]  M. W. Doherty, V. M. Acosta, A. Jarmola, M. S. J. Barson, N. B. Manson, D. Budker, L. C. L. Hollenberg, *Phys. Rev. B* **2014**, *90*, 41201.

[290]  Y. Wu, F. Jelezko, M. Bplenio, T. Weil, *Angew. Chemie - Int. Ed.* **2016**, *55*, 6586.

[291]  J. L. Webb, L. Troise, N. W. Hansen, C. Olsson, A. M. Wojciechowski, J. Achard, O. Brinza, R. Staacke, M. Kieschnick, J. Meijer, A. Thielscher, J.-F. Perrier, K. Berg-Sørensen, A. Huck, U. L. Andersen, *Sci. Rep.* **2021**, *11*, 2412.

[292]  J. F. Barry, M. J. Turner, J. M. Schloss, D. R. Glenn, Y. Song, M. D. Lukin, H. Park, R. L. Walsworth, *PNAS* **2016**, *113*, 14133.

[293]  H. S. Knowles, D. M. Kara, M. Atatüre, *Nat. Mater.* **2014**, *13*, 21.

[294]  C. R. Kagan, L. C. Bassett, C. B. Murray, S. M. Thompson, *Chem. Rev* **2021**, *121*, 3233.

[295]  Y. Léger, L. Besombes, J. Fernández-Rossier, L. Maingault, H. Mariette, *Phys. Rev. Lett.* **2006**, *97*.

[296]  B. Varghese, H. Boukari, L. Besombes, *Phys. Rev. B - Condens. Matter Mater. Phys.* **2014**, *90*, 115307.

[297]  L. Besombes, H. Boukari, V. Tiwari, A. Lafuente-Sampietro, S. Kuroda, K. Makita, *Semicond. Sci. Technol.* **2019**, *34*, 063001.

[298]  A. H. Trojnar, M. Korkusiński, E. S. Kadantsev, P. Hawrylak, M. Goryca, T. Kazimierczuk, P. Kossacki, P. Wojnar, M. Potemski, *Phys. Rev. Lett.* **2011**, *107*.

[299]  J. K. Bindra, K. Singh, J. Van Tol, N. S. Dalal, G. F. Strouse, *J. Phys. Chem. C* **2020**, *124*, 10.

[300]  J.-H. Choi, H. Wang, S. J. Oh, T. Paik, P. Sung, J. Sung, X. Ye, T. Zhao, B. T. Diroll, C. B. Murray, C. R. Kagan, *Science (80-. ).* **2016**, *352*, 205.

[301]  J. Yang, M. K. Choi, D.-H. Kim, T. Hyeon, *Adv. Mater.* **2016**, *28*, 1176.

[302]  F. Moro, A. J. Fielding, L. Turyanska, A. Patanè, *Adv. Quantum Tecchnologies* **2019**, *2*, 1900017.

[303] K. S. Pedersen, A.-M. Ariciu, S. McAdams, H. Weihe, J. Bendix, F. Tuna, S. Piligkos, *J. Am. Chem. Soc* **2016**, *138*, 5801.

[304] J. M. Zadrozny, D. E. Freedman, *Inorg. Chem* **2015**, *54*, 12027.

[305] S. J. Lockyer, A. Chiesa, G. A. Timco, E. J. L. Mcinnes, T. S. Bennett, I. J. Vitorica-Yrezebal, S. Carretta, R. E. P. Winpenny, *Chem. Sci* **2021**, *12*, 9104.

[306] L. Tesi, E. Lucaccini, I. Cimatti, M. Perfetti, M. Mannini, M. Atzori, E. Morra, M. Chiesa, A. Caneschi, L. Sorace, R. Sessoli, *Chem. Sci.* **2016**, *7*, 2074.

[307] M. N. Leuenberger, D. Loss, *Nature* **2001**, *410*, 789.

[308] K. S. Pedersen, A. M. Ariciu, S. McAdams, H. Weihe, J. Bendix, F. Tuna, S. Piligkos, *J. Am. Chem. Soc.* **2016**, *138*, 5801.

[309] J. M. Zadrozny, J. Niklas, O. G. Poluektov, D. E. Freedman, *J. Am. Chem. Soc.* **2014**, *136*, 15841.

[310] S. Chicco, A. Chiesa, G. Allodi, E. Garlatti, M. Atzori, L. Sorace, R. De Renzi, R. Sessoli, S. Carretta, *Chem. Sci.* **2021**, *12*, 12046.

[311] E. Macaluso, M. Rubín, D. Aguià, A. Chiesa, L. A. Barrios, J. I. Martínez, P. J. Alonso, O. Roubeau, F. Luis, G. Aromí, S. Carretta, *Chem. Sci.* **2020**, *11*, 10337.

[312] F. Petiziol, A. Chiesa, S. Wimberger, P. Santini, S. Carretta, *npj Quantum Inf.* **2021**, *7*, 133.

[313] J. M. Zadrozny, J. Niklas, O. G. Poluektov, D. E. Freedman, *ACS Cent. Sci.* **2015**, *1*, 488.

[314] M. Mayländer, S. Chen, E. R. Lorenzo, M. R. Wasielewski, S. Richert, *J. Am. Chem. Soc* **2021**, *143*, 58.

[315] Y.-X. Wang, Z. Liu, Y.-H. Fang, S. Zhou, S.-D. Jiang, S. Gao, *npj Quantum Inf.* **2021**, *7*, 32.

[316] J. Bardeen, L. N. Cooper, J. R. Schrieffer, *Phys. Rev.* **1957**, *108*, 1175.

[317] J. Clarke, F. K. Wilhelm, *Nature* **2008**, *453*, 1031.

[318] P. Krantz, M. Kjaergaard, F. Yan, T. P. Orlando, S. Gustavsson, W. D. Oliver, *Appl. Phys. Rev.* **2019**, *6*, 021318.

[319] J. Koch, T. M. Yu, J. Gambetta, A. A. Houck, D. I. Schuster, J. Majer, A. Blais, M. H. Devoret, S. M. Girvin, R. J. Schoelkopf, *Phys. Rev. A* **2007**, *76*, 042319.

[320] R. Barends, J. Kelly, A. Megrant, D. Sank, E. Jeffrey, Y. Chen, Y. Yin, B. Chiaro, J. Mutus, C. Neill, P. O'malley, P. Roushan, J. Wenner, T. C. White, A. N. Cleland, J. M. Martinis, *Phys. Rev. Lett.* **2013**, *111*, 080502.

[321] J. Q. You, X. Hu, S. Ashhab, F. Nori, *Phys. Rev. B* **2007**, *75*, 140515.

[322] V. E. Manucharyan, J. Koch, L. I. Glazman, M. H. Devoret, *Science (80-. ).* **2009**, *326*, 113.

[323] T. W. Larsen, K. D. Petersson, F. Kuemmeth, T. S. Jespersen, P. Krogstrup, J. Nygård, C. M. Marcus, *Phys. Rev. Lett.* **2015**, *115*, 1.

[324] M. Kjaergaard, M. E. Schwartz, J. Braumüller, P. Krantz, J. I-J Wang, S. Gustavsson, W. D. Oliver, *Annu. Rev. Condens. Matter Phys.* **2020**, *11*, 369.

[325] K. S. Chou, J. Z. Blumoff, C. S. Wang, P. C. Reinhold, C. J. Axline, Y. Y. Gao, L. Frunzio, M. H. Devoret, L. Jiang, J. Schoelkopf, *Nature* **2018**, *561*, 369.

[326] P. Jurcevic, A. Javadi-Abhari, L. S. Bishop, I. Lauer, D. F. Bogorin, M. Brink, L. Capelluto, O. Günlük, T. Itoko, N. Kanazawa, A. Kandala, G. A. Keefe, K. Krsulich, W. Landers, E. P. Lewandowski, D. T. Mcclure, G. Nannicini, A. Narasgond, H. M. Nayfeh, E. Pritchett, M. B. Rothwell, S. Srinivasan, N. Sundaresan, C. Wang, K. X. Wei, C. J. Wood, J.-B. Yau, E. J. Zhang, O. E. Dial, J. M. Chow, J. M. Gambetta, *Quantum Sci. Technol* **2021**, *6*, 25020.

[327] S. Rosenblum, Y. Y. Gao, P. Reinhold, C. Wang, C. J. Axline, L. Frunzio, S. M. Girvin, L. Jiang, M. Mirrahimi, M. H. Devoret, R. J. Schoelkopf, *Nat. Commun.* **2018**, *9*, 652.

[328] R. McDermott, *IEEE Trans. Appl. Supercond.* **2009**, *19*, 2.

[329] J. Lisenfeld, A. Bilmes, A. Megrant, R. Barends, J. Kelly, P. Klimov, G. Weiss, J. M. Martinis, A. V Ustinov, *npj Quantum Inf.* **2019**, *5*, 105.

[330] J. M. Martinis, K. B. Cooper, R. Mcdermott, M. Steffen, M. Ansmann, K. D. Osborn, K. Cicak, S. Oh, D. P. Pappas, R. W. Simmonds, C. C. Yu, *Phys. Rev. Lett.* **2005**, *95*, 210503.

[331] J. I.-J. Wang, D. Rodan-Legrain, L. Bretheau, D. L. Campbell, B. Kannan, D. Kim, M. Kjaergaard, P. Krantz, G. O. Samach, F. Yan, J. L. Yoder, K. Watanabe, T. Taniguchi, T. P. Orlando, S. Gustavsson, P. Jarillo-Herrero, W. D. Oliver, *Nat. Nanotechnol.* **2019**, *14*, 120.

[332] K.-H. Lee, S. Chakram, S. E. Kim, F. Mujid, A. Ray, H. Gao, C. Park, Y. Zhong, D. A. Muller, D. I. Schuster, J. Park, *Nano Lett.* **2019**, *19*, 8287.

[333] M. Ramezani, I. C. Sampaio, K. Watanabe, T. Taniguchi, C. Schönenberger, A. Baumgartner, *Nano Lett.* **2021**, *21*, 5614.

[334] J. I.-J. Wang, M. A. Yamoah, Q. Li, A. Karamlou, T. Dinh, B. Kannan, J. Braumueller, D. Kim, A. J. Melville, S. E. Muschinske, B. M. Niedzielski, K. Serniak, Y. Sung, R. Winik, J. L. Yoder, M. Schwartz, K. Watanabe, T. Taniguchi, T. P. Orlando, S. Gustavsson, P. Jarillo-Herrero, W. D. Oliver, *Hexagonal Boron Nitride (hBN) as a Low-loss Dielectric for Superconducting Quantum Circuits and Qubits*, **2021**.

[335] T. Wakamura, N. J. Wu, A. D. Chepelianskii, S. Guéron, M. Och, M. Ferrier, T. Taniguchi, K. Watanabe, C. Mattevi, H. Bouchiat, *Phys. Rev. Lett.* **2020**, *125*.

[336] T. Zhang, P. Cheng, W.-J. Li, Y.-J. Sun, G. Wang, X.-G. Zhu, K. He, L. Wang, X. Ma, X. Chen, Y. Wang, Y. Liu, H.-Q. Lin, J.-F. Jia, Q.-K. Xue, *Nat. Phys.* **2010**, *6*, 104.

[337] D. Qiu, C. Gong, S. Wang, M. Zhang, C. Yang, X. Wang, J. Xiong, D. Qiu, C. Gong, S. Wang, M. Zhang, C. Yang, X. Wang, J. Xiong, *Adv. Mater.* **2021**, *33*, 2006214.

[338] A. T. Yokoya, T. Kiss, A. Chainani, S. Shin, M. Nohara, H. Takagi, *Science (80-. ).* **2001**, *294*, 2518.

[339] S. Park, S. Young Kim, H. Kug Kim, M. Jeong Kim, T. Kim, H. Kim, G. Seung Choi, C. J. Won, S. Kim, K. Kim, E. F. Talantsev, K. Watanabe, T. Taniguchi, S.-W. Cheong, B. J. Kim, H. W. Yeom, J. Kim, T.-H. Kim, J. Sung Kim, *Nat. Commun.* **2021**, *12*, 3157.

[340] W. Shi, J. Ye, Y. Zhang, R. Suzuki, M. Yoshida, J. Miyazaki, N. Inoue, Y. Saito, Y. Iwasa, *Sci. Rep.* **2015**, *5*, 12534.

[341] B. S. de Lima, R. R. de Cassia, F. B. Santos, L. E. Correa, T. W. Grant, A. L. R. Manesco, G. W. Martins, L. T. F. Eleno, M. S. Torikachvili, A. J. S. Machado, *Solid State Commun.* **2018**, *283*, 27.

[342] R. Zhang, I.-L. Tsai, J. Chapman, E. Khestanova, J. Waters, I. V Grigorieva, *Nano Lett.* **2016**, *16*, 629.

[343] D. J. Trainer, B. Wang, F. Bobba, N. Samuelson, X. Xi, J. Zasadzinski, J. Nieminen, A. Bansil, M. Iavarone, *ACS Nano* **2020**, *14*, 2718.

[344] B. T. Zhou, N. F. Q. Yuan, H.-L. Jiang, K. T. Law, *Phys. Rev. B* **2016**, *93*, 180501.

[345] S. C. De La Barrera, M. R. Sinko, D. P. Gopalan, N. Sivadas, K. L. Seyler, K. Watanabe, T. Taniguchi, A. W. Tsen, X. Xu, D. Xiao, B. M. Hunt, *Nat. Commun.* **2018**, *9*, 1427.

[346] B. Sipos, A. F. Kusmartseva, A. Akrap, H. Berger, L. Forró, F. Forró, E. Tutiš, T. Tutiš, *Nat. Mater.* **2008**, *7*, 960.

[347] E. Liebhaber, S. Acero Gonzaíez, R. Baba, B. W. Heinrich, S. Rohlf, K. Rossnagel, F. von Oppen, K. J. Franke, *Nano Lett.* **2020**, *20*, 339.

[348] A. Majumdar, D. Vangennep, J. Brisbois, D. Chareev, A. V Sadakov, A. S. Usoltsev, M. Mito, A. V Silhanek, T. Sarkar, A. Hassan, O. Karis, R. Ahuja, M. Abdel-Hafiez, *Phys. Rev. Mater.* **2020**, *4*, 084004.

[349] Y. Kvashnin, D. Vangennep, M. Mito, S. A. Medvedev, R. Thiyagarajan, O. Karis, A. N. Vasiliev, O. Eriksson, M. Abdel-Hafiez, *Phys. Rev. Lett.* **2020**, *125*, 186401.

[350] C. Xu, L. Wang, Z. Liu, L. Chen, J. Guo, N. Kang, X.-L. Ma, H.-M. Cheng, W. Ren, *Nat. Mater.* **2015**, *14*, 1135.

[351] S. Tan, Y. Zhang, M. Xia, Z. Ye, F. Chen, X. Xie, R. Peng, D. Xu, Q. Fan, H. Xu, J. Jiang, T. Zhang, X. Lai, T. Xiang, J. Hu, B. Xie, D. Feng, *Nat. Mater.* **2013**, *12*, 634.

[352] Y. Song, Z. Chen, Q. Zhang, H. Xu, X. Lou, X. Chen, X. Xu, X. Zhu, R. Tao, T. Yu, H. Ru, Y. Wang, T. Zhang, J. Guo, L. Gu, Y. Xie, R. Peng, D. Feng, *Nat. Commun.* **2021**, *12*, 5926.

[353] C. Liu, H. Shin, A. Doll, H.-H. Kung, R. P. Day, B. A. Davidson, J. Dreiser, G. Levy, A. Damascelli, C. Piamonteze, K. Zou, *npj Quantum Mater.* **2021**, *6*, 85.

[354] Y. Cao, V. Fatemi, S. Fang, K. Watanabe, T. Taniguchi, E. Kaxiras, P. Jarillo-Herrero, *Nat. Gr.* **2018**, *556*, 43.

[355] G. Chen, A. L. Sharpe, P. Gallagher, I. rosen, eli J. Fox, L. Jiang, B. Lyu, H. Li, K. Watanabe, takashi taniguchi, J. Jung, Z. Shi, D. Goldhaber-Gordon, Y. Zhang, F. Wang, *Nature* **2019**, *572*, 215.

[356] D. I. Indolese, R. Delagrange, P. Makk, J. R. Wallbank, K. Wanatabe, T. Taniguchi, C. Schönenberger, *Phys. Rev. Lett.* **2018**, *121*, 137701.

[357] W. Tian, S. Chen, Z. Xu, D. Li, H. Du, Z. Wei, K. Wu, H. Sun, S. Dong, Y. Lv, Y.-L. Wang, D. Koelle, R. Kleiner, H. Wang, P. Wu, *Supercond. Sci. Technol.* **2021**, *34*, 115015.

[358] C. Xu, S. Song, Z. Liu, L. Chen, L. Wang, D. Fan, N. Kang, X. Ma, H.-M. Cheng, W. Ren, *ACS Nano* **2017**, *11*, 5906.

[359] H. Wang, X. Huang, J. Lin, J. Cui, Y. Chen, C. Zhu, F. Liu, Q. Zeng, J. Zhou, P. Yu, X. Wang, H. He, S. H. Tsang, W. Gao, K. Suenaga, F. Ma, C. Yang, L. Lu, T. Yu, E. Hang, T. Teo, G. Liu, Z. Liu, *Nat. Commun.* **2017**, *8*, 394.

[360] T. Dvir, A. Zalic, E. H. Fyhn, M. Amundsen, T. Taniguchi, K. Watanabe, J. Linder, H. Steinberg, *Phys. Rev. B* **2021**, *103*, 115401.

[361] M. Kim, G.-H. Park, J. Lee, J. H. Lee, J. Park, H. Lee, G.-H. Lee, H.-J. Lee, *Nano Lett.* **2017**, *17*, 6125.

[362] L. Ai, E. Zhang, J. Yang, X. Xie, Y. Yang, Z. Jia, Y. Zhang, S. Liu, Z. Li, P. Leng, X. Cao, X. Sun, T. Zhang, X. Kou, Z. Han, F. Xiu, S. Dong, *Nat. Commun.* **2021**, *12*, 6580.

[363] A.Yu. Kitaev, *Ann. Phys. (N. Y).* **2002**, *303*, 2.

[364] H. Bartolomei, M. Kumar, R. Bisognin, A. Marguerite, J. M. Berroir, E. Bocquillon, B. Plaçais, A. Cavanna, Q. Dong, U. Gennser, Y. Jin, G. Fève, *Science (80-. ).* **2020**, *368*, 173.

[365] J. Nakamura, S. Liang, G. C. Gardner, M. J. Manfra, *Nat. Phys.* **2020**, *16*, 931.

[366] S. Das Sarma, C. Nayak, S. Tewari, *Phys. Rev. B* **2006**, *73*, 220502.

[367] L. Fu, C. L. Kane, *Phys. Rev. Lett.* **2008**, *100*, 96407.

[368] R. M. Lutchyn, J. D. Sau, S. Das Sarma, *Phys. Rev. Lett.* **2010**, *105*, 077001.

[369] G. Moore, N. Read, *Nucl. Physics, Sect. B* **1991**, *360*, 362.

[370] S. Frolov, *Nature* **2021**, *592*, 350.

[371] C. X. Liu, J. D. Sau, T. D. Stanescu, S. Das Sarma, *Phys. Rev. B* **2017**, *96*, 075161.

[372] N. M. Chtchelkatchev, Y. V Nazarov, *Phys. Rev. Lett.* **2003**, *90*, 226806.

[373] M. Hays, V. Fatemi, D. Bouman, J. Cerrillo, S. Diamond, K. Serniak, T. Connolly, P. Krogstrup, J. Nygård, A. L. Yeyati, A. Geresdi, M. H. Devoret, *Science (80-. ).* **2021**, *373*, 430.

[374] P. Krogstrup, N. L. B. Ziino, W. Chang, S. M. Albrecht, M. H. Madsen, E. Johnson, J. Nygård, C. M. Marcus, T. S. Jespersen, *Nat. Mater. 2014 144* **2015**, *14*, 400.

[375] M. T. Deng, C. L. Yu, G. Y. Huang, M. Larsson, P. Caroff, H. Q. Xu, *Nano Lett* **2012**, *12*, 6414.

[376] V. Mourik, K. Zuo, S. M. Frolov, S. R. Plissard, E. P. A. M. Bakkers, L. P. Kouwenhoven, *Science (80-.)*. **2012**, *336*, 1003.

[377] F. Maier, J. Klinovaja, D. Loss, *Phys. Rev. B - Condens. Matter Mater. Phys.* **2014**, *90*, 195421.

[378] J. Sun, R. S. Deacon, R. Wang, J. Yao, C. M. Lieber, K. Ishibashi, *Nano Lett.* **2018**.

[379] J. Xiang, A. Vidan, M. Tinkham, R. M. Westervelt, † And, C. M. Lieber, *Nat. Nanotechnol.* **2006**, *1*, 208.

[380] J. Ridderbos, M. Brauns, A. Li, E. P. A. M. Bakkers, A. Brinkman, W. G. Van Der Wiel, F. A. Zwanenburg, *Phys. Rev. Mater.* **2019**, *3*, 084803.

[381] M. Marganska, L. Milz, W. Izumida, C. Strunk, M. Grifoni, *Phys. Rev. B* **2018**, *97*, 75141.

[382] J. D. Sau, S. Tewari, *Phys. Rev. B* **2013**, *88*, 54503.

[383] R. Egger, K. Flensberg, *Phys. Rev. B - Condens. Matter Mater. Phys.* **2012**, *85*, 235462.

[384] J. Klinovaja, S. Gangadharaiah, D. Loss, *Phys. Rev. Lett.* **2012**, *108*, 196804.

[385] M. M. Desjardins, L. C. Contamin, M. R. Delbecq, M. C. Dartiailh, L. E. Bruhat, T. Cubaynes, J. J. Viennot, F. Mallet, S. Rohart, A. Thiaville, A. Cottet, T. Kontos, *Nat. Mater.* **2019**, *18*, 1060.

[386] C. Bäuml, L. Bauriedl, M. Marganska, M. Grifoni, C. Strunk, N. Paradiso, *Nano Lett.* **2021**, *21*, 8627.

[387] C. L. Kane, E. J. Mele, *Phys. Rev. Lett.* **2005**, *95*, 226801.

[388] X.-T. An, Y.-Y. Zhang, J.-J. Liu, *Appl. Phys. Lett* **2013**, *102*, 43113.

[389] C.-C. Liu, W. Feng, Y. Yao, *Phys. Rev. Lett.* **2011**, *107*, 076802.

[390] M. Tahir, A. Manchon, K. Sabeeh, *Appl. Phys. Lett* **2013**, *102*, 162412.

[391] I. K. Drozdov, A. Alexandradinata, S. Jeon, S. Nadj-Perge, H. Ji, R. J. Cava, B. Andrei Bernevig, A. Yazdani, *Nat. Phys.* **2014**, *10*, 664.

[392] S.-Y. Zhu, Y. Shao, E. Wang, L. Cao, X.-Y. Li, Z.-L. Liu, C. Liu, L.-W. Liu, J.-O. Wang, K. Ibrahim, J.-T. Sun, Y.-L. Wang, S. Du, H.-J. Gao, *Nano Lett.* **2019**, *19*, 6323.

[393] P. Chen, W. W. Pai, Y.-H. Chan, W.-L. Sun, C.-Z. Xu, D.-S. Lin, M. Y. Chou, A.-V. Fedorov, T.-C. Chiang, *Nat. Commun.* **2018**, *9*, 2003.

[394] D. V Gruznev, S. V Eremeev, L. V Bondarenko, A. Y. Tupchaya, A. A. Yakovlev, A. N. Mihalyuk, J.-P. Chou, A. V Zotov, A. A. Saranin, *Nano Lett.* **2018**, *18*, 4338.

[395] R. Wu, J.-Z. Ma, S.-M. Nie, L.-X. Zhao, X. Huang, J.-X. Yin, B.-B. Fu, P. Richard, G.-F. Chen, Z. Fang, X. Dai, H.-M. Weng, T. Qian, H. Ding, S. H. Pan, *Phys. Rev. X* **2016**, *6*, 021017.

[396] C. Si, K. H. Jin, J. Zhou, Z. Sun, F. Liu, *Nano Lett.* **2016**, *16*, 6584.

[397] M. Zhao, X. Zhang, L. Li, *Sci. Rep.* **2015**, *5*, 16108.

[398] Y. Ma, Y. Dai, L. Kou, T. Frauenheim, T. Heine, *Nano Lett.* **2015**, *15*, 1083.

[399] X. Hu, N. Mao, H. Wang, Y. Dai, B. Huang, C. Niu, *Phys. Rev. B* **2021**, *103*, 85109.

[400] Q. Liu, X. Zhang, L. B. Abdalla, A. Fazzio, A. Zunger, *Nano Lett.* **2015**, *15*, 1222.

[401] F. Zhao, T. Cao, S. G. Louie, *Phys. Rev. Lett.* **2021**, *127*, 166401.

[402] F. Lüpke, D. Waters, S. C. de la Barrera, M. Widom, D. G. Mandrus, J. Yan, R. M. Feenstra, B. M. Hunt, *Nat. Phys. 2020 165* **2020**, *16*, 526.

[403] H.-H. Sun, M.-X. Wang, F. Zhu, G.-Y. Wang, H.-Y. Ma, Z.-A. Xu, Q. Liao, Y. Lu, C.-L. Gao, Y.-Y. Li, C. Liu, D. Qian, D. Guan, J.-F. Jia, *Nano Lett.* **2017**, *17*, 3035.

[404] T. Kawakami, X. Hu, *Phys. Rev. Lett.* **2015**, *115*, 177001.

[405]  Z. F. Wang, H. Zhang, D. Liu, C. Liu, C. Tang, C. Song, Y. Zhong, J. Peng, F. Li, C. Nie, L. Wang, X. J. Zhou, X. Ma, Q. K. Xue, F. Liu, *Nat. Mater.* **2016**, *15*, 968.

[406]  A. Aspuru-Guzik, P. Walther, *Nat. Phys.* **2012**, *8*, 285.

[407]  S. Pirandola, B. R. Bardhan, T. Gehring, C. Weedbrook, S. Lloyd, *Nat. Photonics* **2018**, *12*, 724.

[408]  I. Aharonovich, D. Englund, M. Toth, *Nat. Photonics* **2016**, *10*, 631.

[409]  A. Dietrich, M. W. Doherty, I. Aharonovich, A. Kubanek, *Phys. Rev. B* **2020**, *101*, 81401.

[410]  T. Grange, G. Hornecker, D. Hunger, J. P. Poizat, J. M. Gérard, P. Senellart, A. Auffèves, *Phys. Rev. Lett.* **2015**, *114*, 193601.

[411]  H. Choi, D. Zhu, Y. Yoon, D. Englund, *Phys. Rev. Lett.* **2019**, *122*, 183602.

[412]  P. Lodahl, S. Mahmoodian, S. Stobbe, *Rev. Mod. Phys.* **2015**, *87*, 347.

[413]  J. A. Schuller, E. S. Barnard, W. Cai, Y. C. Jun, J. S. White, M. L. Brongersma, *Nat. Mater.* **2010**, *9*, 193.

[414]  K. J. Russell, T. L. Liu, S. Cui, E. L. Hu, *Nat. Photonics* **2012**, *6*, 459.

[415]  G. Zhang, Y. Cheng, J. P. Chou, A. Gali, *Appl. Phys. Rev.* **2020**, *7*, 031308.

[416]  I. Aharonovich, S. Castelletto, D. A. Simpson, C.-H. Su, A. D. Greentree, S. Prawer, *Rep. Prog. Phys.* **2011**, *74*, 076501.

[417]  P. G. Kwiat, K. Mattle, H. Weinfurter, A. Zeilinger, A. V. Sergienko, Y. Shih, *Phys. Rev. Lett.* **1995**, *75*, 4337.

[418]  P. G. Kwiat, E. Waks, A. G. White, I. Appelbaum, P. H. Eberhard, *Phys. Rev. A* **1999**, *60*, R773.

[419]  C. Couteau, *Contemp. Phys.* **2018**, *59*, 291.

[420]  D. Huber, M. Reindl, J. Aberl, A. Rastelli, R. Trotta, *J. Opt.* **2018**, *20*, 073002.

[421]  R. J. Young, R. M. Stevenson, A. J. Hudson, C. A. Nicoll, D. A. Ritchie, A. J. Shields, *Phys. Rev. Lett.* **2009**, *102*, 030406.

[422]  D. Huber, M. Reindl, S. Filipe Covre da Silva, C. Schimpf, J. Martín-Sánchez, H. Huang, G. Piredda, J. Edlinger, A. Rastelli, R. Trotta, *Phys. Rev. Lett.* **2018**, *121*, 33902.

[423]  K. Kowalik, O. Krebs, A. Lemaître, S. Laurent, P. Senellart, P. Voisin, J. A. Gaj, *Appl. Phys. Lett.* **2005**, *86*, 041907.

[424]  N. Nikolay, N. Mendelson, E. Özelci, B. Sontheimer, F. Böhm, G. Kewes, M. Toth, I. Aharonovich, O. Benson, *Optica* **2019**, *6*, 1084.

[425]  G. Noh, D. Choi, J.-H. Kim, D.-G. Im, Y.-H. Kim, H. Seo, J. Lee, *Nano Lett* **2018**, *18*, 4710.

[426]  S. Lazić, A. Espinha, S. P. Yanguas, C. Gibaja, F. Zamora, P. Ares, M. Chhowalla, W. S. Paz, J. José, P. Burgos, A. Hernández-Mínguez, P. V Santos, H. P. Van Der Meulen, *Communicat* **2019**, *2*, 113.

[427]  N. Mendelson, M. Doherty, M. Toth, I. Aharonovich, T. Trong Tran, N. Mendelson, M. Toth, I. Aharonovich, T. T. Tran, M. Doherty, *Adv. Mater.* **2020**, *32*, 1908316.

[428]  J. Ziegler, R. Klaiss, A. Blaikie, D. Miller, V. R. Horowitz, B. J. A. Alemán, *Nano Lett.* **2019**, *19*, 2121.

[429]  H. Ngoc My Duong, M. Anh Phan Nguyen, M. Kianinia, T. Ohshima, H. Abe, K. Watanabe, T. Taniguchi, J. H. Edgar, I. Aharonovich, M. Toth, *ACS Appl. Mater. Interfaces* **2018**, *10*, 24886.

[430]  C. Fournier, A. Plaud, S. Roux, A. Pierret, M. Rosticher, K. Watanabe, T. Taniguchi, S. Buil, X. Quélin, J. Barjon, J.-P. Hermier, A. Deltenil, *Nat. Commun.* **2021**, *12*, 3779.

[431]  N. V Proscia, Z. Shotan, H. Jayakumar, P. Reddy, C. Cohen, M. Dollar, A. Alkauskas, M. Doherty, C. A. Meriles, V. M. Menon, *Optica* **2018**, *5*, 1128.

[432] C. Li, N. Mendelson, R. Ritika, Y. Chen, Z.-Q. Xu, M. Toth, I. Aharonovich, *Nano Lett.* **2021**, *21*, 3632.

[433] J. A. Preuß, E. Rudi, J. Kern, R. Schmidt, R. Bratschitsch, S. Michaelis De Vasconcellos, *2D Mater.* **2021**, *8*, 35005.

[434] A. L. Exarhos, D. A. Hopper, R. R. Grote, A. Alkauskas, L. C. Bassett, *ACS Nano* **2017**, *11*, 3328.

[435] T. T. Tran, K. Bray, M. J. Ford, M. Toth, I. Aharonovich, *Nat. Nanotechnol.* **2016**, *11*, 37.

[436] N. R. Jungwirth, B. Calderon, Y. Ji, M. G. Spencer, M. E. Flatté,Flatté,, G. D. Fuchs, *Nano Lett* **2016**, *16*, 22.

[437] V. Ivády, G. Barcza, G. Thiering, S. Li, H. Hamdi, J.-P. Chou, Ö. Legeza, A. Gali, *npj Comput. Mater.* **2020**, *6*, 41.

[438] M. E. Turiansky, A. Alkauskas, L. C. Bassett, C. G. Van De Walle, *Phys. Rev. Lett.* **2019**, *123*, 127401.

[439] M. E. Turiansky, C. G. Van De Walle, *J. Appl. Phys* **2021**, *129*, 64301.

[440] N. Mendelson, D. Chugh, J. R. Reimers, T. S. Cheng, A. Gottscholl, H. Long, C. J. Mellor, A. Zettl, V. Dyakonov, P. H. Beton, S. V Novikov, C. Jagadish, H. Hoe Tan, M. J. Ford, M. Toth, C. Bradac, I. Aharonovich, *Nat. Mater.* **2021**, *20*, 321.

[441] M. Mackoit-Sinkeviciene, M. Maciaszek, C. G. Van de Walle, *Appl. Phys. Lett.* **2019**, *115*, 212101.

[442] T. Vogl, R. Lecamwasam, B. C. Buchler, Y. Lu, P. K. Lam, *ACS Photonics* **2019**, *6*, 1955.

[443] X. Li, R. A. Scully, K. Shayan, Y. Luo, S. Strauf, *ACS Nano* **2019**, *13*, 6992.

[444] J. E. Fröch, L. P. Spencer, M. Kianinia, D. D. Totonjian, M. Nguyen, A. Gottscholl, V. Dyakonov, M. Toth, S. Kim, I. Aharonovich, *Nano Lett.* **2021**, *21*, 6549.

[445] Y. Wang, J. Lee, J. Berezovsky, P. X-L Feng, *Appl. Phys. Lett* **2021**, *118*, 244003.

[446] Y. You, X.-X. Zhang, T. C. Berkelbach, M. S. Hybertsen, D. R. Reichman, T. F. Heinz, *Nat. Phys.* **2015**, *11*, 477.

[447] C. Chakraborty, L. Kinnischtzke, K. M. Goodfellow, R. Beams, A. N. Vamivakas, *Nat. Nanotechnol.* **2015**, *10*, 507.

[448] A. Srivastava, M. Sidler, A. V. Allain, D. S. Lembke, A. Kis, A. Imamoglu, *Nat. Nanotechnol.* **2015**, *10*, 491.

[449] J. Dang, S. Sun, X. Xie, Y. Yu, K. Peng, C. Qian, S. Wu, F. Song, J. Yang, S. Xiao, L. Yang, Y. Wang, M. A. Rafiq, C. Wang, X. Xu, *npj 2D Mater. Appl.* **2020**, *4*, 2.

[450] J. Klein, M. Lorke, M. Florian, F. Sigger, L. Sigl, S. Rey, J. Wierzbowski, J. Cerne, K. Müller, E. Mitterreiter, P. Zimmermann, T. Taniguchi, K. Watanabe, U. Wurstbauer, M. Kaniber, M. Knap, R. Schmidt, J. J. Finley, A. W. Holleitner, *Nat. Commun.* **2019**, *10*, 2755.

[451] E. Mitterreiter, B. Schuler, A. Micevic, D. Hernangómez-Pérez, K. Barthelmi, K. A. Cochrane, J. Kiemle, F. Sigger, J. Klein, E. Wong, E. S. Barnard, K. Watanabe, T. Taniguchi, M. Lorke, F. Jahnke, J. J. Finley, A. M. Schwartzberg, D. Y. Qiu, S. Refaely-Abramson, A. W. Holleitner, A. Weber-Bargioni, C. Kastl, *Nat. Commun.* **2021**, *12*, 3822.

[452] A. Hötger, J. Klein, K. Barthelmi, L. Sigl, F. Sigger, W. Mä, S. Gyger, M. Florian, M. Lorke, F. Jahnke, T. Taniguchi, K. Watanabe, K. D. Jö, U. Wurstbauer, C. Kastl, K. Mü, J. J. Finley, A. W. Holleitner, *Nano Lett.* **2021**, *21*, 1040.

[453] K. Barthelmi, J. Klein, A. Hötger, L. Sigl, F. Sigger, E. Mitterreiter, S. Rey, S. Gyger, M. Lorke, M. Florian, F. Jahnke, T. Taniguchi, K. Watanabe, V. Zwiller, K. D. Jöns, U. Wurstbauer, C. Kastl, A. Weber-Bargioni, J. J. Finley, K. Müller, A. W. Holleitner, *Appl. Phys. Lett* **2020**, *117*, 070501.

[454] L. Yu, M. Deng, J. L. Zhang, S. Borghardt, B. Kardynal, J. Vučković, T. F. Heinz, *Nano Lett.* **2021**, *21*, 2376.

[455] L. Peng, H. Chan, P. Choo, T. W. Odom, S. K. R S Sankaranarayanan, X. Ma, *Nano Lett.* **2020**, *20*,

5866.

[456] M. Paur, A. J. Molina-Mendoza, D. Polyushkin, S. Michaelis de Vasconcellos, R. Bratschitsch, T. Mueller, *2D Mater.* **2020**, *7*, 045021.

[457] N. Liu, C. M. Gallaro, K. Shayan, A. Mukherjee, B. Kim, J. Hone, N. Vamivakas, S. Strauf, V. der Waals, *Nanoscale* **2021**, *13*, 832.

[458] F. Mahdikhanysarvejahany, D. N. Shanks, C. Muccianti, B. H. Badada, I. Idi, A. Alfrey, S. Raglow, M. R. Koehler, D. G. Mandrus, T. Taniguchi, K. Watanabe, O. L. A. Monti, H. Yu, B. J. Leroy, J. R. Schaibley, *npj 2D Mater. Appl.* **2021**, *5*, 67.

[459] H. Baek, M. Brotons-Gisbert, Z. X. Koong, A. Campbell, M. Rambach, K. Watanabe, T. Taniguchi, B. D. Gerardot, *Sci. Adv* **2020**, *6*, 8526.

[460] K. Parto, S. I. Azzam, K. Banerjee, G. Moody, *Nat. Commun.* **2021**, *12*, 3585.

[461] L. Linhart, M. Paur, V. Smejkal, J. Burgdörfer, T. Mueller, F. Libisch, *Phys. Rev. Lett.* **2019**.

[462] C. Zeng, J. Zhong, Y.-P. Wang, J. Yu, L. Cao, Z. Zhao, J. Ding, C. Cong, X. Yue, Z. Liu, Y. Liu, *Appl. Phys. Lett* **2020**, *117*, 153103.

[463] K. Barthelmi, J. Klein, A. Hötger, L. Sigl, F. Sigger, E. Mitterreiter, S. Rey, S. Gyger, M. Lorke, M. Florian, F. Jahnke, T. Taniguchi, K. Watanabe, V. Zwiller, K. D. Jöns, U. Wurstbauer, C. Kastl, A. Weber-Bargioni, J. J. Finley, K. Müller, A. W. Holleitner, *Appl. Phys. Lett* **2020**, *117*, 70501.

[464] H. Moon, E. Bersin, C. Chakraborty, A. Y. Lu, G. Grosso, J. Kong, D. Englund, *ACS Photonics* **2020**, *7*, 1135.

[465] M. R. Rosenberger, ∥ Chandriker, K. Dass, H.-J. Chuang, S. V Sivaram, K. M. Mccreary, J. R. Hendrickson, B. T. Jonker, *ACS Nano* **2019**, *13*, 51.

[466] H. Zhao, M. T. Pettes, Y. Zheng, H. Htoon, *Nat. Commun.* **2021**, *12*, 6753.

[467] K. Tran, G. Moody, F. Wu, X. Lu, J. Choi, K. Kim, A. Rai, D. A. Sanchez, J. Quan, A. Singh, J. Embley, A. Zepeda, M. Campbell, T. Autry, T. Taniguchi, K. Watanabe, N. Lu, S. K. Banerjee, K. L. Silverman, S. Kim, E. Tutuc, L. Yang, A. H. Macdonald, X. Li, *Nature* **2019**, *567*, 71.

[468] S. I. Azzam, K. Parto, G. Moody, *Cite as Appl. Phys. Lett* **2021**, *118*, 240502.

[469] K. Hao, J. F. Specht, P. Nagler, L. Xu, K. Tran, A. Singh, C. K. Dass, C. Schüller, T. Korn, M. Richter, A. Knorr, X. Li, G. Moody, *Nat. Commun.* **2017**, *8*, 15552.

[470] C. C. Price, N. C. Frey, D. Jariwala, V. B. Shenoy, *ACS Nano* **2019**, *13*, 8303.

[471] P. Tonndorf, S. Schwarz, J. Kern, I. Niehues, O. Del Pozo-Zamudio, A. I. Dmitriev, A. P. Bakhtinov, D. N. Borisenko, N. N. Kolesnikov, A. I. Tartakovskii, S. Michaelis de Vasconcellos, R. Bratschitsch, *2D Mater.* **2017**, *4*, 021010.

[472] P. Tonndorf, O. Del Pozo-Zamudio, N. Gruhler, J. Kern, R. Schmidt, A. I. Dmitriev, A. P. Bakhtinov, A. I. Tartakovskii, W. Pernice, S. Michaelis De Vasconcellos, R. Bratschitsch, *Nano Lett.* **2017**, *17*, 5446.

[473] A. Ishii, M. Yoshida, Y. K. Kato, *Phys. Rev. B - Condens. Matter Mater. Phys.* **2015**, *91*, 125427.

[474] A. H. Brozena, M. Kim, L. R. Powell, Y. Wang, *Nat. Rev. Chem.* **2019**, *3*, 375.

[475] Z. Li, K. Otsuka, D. Yamashita, D. Kozawa, Y. K. Kato, *ACS Photonics* **2021**, *8*, 2367.

[476] X. He, N. F. Hartmann, X. Ma, Y. Kim, R. Ihly, J. L. Blackburn, W. Gao, J. Kono, Y. Yomogida, A. Hirano, T. Tanaka, H. Kataura, H. Htoon, S. K. Doorn, *Nat. Photonics* **2017**, *11*, 577.

[477] A. Ishii, T. Uda, Y. K. Kato, *Phys. Rev. Appl.* **2017**, *8*, 054039.

[478] V. A. Shahnazaryan, V. A. Saroka, I. A. Shelykh, W. L. Barnes, M. E. Portnoi, *ACS Photonics* **2019**, *6*, 904.

[479] J. M. Lü, F. J. Berger, J. Zaumseil, *ACS Photonics* **2021**, *8*, 182.

[480] A. Dhavamani, L. Haeberlé, J. Wang, S. Kéna-Cohen, M. S. Arnold, *ACS Photonics* **2021**, *8*, 2375.

[481] X. Wu, M. Kim, H. Kwon, Y. H. Wang, *Angew. Chemie - Int. Ed.* **2018**, *57*, 648.

[482] Z. Yuan, B. E. Kardynal, R. M. Stevenson, A. J. Shields, C. J. Lobo, K. Cooper, N. S. Beattie, D. A. Ritchie, M. Pepper, *Science (80-. ).* **2002**, *295*, 102.

[483] C. L. Salter, R. M. Stevenson, I. Farrer, C. A. Nicoll, D. A. Ritchie, A. J. Shields, *Nature* **2010**, *465*, 594.

[484] M. Paul, F. Olbrich, J. Höschele, S. Schreier, J. Kettler, S. L. Portalupi, M. Jetter, P. Michler, *Appl. Phys. Lett* **2017**, *111*, 33102.

[485] K. D. Zeuner, K. D. Jöns, L. Schweickert, C. Reuterskiöld Hedlund, C. Nuñez Lobato, T. Lettner, K. Wang, S. Gyger, E. Schöll, S. Steinhauer, M. Hammar, V. Zwiller, *ACS Photonics* **2021**, *8*, 2337.

[486] F.-Y. Tsai, C. P. Lee, *J. Appl. Phys.* **1998**, *84*, 2627.

[487] T. Mano Marco Abbarchi Takashi Kuroda Brian McSkimming Akihiro Ohtake Kazutaka Mitsuishi Kazuaki Sakoda, *Appl. Phys. Express* **2010**, *3*, 065203.

[488] M. V Durnev, M. M. Glazov, E. L. Ivchenko, M. Jo, T. Mano, T. Kuroda, K. Sakoda, S. Kunz, G. Sallen, L. Bouet, X. Marie, D. Lagarde, T. Amand, B. Urbaszek, *Phys. Rev. B* **2013**, *87*, 85315.

[489] D. Huber, M. Reindl, Y. Huo, H. Huang, J. S. Wildmann, O. G. Schmidt, A. Rastelli, R. Trotta, *Nat. Commun.* **2017**, *8*, 15506.

[490] F. Basso Basset, F. Salusti, L. Schweickert, M. B. Rota, D. Tedeschi, S. F. Covre da Silva, E. Roccia, V. Zwiller, K. D. Jöns, A. Rastelli, R. Trotta, *npj Quantum Inf.* **2021**, *7*, 7.

[491] L. Schweickert, K. D. Jöns, K. D. Zeuner, S. F. C. da Silva, H. Huang, T. Lettner, M. Reindl, J. Zichi, R. Trotta, A. Rastelli, V. Zwiller, *Appl. Phys. Lett* **2018**, *112*, 93106.

[492] C. H. Van der Wal, M. D. Eisaman, A. André, R. L. Walsworth, D. F. Phillips, A. S. Zibrov, M. D. Lukin, *Science (80-. ).* **2003**, *301*, 196.

[493] J. Appel, E. Figueroa, D. Korystov, M. Lobino, A. I. Lvovsky, *Phys. Rev. Lett.* **2008**, *100*, 093602.

[494] Y. Wang, J. Li, S. Zhang, K. Su, Y. Zhou, K. Liao, S. Du, H. Yan, S. L. Zhu, *Nat. Photonics* **2019**, *13*, 346.

[495] S. Deshpande, T. Frost, A. Hazari, P. Bhattacharya, *Appl. Phys. Lett* **2014**, *105*, 141109.

[496] S. Lazić, E. Chernysheva, Ž. Gačević, N. García-Lepetit, H. van der Meulen, In *Proceedings of SPIE*, **2015**.

[497] M. Arita, F. Le Roux, M. J. Holmes, S. Kako, Y. Arakawa, *Nano Lett* **2017**, *17*, 2902.

[498] S. Tamariz, G. Callsen, J. Stachurski, K. Shojiki, N. Grandjean, *ACS Photonics* **2020**, *7*, 1515.

[499] S. T. Tomić, J. Pal, M. A. Migliorato, R. J. Young, N. Vukmirović‖vukmirović‖, *ACS Photonics* **2015**, *2*, 958.

[500] M. J. Holmes, T. Zhu, F. C-P Massabuau, J. Jarman, R. A. Oliver, Y. Arakawa, *APL Mater.* **2021**, *9*, 61106.

[501] C. Kindel, G. Callsen, S. Kako, T. Kawano, H. Oishi, G. Hönig, A. Schliwa, A. Hoffmann, Y. Arakawa, *Phys. Status Solidi RRL* **2013**, *8*, 408.

[502] S. Xia, T. Aoki, K. Gao, M. Arita, Y. Arakawa, M. J. Holmes, *ACS Photonics* **2021**, *8*, 1656.

[503] G. Hönig, G. Callsen, A. Schliwa, S. Kalinowski, C. Kindel, S. Kako, Y. Arakawa, D. Bimberg, A. Hoffmann, *Nat. Commun.* **2014**, *5*, 5721.

[504] J. Claudon, J. Bleuse, N. S. Malik, M. Bazin, P. Jaffrennou, N. Gregersen, C. Sauvan, P. Lalanne, J.-M. Gérard, *Nat. Photonics* **2010**, *4*, 174.

[505] Y. Zhang, A. V. Velichko, H. A. Fonseka, P. Parkinson, J. A. Gott, G. Davis, M. Aagesen, A. M. Sanchez, D. Mowbray, H. Liu, *Nano Lett.* **2021**, *21*, 5722.

[506] D. Dalacu, K. Mnaymneh, J. Lapointe, X. Wu, P. J. Poole, G. Bulgarini, V. Zwiller, M. E. Reimer, *Nano Lett.* **2012**, *12*, 5919.

[507] S. Haffouz, K. D. Zeuner, D. Dalacu, P. J. Poole, J. Lapointe, D. Poitras, K. Mnaymneh, X. Wu, M. Couillard, M. Korkusinski, E. Schö, K. D. Jö, V. Zwiller, R. L. Williams, *Nano Lett* **2018**, *18*, 3047.

[508] P. Laferrière, E. Yeung, L. Giner, S. Haffouz, J. Lapointe, G. C. Aers, P. J. Poole, R. L. Williams, D. Dalacu, *Nano Lett.* **2020**, *20*, 3688.

[509] B. Lounis, H. A. Bechtel, D. Gerion, P. Alivisatos, W. E. Moerner, *Chem. Phys. Lett.* **2000**, *329*, 399.

[510] A. Saxena, Y. Chen, A. Ryou, C. G. Sevilla, P. Xu, A. Majumdar, *ACS Photonics* **2019**, *6*, 3166.

[511] S. Morozov, E. L. Pensa, A. H. Khan, A. Polovitsyn, E. Cortés, S. A. Maier, S. Vezzoli, I. Moreels, R. Sapienza, *Sci. Adv.* **2020**, *6*, 1.

[512] Q. Jiang, P. Roy, J.-B. Claude, J. Me Wenger, *Nano Lett.* **2021**, *21*, 7030.

[513] T. B. Hoang, G. M. Akselrod, M. H. Mikkelsen, *Nano Lett.* **2016**, *16*, 270.

[514] X. Lin, X. Dai, C. Pu, Y. Deng, Y. Niu, L. Tong, W. Fang, Y. Jin, X. Peng, *Nat. Commun.* **2021**, *8*, 1132.

[515] Q. A. Akkerman, G. Rainò, M. V. Kovalenko, L. Manna, *Nat. Mater.* **2018**, *17*, 394.

[516] J. C. Blancon, J. Even, C. C. Stoumpos, M. G. Kanatzidis, A. D. Mohite, *Nat. Nanotechnol.* **2020**, *15*, 969.

[517] J. Shamsi, G. Rainò, M. V. Kovalenko, S. D. Stranks, *Nat. Nanotechnol.* **2021**, *16*, 1164.

[518] H. Utzat, W. Sun, A. E. K. Kaplan, F. Krieg, M. Ginterseder, B. Spokoyny, N. D. Klein, K. E. Shulenberger, C. F. Perkinson, M. V Kovalenko, M. G. Bawendi, *Science (80-. ).* **2019**, *363*, 1068.

[519] C. Huo, C. F. Fong, M.-R. Amara, Y. Huang, B. Chen, H. Zhang, L. Guo, H. Li, W. Huang, C. Diederichs, Q. Xiong, *Nano Lett.* **2020**, *20*, 3673.

[520] M. A. Becker, R. Vaxenburg, G. Nedelcu, P. C. Sercel, A. Shabaev, M. J. Mehl, G. Michopoulos, S. G. Lambrakos, N. Bernstein, L. Lyons, T. Stöferle, F. Mahrt, M. V Kovalenko, D. J. Norris, G. Rainò, A. L. Efros, *Nature* **2018**, *553*, 189.

[521] K. Cho, T. Yamada, H. Tahara, T. Tadano, H. Suzuura, M. Saruyama, R. Sato, T. Teranishi, Y. Kanemitsu, *Nano Lett.* **2021**, *21*, 7206.

[522] S. B. Anantharaman, K. Jo, D. Jariwala, *ACS Nano* **2021**, *15*, 12628.

[523] M. Liu, N. Yazdani, M. Yarema, M. Jansen, V. Wood, E. H. Sargent, *Nat. Electron.* **2021**, *4*, 548.

[524] Y.-S. Park, S. Guo, N. S. Makarov, V. I. Klimov, *ACS Nano* **2015**, *9*, 10386.

[525] G. Rainò, G. Nedelcu, L. Protesescu, M. I. Bodnarchuk, M. V Kovalenko, R. F. Mahrt, T. Stö, *ACS Nano* **2016**, *10*, 2485.

[526] L. Hou, C. Zhao, X. Yuan, J. Zhao, F. Krieg, P. Tamarat, M. V Kovalenko, C. Guo, B. Lounis, *Nanoscale* **2020**, *12*, 6795.

[527] M. Gerhard, B. Louis, R. Camacho, A. Merdasa, J. Li, A. Kiligaridis, A. Dobrovolsky, J. Hofkens, I. G. Scheblykin, *Nat. Commun.* **2019**, *10*, 1698.

[528] I. M. Palstra, I. Maillette De Buy Wenniger, B. K. Patra, E. C. Garnett, A. Femius Koenderink, *J. Phys. Chem. C* **2021**, *125*, 12061.

[529] T. Ahmed, S. Seth, A. Samanta, *ACS Nano* **2021**, *13*, 13537.

[530] K. Kundu, P. Acharyya, K. Maji, R. Sasmal, S. S. Agasti, K. Biswas, *Angew. Chem. Int. Ed.* **2020**, *59*, 13093.

[531] L. Chouhan, S. Ito, E. M. Thomas, Y. Takano, S. Ghimire, H. Miyasaka, V. Biju, *ACS Nano* **2021**, *15*, 2831.

[532] T. Guo, R. Bose, X. Zhou, Y. N. Gartstein, H. Yang, S. Kwon, M. J. Kim, M. Lutfullin, L. Sinatra, I. Gereige, A. Al-Saggaf, O. M. Bakr, O. F. Mohammed, A. V Malko, *J. Phys. Chem. Lett.* **2019**, *10*, 6780.

[533] G. Rainò, A. Landuyt, F. Krieg, C. Bernasconi, S. T. Ochsenbein, D. N. Dirin, M. I. Bodnarchuk, M. V Kovalenko, *Nano Lett.* **2019**, *19*, 3648.

[534] P. Borri, W. Langbein, S. Schneider, U. Woggon, R. L. Sellin, D. Ouyang, D. Bimberg, *Phys. Rev. Lett.* **2001**, *87*, 157401.

[535] A. V Kuhlmann, J. Houel, A. Ludwig, L. Greuter, D. Reuter, A. D. Wieck, M. Poggio, R. J. Warburton, *Nat. Phys.* **2013**, *9*, 570.

[536] S. A. Empedocles, M. G. Bawendi, *J. Phys. Chem. B* **1999**, *103*, 1826.

[537] A. Saxena, Y. Chen, A. Ryou, C. G. Sevilla, P. Xu, A. Majumdar, *ACS Photonics* **2019**, *6*, 3166.

[538] P. B. Deotare, T. S. Mahony, V. Bulovi, *ACS Nano* **2014**, *8*, 11080.

[539] B. van Dam, C. I. Osorio, M. A. Hink, R. Muller, A. F. Koenderink, K. Dohnalova, *ACS Photonics* **2018**, *5*, 2129.

[540] W. Y. So, Q. Li, C. M. Legaspi, B. Redler, K. M. Koe, R. Jin, L. A. Peteanu, *ACS Nano* **2018**, *12*, 7232.

[541] W. Y. So, S. Abbas, Q. Li, R. Jin, L. A. Peteanu, *Nanoscale* **2021**, *13*, 15238.

[542] S. Zhao, J. Lavie, L. Rondin, L. Orcin-Chaix, C. Diederichs, P. Roussignol, Y. Chassagneux, C. Voisin, K. Müllen, A. Narita, S. Campidelli, J.-S. Lauret, *Nat. Commun.* **2018**, *9*, 3470.

[543] S. Lohrmann, A; Johnson, B C; McCallum, J C; Castelletto, *Reports Prog. Phys.* **2017**, *80*, 034502.

[544] A. Branny, S. Kumar, R. Proux, B. D. Gerardot, *Nat. Commun.* **2017**, *8*, 15053.

[545] S. Settele, F. J. Berger, S. Lindenthal, S. Zhao, A. Ali, E. Yumin, N. F. Zorn, A. Asyuda, M. Zharnikov, A. Högele, J. Zaumseil, *Nat. Commun.* **2021**, *12*, 2119.

[546] Y. Zheng, Y. Kim, A. C. Jones, G. Olinger, E. R. Bittner, S. M. Bachilo, S. K. Doorn, R. B. Weisman, A. Piryatinski, H. Htoon, *ACS Nano* **2021**, *15*, 10406.

[547] L. Zhai, M. C. Löbl, G. N. Nguyen, J. Ritzmann, A. Javadi, C. Spinnler, A. D. Wieck, A. Ludwig, R. J. Warburton, *Nat. Commun.* **2020**, *11*, 4745.

[548] J. Liu, R. Su, Y. Wei, B. Yao, S. Filipe Covre da Silva, Y. Yu, J. Iles-Smith, K. Srinivasan, Rastelli, Armando, J. Li, X. Wang, *Nat. Nanotechnol.* **2019**, *14*, 586.

[549] C. Tai Trinh, D. Nguyen Minh, K. Jun Ahn, Y. Kang, K.-G. Lee, *ACS Photonics* **2018**, *5*, 4937.

[550] D. Magde, H. Mahr, *Phys. Rev. Lett.* **1967**, *18*, 905.

[551] S. E. Harris, M. K. Oshman, R. L. Byer, *Phys. Rev. Lett.* **1967**, *18*, 732.

[552] R. Ghosh, L. Mandel, *Phys. Rev. Lett.* **1987**, *59*, 1903.

[553] A. Davoyan, H. Atwater, *Optica* **2018**, *5*, 608.

[554] X. H. Bao, Y. Qian, J. Yang, H. Zhang, Z. B. Chen, T. Yang, J. W. Pan, *Phys. Rev. Lett.* **2008**, *101*, 190501.

[555] M. Bock, A. Lenhard, C. Chunnilall, C. Becher, A. Clausen, B. Tiranov, V. B. Korzh, S. W. Verma, F. Nam, A. Marsili, P. Ferrier, H. Goldner, C. Herrmann, W. Silberhorn, M. Sohler, N. Afzelius, *Opt. Express* **2016**, *24*, 23996.

[556] M. Tokman, Z. Long, S. Almutairi, Y. Wang, V. Vdovin, M. Belkin, A. Belyanin, *APL Photonics* **2018**, *4*, 034403.

[557] C. Zou, J. Sautter, F. Setzpfandt, al -, C.-X. Wang, C. Zhang, J.-W. Jiang, M. I. Petrov, A. A. Nikolaeva, K. S. Frizyuk, N. A. Olekhno, *J. Phys. Conf. Ser.* **2018**, *1124*, 051021.

[558] M. D. Eisaman, J. Fan, A. Migdall, *Rev. Sci. Instrum* **2011**, *82*, 71101.

[559] J. Zhang, M. A. Itzler, H. Zbinden, J.-W. Pan, *Light Sci. Appl. 2015 45* **2015**, *4*, e286.

[560] M. Ghioni, A. Gulinatti, I. Rech, F. Zappa, S. Cova, *IEEE J. Sel. Top. Quantum Electron.* **2007**, *13*, 852.

[561] H. Wang, J. Guo, J. Miao, W. Luo, Y. Gu, R. Xie, F. Wang, L. Zhang, P. Wang, W. Hu, *Small* **2021**, 2103963.

[562] L. You, *Nanophotonics* **2020**, *9*, 2673.

[563] G. N. Gol'tsman, O. Okunev, G. Chulkova, A. Lipatov, A. Semenov, K. Smirnov, B. Voronov, A. Dzardanov, C. Williams, R. Sobolewski, *Appl. Phys. Lett.* **2001**, *79*, 705.

[564] A. J. Miller, A. E. Lita, S. W. Nam, *Opt. Express, Vol. 16, Issue 5, pp. 3032-3040* **2008**, *16*, 3032.

[565] H. Fang, W. Hu, *Adv. Sci.* **2017**, *4*, 1700323.

[566] J. Miao, C. Wang, *Nano Res.* **2021**, *14*, 1878.

[567] A. Gao, J. Lai, Y. Wang, Z. Zhu, J. Zeng, G. Yu, N. Wang, W. Chen, T. Cao, W. Hu, D. Sun, X. Chen, F. Miao, Y. Shi, X. Wang, *Nat. Nanotechnol.* **2019**, *14*, 217.

[568] H. Wang, S. Gao, F. Zhang, F. Meng, Z. Guo, R. Cao, Y. Zeng, J. Zhao, S. Chen, H. Hu, Y. J. Zeng, S. J. Kim, D. Fan, H. Zhang, P. N. Prasad, *Adv. Sci.* **2021**, *8*, 2100503.

[569] B. Miller, E. Parzinger, A. Vernickel, A. W. Holleitner, U. Wurstbauer, *Appl. Phys. Lett.* **2015**, *106*, 122103.

[570] M. Yamamoto, K. Ueno, K. Tsukagoshi, *Appl. Phys. Lett.* **2018**, *112*, 181902.

[571] F. Luo, M. Zhu, Y. Tan, *AIP Adv.* **2018**, *8*, 115106.

[572] K. Roy, T. Ahmed, H. Dubey, T. P. Sai, R. Kashid, S. Maliakal, K. Hsieh, S. Shamim, A. Ghosh, *Adv. Mater.* **2018**, *30*, 1704412.

[573] P. Kumar, J. Lynch, B. Song, H. Ling, F. Barrera, H. Zhang, S. B. Anantharaman, J. Digani, H. Zhu, T. H. Choudhury, C. Mcaleese, X. Wang, B. R. Conran, O. Whear, M. J. Motala, M. Snure, C. Muratore, J. M. Redwing, N. R. Glavin, E. A. Stach, A. R. Davoyan, D. Jariwala, *Light-Matter Coupling in Scalable Van der Waals Superlattices*, **2021**.

[574] E. D. Walsh, W. Jung, G.-H. Lee, D. K. Efetov, B.-I. Wu, K.-F. Huang, T. A. Ohki, T. Taniguchi, K. Watanabe, P. Kim, D. Englund, K. C. Fong, *Science (80-. ).* **2021**, *372*, 409.

[575] L. Cao, J. S. White, J.-S. Park, J. A. Schuller, B. M. Clemens, M. L. Brongersma, *Nat. Mater. 2009 88* **2009**, *8*, 643.

[576] W. Luo, Q. Weng, M. Long, P. Wang, F. Gong, H. Fang, M. Luo, W. Wang, Z. Wang, D. Zheng, W. Hu, X. Chen, W. Lu, *Nano Lett.* **2018**, *18*, 5439.

[577] A. C. Farrell, X. Meng, D. Ren, H. Kim, P. Senanayake, N. Y. Hsieh, Z. Rong, T.-Y. Chang, K. M. Azizur-Rahman, D. L. Huffaker, *Nano Lett.* **2018**, *19*, 582.

[578] A. J. Shields, M. P. O'Sullivan, I. Farrer, D. A. Ritchie, R. A. Hogg, M. L. Leadbeater, C. E. Norman, M. Pepper, *Appl. Phys. Lett.* **2000**, *76*, 3673.

[579] B. E. Kardynał, S. S. Hees, A. J. Shields, C. Nicoll, I. Farrer, D. A. Ritchie, *Appl. Phys. Lett.* **2007**, *90*, 181114.

[580] J. C. Blakesley, P. See, A. J. Shields, B. E. Kardynał, P. Atkinson, I. Farrer, D. A. Ritchie, *Phys. Rev. Lett.* **2005**, *94*, 067401.

[581] V. S. Pribiag, S. Nadj-Perge, S. M. Frolov, J. W. G. Van Den Berg, I. Van Weperen, S. R. Plissard, E. P. A. M. Bakkers, L. P. Kouwenhoven, *Nat. Nanotechnol.* **2013**, *8*, 170.

[582] X. Lu, X. Chen, S. Dubey, Q. Yao, W. Li, X. Wang, Q. Xiong, A. Srivastava, *Nat. Nanotechnol.* **2019**, *14*, 426.

[583] T. Rendler, J. Neburkova, O. Zemek, J. Kotek, A. Zappe, Z. Chu, P. Cigler, J. Wrachtrup, *Nat. Commun.* **2017**, *8*, 14701.

[584] C. Clausen, I. Usmani, F. Bussières, N. Sangouard, M. Afzelius, H. De Riedmatten, N. Gisin, *Nature* **2011**, *469*, 508.

[585] F. Muckel, S. Delikanli, P. L. Hernándezhernández-Martínez, T. Priesner, S. Lorenz, J. Ackermann, M. Sharma, H. V. Demir, G. Bacher, *Nano Lett.* **2018**, *18*, 2047.

[586] Y. Zhou, J. Chen, O. M. Bakr, H. T. Sun, *Chem. Mater.* **2018**, *30*, 6589.

[587] W. Ning, J. Bao, Y. Puttisong, F. Moro, L. Kobera, S. Shimono, L. Wang, F. Ji, M. Cuartero, S. Kawaguchi, S. Abbrent, H. Ishibashi, R. de Marco, I. A. Bouianova, G. A. Crespo, Y. Kubota, J. Brus, D. Y. Chung, L. Sun, W. M. Chen, M. G. Kanatzidis, F. Gao, *Sci. Adv.* **2020**, *6*, 5381.

[588] T. Neumann, S. Feldmann, P. Moser, A. Delhomme, J. Zerhoch, T. van de Goor, S. Wang, M. Dyksik, T. Winkler, J. J. Finley, P. Plochocka, M. S. Brandt, C. Faugeras, A. V. Stier, F. Deschler, *Nat. Commun.* **2021**, *12*, 3489.

[589] S. M. Ferro, M. Wobben, B. Ehrler, *Mater. Horizons* **2021**, *8*, 1072.

[590] X. Xi, Z. Wang, W. Zhao, J.-H. Park, K. Tuen Law, H. Berger, L. Forró, J. Shan, K. Fai Mak, *Nat. Phys.* **2016**, *12*, 139.

[591] S. Tan, Y. Zhang, M. Xia, Z. Ye, F. Chen, X. Xie, R. Peng, D. Xu, Q. Fan, H. Xu, J. Jiang, T. Zhang, X. Lai, T. Xiang, J. Hu, B. Xie, D. Feng, *Nat. Mater.* **2013**, *12*, 634.

[592] S. Iljima, T. Ichihashi, *Nature* **1993**, *363*, 603.

[593] K. Hata, D. N. Futaba, K. Mizuno, T. Namai, M. Yumura, S. Iijima, *Science (80-. ).* **2004**, *306*, 1362.

[594] J. Kong, A. M. Cassell, H. Dai, *Chem. Phys. Lett.* **1998**, *292*, 567.

[595] M. S. Arnold, A. A. Green, J. F. Hulvat, S. I. Stupp, M. C. Hersam, *Nat. Nanotechnol.* **2006**, *1*, 60.

[596] M. C. Hersam, *Nat. Nanotechnol.* **2008**, *3*, 387.

[597] M. Zheng, A. Jagota, M. S. Strano, A. P. Santos, P. Barone, S. G. Chou, B. A. Diner, M. S. Dresselhaus, R. S. McLean, G. B. Onoa, G. G. Samsonidze, E. D. Semke, M. Usrey, D. J. Watts, *Science (80-. ).* **2003**, *302*, 1545.

[598] D. Tasis, N. Tagmatarchis, A. Bianco, M. Prato, *Chem. Rev.* **2006**, *106*, 1105.

[599] C. A. Dyke, J. M. Tour, *J. Phys. Chem. A* **2004**, *108*, 11151.

[600] S. I. Azzam, K. Parto, G. Moody, *Appl. Phys. Lett* **2021**, *118*, 240502.